\documentclass{svmult}

\usepackage{makeidx}         
\usepackage{graphicx}        
\usepackage{multicol}        
\usepackage[bottom]{footmisc}

\usepackage{color}
\definecolor{darkgreen}{rgb}{0,0.5,0}
\definecolor{purple}{rgb}{0.5,0,0.5}
\definecolor{nblue}{rgb}{0.0,0.0,0.50}
\definecolor{scarlet}{rgb}{1.0,0.2,0}

\makeindex             

\newcommand{\sfrac}[2]{\mbox{\footnotesize $\displaystyle \frac{#1}{#2}$}} 
\newcommand{\lsim}{\mathrel{\rlap{\lower4pt\hbox{\hskip0pt$\sim$}} 
\raise1pt\hbox{$<$}}}           
\newcommand{\gsim}{\mathrel{\rlap{\lower4pt\hbox{\hskip0pt$\sim$}} 
\raise1pt\hbox{$>$}}}           

\begin{document}

\title*{Aspects of Hadron Physics}
\author{C.\,D.\ Roberts\inst{1} \and M.\,S.\ Bhagwat\inst{1} \and A.\ H\"oll\inst{2} \and S.\,V.\ Wright\inst{1} 
}
\institute{%
Physics Division, Argonne National Laboratory, Argonne IL 60439, USA \texttt{mbhagwat@phy.anl.gov,\,cdroberts@anl.gov,\,svwright@anl.gov}
\and 
Institut f\"ur Physik, Universit\"at Rostock, D-18051 Rostock, Germany, \texttt{arne.hoell@uni-rostock.de} 
}
%
%
%
\maketitle

\renewcommand{\theequation}{\arabic{section}.\arabic{equation}}
%

\section{Hadron Physics}
\index{Hadron physics}
\label{hadronphysics}
\setcounter{equation}{0}
Detailed investigations of the structure of hadrons are essential for understanding how matter is constructed from the quarks and gluons of Quantum chromodynamics (QCD), and amongst the questions posed to modern hadron physics, three stand out.  What is the rigorous, quantitative mechanism responsible for confinement?  What is the connection between confinement and dynamical chiral symmetry breaking?  And are these phenomena together sufficient to explain the origin of more than 98\% of the mass of the observable universe?  Such questions may only be answered using the full machinery of nonperturbative relativistic quantum field theory.  

This contribution provides a perspective on progress toward answering these key questions.  In so doing it will provide an overview of the contemporary application of Dyson-Schwinger equations in Hadron Physics, additional information on which may be found in Refs.\,\cite{cdragw,cdrsms,ralvs,pmcdr,hrw,cf06}.  The presentation assumes that the reader is familiar with the concepts and notation of relativistic quantum mechanics, with the functional integral formulation of quantum field theory and with regularisation and renormalisation in its perturbative formulation.  For these topics, in order of appearance, Refs.\,\cite{bd1,bd2,iz80,pt84} are useful.  In addition, Chaps.~1 and 2 of Ref.\,\cite{hrw} review the bulk of the necessary concepts.

Hadron physics is a key part of the international effort in basic science.  For example, in the USA we currently have the Thomas Jefferson National Accelerator Facility (JLab) and the Relativistic Heavy Ion Collider (RHIC) while in Europe hadron physics is studied at the Frascati National Laboratory and is an important part of a forthcoming pan-European initiative; namely, the Facility for Antiproton and Ion Research (FAIR) at GSI-Darmstadt.  Progress in this field is gauged via the successful completion of precision measurements of fundamental properties of hadrons; e.g., the pion, proton and neutron, and simple nuclei, for comparison with theoretical calculations to provide a quantitative understanding of their quark substructure.

The proton and neutron (collectively termed \emph{nucleons}) are fermions.  They are characterised by two static properties: an electric charge and a magnetic moment.  In Dirac's theory of pointlike relativistic fermions the magnetic moment is
\begin{equation}
\mu_D = \frac{e}{2 M}\,.
\end{equation}
The proton's magnetic moment was discovered in 1933 by Otto Stern, who was awarded the Nobel Prize in 1943 for this work \cite{nobelStern}.  He found, however, that 
\begin{equation}
\mu_p = (1 + \mbox{\underline{$1.79$}}) \, \mu_D\,.
\end{equation}
This was the first indication that the proton is not a point particle.  Of course, the proton is now explained as a bound state of quarks and gluons, a composition which is seeded and determined by the properties of three valence quarks.  These quarks and gluons are the elementary quanta of QCD.

Important hadron physics experiments measure hadron and nuclear form factors via electron scattering.  The electron current is known from QED:
\begin{eqnarray}
\label{jmue}
j_\mu(P^\prime,P)  &=& ie\,\bar u_e(P^\prime)\, { \Lambda_\mu(Q,P)} \,u_e(P)\,, \; Q=P^\prime-P  \\
& = &  i e \,\bar u_e(P^\prime)\,{ \gamma_\mu (-1)} \, u_e(P)\,, \label{eborn}
\end{eqnarray}
where $u_e$, $\bar u_e$ are Dirac spinors for a real (on-mass-shell) electron.  Equation (\ref{eborn}) is the Born approximation result, in which the dressed-electron-photon vertex is just $(- \gamma_\mu)$ and where the negative sign merely indicates the sign of the electron charge.  For the nucleons, Eq.\,(\ref{jmue}) is written
\begin{eqnarray}
\lefteqn{ J_\mu^N(P^\prime,P)  = ie\,\bar u_N(P^\prime)\, { \Lambda_\mu(Q,P)} \,u_N(P) } \\
& = &  i e \,\bar u_N(P^\prime)\,\left( \gamma_\mu F_1^N(Q^2) +
\frac{1}{2M}\, \sigma_{\mu\nu}\,Q_\nu\,F_2^N(Q^2)\right) u_N(P) \label{jmuN}
\end{eqnarray}
where $u_N$, $\bar u_N$ are on-shell nucleon spinors, and $F_{1}^N(Q^2)$ is the Dirac form factor and $F_2^N(Q^2)$ is the Pauli form factor.  The so-called Sachs form factors are defined via \index{Electromagnetic form factors: nucleon}
\begin{equation}
\label{defGEGM}
G_E(Q^2)  =  F_1(Q^2) - \frac{Q^2}{4 M^2} F_2(Q^2)\,,\; 
G_M(Q^2)  =  F_1(Q^2) + F_2(Q^2)\,.
\end{equation}
In the Breit frame in the nonrelativistic limit, the three-dimensional Fourier transform of $G_E(Q^2)$ provides the electric-charge-density distribution within the nucleon, while that of $G_M(Q^2)$ gives the magnetic-current-density distribution.  This explains their names: $G_E^N(Q^2)$ is the nucleon's electric form factor and $G_M^N(Q^2)$ is the magnetic form factor.  It is apparent via a comparison between Eqs.\,(\ref{eborn}) and (\ref{jmuN}) that $F_2\equiv 0$ for a point particle, in which case $G_E = G_M$.  This means, of course, that if the neutron is a point particle then it possesses neither a charge nor a magnetic moment; i.e., they are both zero.

There are six quarks in the Standard Model of particle physics: $u$ (up), $d$ (down), $s$ (strange), $c$ (charm), $b$ (bottom) and $t$ (top).  The first three are most important in hadron physics.  A central goal of nuclear physics is to understand the structure and properties of hadrons, and ultimately atomic nuclei, in terms of the quarks and gluons of QCD.  So, why don't we just go ahead and do it?  One of the answers is \index{Confinement} \emph{confinement}: no quark or gluon has ever been seen in isolation.  Another is dynamical chiral symmetry breaking (DCSB); e.g., the masses of the $u$, $d$ quarks in perturbative QCD provide no explanation for $\simeq 98$\% of the proton's mass.  One therefore has to ask, with quarks and gluons are we dealing with the right degrees of freedom?

The search for patterns in the hadron spectrum adds emphasis to this question.  Let's consider the proton, for example.  It has a mass of approximately $1\,$GeV.  Suppose it to be composed of three (two $u$ and one $d$) \index{Constituent-quark} \emph{constituent}-quarks, as in the \emph{eightfold way} classification of hadrons into groups on the basis of their symmetry properties \cite{gellmannNobel}.  A first guess would place the mass of these constituents at $\sim 350\,$MeV.  In the same approach, the pseudoscalar $\pi$-meson is composed of a constituent-quark and a constituent-antiquark.  It should therefore have a mass of $\sim 700\,$MeV.  However, its true mass is $\sim 140\,$MeV!  On the other hand, the mass of the vector $\rho$-meson is correctly estimated in this way: $m_\rho = 770\,$MeV.  Such mismatches are repeated in the spectrum.  \label{eightfold} 

\begin{figure}[t]
\centerline{\includegraphics[width=0.7\textwidth,angle=270]{%
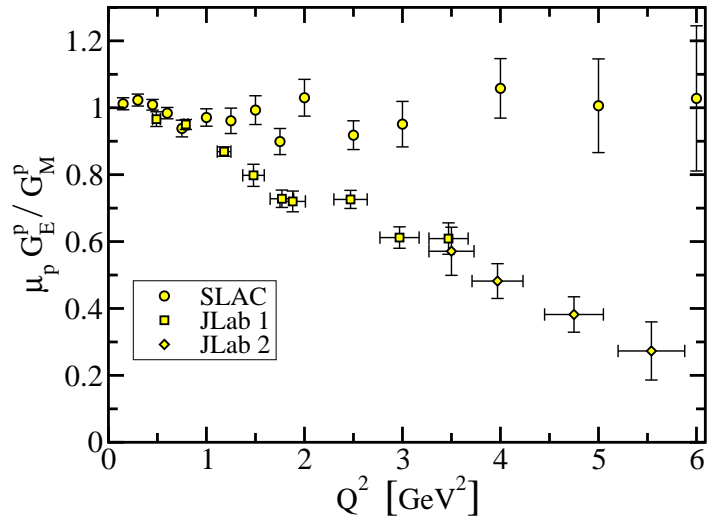}}
\caption{\label{gepgmpdata} Proton \index{Electromagnetic form factors: nucleon} form factor ratio: $\mu_p\, G_E^p(Q^2)/G_M^p(Q^2)$. The data are obtained via Rosenbluth separation, \textit{circles} - Ref.\,\protect\cite{walker}; and polarisation transfer, \textit{squares} - Ref.\,\protect\cite{jones}; \textit{diamonds} - Ref.\,\protect\cite{gayou}. (Adapted from Ref.\,\protect\cite{hrw}.)}
\end{figure}

Furthermore, modern high-luminosity experimental facilities that employ large momentum transfer reactions are providing remarkable and intriguing new information on nucleon structure \cite{gao,leeburkert}.  For an example one need only look so far as the discrepancy between the ratio of electromagnetic proton form factors, $\mu_p G_E^p(Q^2)/G_M^p(Q^2)$, extracted via the Rosenbluth separation method\,\cite{walker} and that inferred from polarisation transfer experiments \cite{jones,roygayou,gayou,arrington,qattan}, Fig.\,\ref{gepgmpdata}.  This discrepancy is marked for $Q^2\gsim 2\,$GeV$^2$ and grows with increasing $Q^2$.\footnote{It is currently believed that the discrepancy between the two classes of experiment may be removed by the inclusion of two-photon exchange contributions in the analysis of the $e p$ scattering cross-section; i.e., by improving on the Born approximation.  Estimates indicate that such effects materially reduce the magnitude of the ratio inferred from Rosenbluth separation and slightly increase that inferred from the polarisation transfer measurements.  See, e.g., Ref.\,\protect\cite{JLabRosenbluth} \label{fnrosenbluth}}  Before the JLab data were analysed it was assumed that $\mu_p \,G_E^p(Q^2)/G_M^p(Q^2)=1$ based on the seemingly sensible argument that the distribution of quark charge and the distribution of quark current should be the same.  However, if the discrepancy is truly resolved in favour of the JLab results, then we must dramatically rethink that picture.

An immediate question to ask is where does the modern theory of hadron structure stand on this issue?  A theoretical understanding might begin with a calculation of the proton's Poincar\'e covariant wave function.  (Remember, the discrepancy is marked at larger momentum transfer; viz., on the relativistic domain.)  One might think that is not a problem.  After all, the wave functions of few electron atoms can be calculated very reliably.  However, there are some big differences.  One is that the ``potential'' between light-quarks is \textbf{completely} \textbf{unknown} throughout $\simeq 98$\% of the proton's volume because in QCD perturbation theory is unreliable at length-scales $\gsim 0.2\,$fm.  Moreover, as we shall see, a reliable description of the proton's wave function will require an accurate treatment of virtual particle effects, which are a core feature of relativistic quantum field theory.  In fact, a computation of the proton's wave function requires the \emph{ab initio} solution of a fully-fledged relativistic quantum field theory.  That is yet far beyond the capacity of modern physics and mathematics.

\subsection{Aspects of QCD}
\index{Quantum Chromodynamics}
\label{formulationofqcd}
The theory one must explore is quantum chromodynamics.  It is a local, renormalisable, non-Abelian gauge theory, in which each flavour of quark comes in three colours and there are eight gauge bosons, called gluons.  It is a peculiar feature of non-Abelian gauge theories that the gauge bosons each carry the gauge charge, \emph{colour} in this case, and hence self-interact.  This is the key difference between QED, an Abelian gauge theory wherein the photons are neutral, and QCD.  The gluon self-interaction is primarily responsible for the marked difference between the running coupling in QCD and that in QED; namely, that the running coupling in QCD decreases relatively rapidly with increasing momentum transfer -- the theory is asymptotically free, whereas the QED coupling increases very slowly with growing momentum transfer.  (See, e.g., Sect.\,1.2 in Ref.\,\cite{cdrANU}.)  

The QCD action is expressed through a local Lagrangian density; viz.,
\begin{eqnarray}
\nonumber S[A_\mu^a, q, \bar{q}\,] & = & \int d^4x\,\bigg\{ \bar q(x)\left[ \gamma_\mu D_\mu + M \right]q(x)\\
&&  + \frac{1}{4} F_{\mu\nu}^a(x) F_{\mu\nu}^a(x) + \frac{1}{2\xi} \, \partial_\mu A_\mu^a(x) \, \partial_\nu A_\nu^a(x) \bigg\},
\label{Sqcd}
\end{eqnarray}
in which the first term of the second line is the chromomagnetic field strength tensor
\begin{equation}
F_{\mu\nu}^a(x)  = \partial_\mu A_\nu^a(x) -\partial_\nu A_\mu^a(x) + g f^{abc} A_\mu^b(x) A_\nu^c(x),
\end{equation}
where $\{f^{abc}:a,b,c=1,\ldots,8\}$ are the structure constants of $SU(3)$, and the second term is a covariant gauge fixing term, with $\xi$ the gauge fixing parameter: $\xi=1$ defines Feynman gauge and the limit $\xi\to 0$ implements Landau gauge in the generating functional.  The first term in Eq.\,(\ref{Sqcd}) involves the covariant derivative:
\begin{equation}
\displaystyle D_\mu = \partial_\mu - i g \frac{\lambda^a}{2} A^a_\mu(x)\,,
\end{equation}
where $\{ \mbox{\small $\frac{1}{2}$}\lambda^a;a=1,\ldots,8\}$ are the generators of $SU(3)$ for the fundamental representation; and, in addition, the current-quark mass matrix
\begin{equation}
M= \left(\begin{array}{cccc}
m_u & 0 & 0 & \ldots \\
0 & m_d & 0 & \ldots \\
0 & 0 & m_s & \ldots \\
\vdots & \vdots & \vdots & 
\end{array}\right)\,,
\end{equation}
wherein only the light-quark elements are indicated.

Understanding hadron observables means knowing all that the quantum field theory based on Eq.\,(\ref{Sqcd}) predicts.  As has been emphasised, perturbation theory is inadequate to that task because confinement, DCSB, and the formation and structure of bound states are all essentially nonperturbative phenomena.  An overview of a nonperturbative approach to exploring strong QCD in the continuum will subsequently be provided.\footnote{Almost all nonperturbative studies in relativistic quantum field theory employ a Euclidean Metric.  A Euclidean metric was used in writing Eq.\,(\protect\ref{Sqcd}).  Appendix~\ref{Appendix1} provides some background and describes the Euclidean conventions employed herein.}

\begin{table}[t]
\caption{Comparison between various hadron mass ratios, involving the $\pi$, $\rho$, $\sigma$ and $a_1$ mesons, the nucleon -- $N$, and the first radial excitations of: the $\pi$ $=\,\pi_1$; and the $\rho$ $=\,\rho_1$.  The individual masses can be found in Ref.\,\protect\cite{pdg}.}
{\normalsize
\begin{tabular*}{\hsize} {
l@{\extracolsep{0ptplus1fil}}%
l@{\extracolsep{0ptplus1fil}}%
l@{\extracolsep{0ptplus1fil}}%
}
{ $\bullet$} $\displaystyle \frac{m_\rho^2}{m_\pi^2} = 30$ & 
{ $\bullet$} $\displaystyle \frac{m_{a_1}^2}{m_{\sigma}^2} = 2.1$ & 
{ ?} Hyper{f}ine Splitting\\[2ex]
{ $\bullet$} $\displaystyle \frac{m_{\pi_1}^2}{m_\pi^2} = 86$ &
{ $\bullet$} $\displaystyle \frac{m_{\rho_1}^2}{m_{\rho}^2} = 3.5$ &
{ ?} Excitation Energy \\[2ex]
{ $\bullet$} $\displaystyle \frac{m_N}{m_\pi} \approx 7$ &
{ $\bullet$} $\displaystyle \frac{m_N}{m_\rho} = \frac{5}{4} \approx
\frac{3}{2} $ & { ?} Quark Counting\\
\end{tabular*}\label{tableSpectrum}}
\end{table}

In Table~\ref{tableSpectrum} we return to the hadron spectrum and focus on some of its features.  In a constituent-quark model the $J^{PC}=1^{--}$ $\rho$-meson is obtained from the $0^{-\,+}$ $\pi$-meson by a spin flip, yielding a vector meson state in which the spins of the constituent-quark and -antiquark are aligned.  The same procedure would yield the $1^{++}$ $a_1$-meson from the $0^{++}$ $\sigma$-meson.  Thus the difference between the masses of the $\rho$ and $\pi$, and the $a_1$ and $\sigma$ would appear to owe to an hyperfine interaction.  As the Table asks in row one, why is this interaction so much greater in the $\pi$ channel?  Another question is raised in row two: why is the radial excitation energy in the pseudoscalar channel so much greater than that in the vector channel?  And row three asks the question that was posed on page~\pageref{eightfold}: why doesn't constituent-quark counting work for the pion's mass?  

Additional questions can be posed.  The range of an interaction is inversely proportional to the mass of the boson that mediates the force.  The nucleon-nucleon interaction has a long-range component generated by the $\pi$.  The fact that the pion is so much lighter than all other hadrons composed of $u$- and $d$-quarks is crucially important in nuclear physics.   If this were not the case; viz., were the pion roughly as massive as all like-constituted hadrons, then the domain of stable nuclei would be much reduced.  In such a universe the Coulomb force would prevent the formation of elements like Fe, and planets such as ours and we, ourselves, would not exist.

\subsection{Emergent Phenomena}
\index{Emergent phenomena}
\label{emergentQCD}
A true understanding of the visible universe thus requires that we learn just what it is about QCD which enables the formation of an unnaturally light pseudoscalar meson from two rather massive constituents.  The correct understanding of hadron observables must explain why the pion is light but the $\rho$-meson and the nucleon are heavy.  The keys to this puzzle are QCD's \emph{emergent phenomena}: \textbf{confinement} \index{Confinement} and \textbf{dynamical chiral symmetry breaking}.  \index{Dynamical chiral symmetry breaking (DCSB)} Confinement is the feature that no matter how hard one strikes a hadron, it never breaks apart into quarks and/or gluons that reach a detector.  DCSB is signalled by the very unnatural pattern of bound state masses, something that is partly illustrated via Table~\ref{tableSpectrum} and the associated discussion.  Neither of these phenomena is apparent in QCD's action and yet they are the dominant determining characteristics of real-world QCD.  Attaining an understanding of these phenomena is one of the greatest intellectual challenges in physics.

In order to come to grips with DCSB it is first necessary to know the meaning of \index{Chiral symmetry} chiral symmetry.  It is a fact that, at the Lagrangian level, local gauge theories with massless fermions possess chiral symmetry.   Consider then helicity, which may be viewed as the projection of an object's spin, $\vec{j}$, onto its direction of motion, $\vec p\,$; viz., $\lambda \propto \vec{j}\cdot \vec p$.  For massless fermions, helicity is a Lorentz invariant \emph{spin observable}.  Plainly, it is either parallel or anti-parallel to the direction of motion.  

In the Dirac basis, $\gamma_5$ is the chirality operator and one may represent a positive helicity (right-handed) fermion via \begin{equation}
q_+(x) = \frac{1}{2} \left(\mbox{\boldmath $I$}_{\rm D} + \gamma_5 \right) q(x) =: P_+ \,q(x)
\end{equation}
and a left-handed fermion through
\begin{equation}
q_-(x) = \frac{1}{2} \left(\mbox{\boldmath $I$}_{\rm D} - \gamma_5 \right) q(x) =: P_-\, q(x)\,.
\end{equation}
A global \index{Chiral transformation} chiral transformation is enacted by\footnote{For this illustrative purpose it is not necessary to consider complications that arise in connection with $U(1)$ chiral anomalies, which appear via quantisation.}
\begin{equation}
\label{globalchiral} 
q(x) \to q(x)^\prime = {\rm e}^{i \gamma_5 \theta} q(x)\,,\;
\bar q(x) \to \bar q(x)^\prime = \bar q(x)\, {\rm e}^{i \gamma_5 \theta} ,
\end{equation}
and with the choice $\theta=\pi/2$ it is evident that this transformation maps $q_+ \to q_+$ and $q_- \to - q_-$.  Hence, a theory that is invariant under chiral transformations can only contain interactions that are insensitive to a particle's helicity; namely, helicity is conserved.

Consider now a composite local pseudoscalar: $\bar q(x) i \gamma_5 q(x)$.  According to Eq.\,(\ref{globalchiral}), a chiral rotation through an angle $\theta = \pi/4$ effects the transformation
\begin{equation}
\bar q(x) i \gamma_5 q(x) \to - \bar q(x) \mbox{\boldmath $I$}_{\rm D}\, q(x)\,;
\end{equation}
i.e., it turns a pseudoscalar into a scalar.  Thus the spectrum of a theory invariant under chiral transformations should exhibit degenerate parity doublets.  Is such a prediction borne out in the hadron spectrum?  Let's check (Ref.\,\cite{pdg}, masses in MeV):
\begin{equation}
\begin{array}{lcr}
\pi(J^P= 0^-,m=140) & \mbox{cf.} & \sigma(J^P= 0^+,m=600)\\
\rho(1^-,770) & \mbox{cf.} & a_1(1^+,1260)\\
N(\frac{1}{2}^+,938) & \;\mbox{cf.} \;& N(\frac{1}{2}^-,1535) 
\end{array}
\end{equation}
Quite clearly, it is not: the difference in masses between parity partners is very large, which forces a conclusion that chiral symmetry is badly broken.  Since the current-quark mass term is the only piece of the QCD Lagrangian that breaks chiral symmetry, this appears to suggest that the quarks are quite massive.  The conundrum reappears again: how can the pion be so light if the quarks are so heavy?

The extraordinary phenomena of \index{Confinement} confinement and \index{Dynamical chiral symmetry breaking (DCSB)} DCSB can be identified with properties of dressed-quark and -gluon propagators.  These two-point functions describe the in-medium propagation characteristics of QCD's elementary excitations.  Here the medium is QCD's ground state; viz., the interacting vacuum.  In propagating through this medium the quark and gluon propagators acquire momentum-dependent modifications that fundamentally alter the spectral properties of these elementary excitations.  

A mass term in the QCD Lagrangian explicitly breaks chiral symmetry.  The effect can be discussed in terms of the \index{Quark propagator} quark propagator.  For illustration it is sufficient to consider that of a noninteracting fermion of mass $m$:
\begin{equation}
S(p) = \frac{- i \gamma\cdot p + m}{p^2 + m^2}\,.
\end{equation}
On this propagator, the chiral rotation of Eq.\,(\ref{globalchiral}) is effected through
\begin{equation}
S(p) \rightarrow {\rm e}^{i \gamma_5 \theta} S(p) {\rm e}^{i \gamma_5 \theta} 
= \frac{- i \gamma\cdot p }{p^2 + m^2} + {\rm e}^{2i \gamma_5 \theta} \frac{ m}{p^2 + m^2} .
\label{chiralT}
\end{equation}
It is therefore clear that the symmetry violation is proportional to the current-quark mass and hence that the theory is chirally symmetric for $m=0$.   Another way of looking at this is to consider the fermion condensate:
\begin{equation}
\label{condensatesimple}
\langle \bar q q \rangle = \,- \,{\rm tr} \int \frac{d^4 p }{(2\pi)^4} \, S(p) \propto 
\,- \int \frac{d^4 p }{(2\pi)^4} \, \frac{m}{p^2 + m^2} .
\end{equation}
This is a quantity that can rigorously be defined in quantum field theory\,\cite{kurtcondensate} and whose strength measures the violation of chiral symmetry.  It is a standard \emph{order parameter} for chiral symmetry breaking, playing a role analogous to that of the magnetisation in a ferromagnet.

\begin{figure}[t]

\vspace*{-7em}

\centerline{\includegraphics[clip,height=1.0\textwidth]{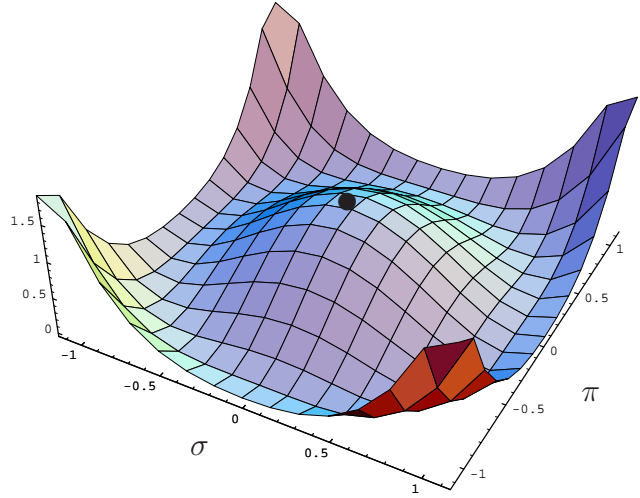}}
\vspace*{-20.7em}

\hspace*{18.3em}%
{\LARGE $\bullet$}\vspace*{6.5em}

\hspace*{28em} {\large $\pi$}\vspace*{1em}

\hspace*{12em} {\large $\sigma$}\vspace*{2em}

\caption{\label{mexicanhat} A rotationally invariant but unstable extremum of the Hamiltonian obtained with the potential in Eq.\,(\protect\ref{Emexican}). (Adapted from Ref.\,\protect\cite{cdrANU}.)}

\end{figure}

This connects immediately to dynamical symmetry breaking.  Consider a point-particle in the rotationally invariant potential 
\begin{equation}
\label{Emexican}
V(\sigma,\pi) = (\sigma^2+\pi^2-1)^2,
\end{equation}
which is illustrated in Fig. \ref{mexicanhat}.  The figure depicts a state wherein the particle is stationary at an extremum of the action.  That state is rotationally invariant but unstable.  On the other hand, in the ground state of the system the particle is stationary at any point $(\sigma,\pi)$ in the trough of the potential, for which $\sigma^2+\pi^2=1$.  There is an uncountable infinity of such vacua, $|\theta\rangle$, which are related, one to another, by rotations in the $(\sigma,\pi)$-plane.  The vacua are degenerate but not rotationally invariant and hence, in general, $\langle \theta | \sigma | \theta\rangle \neq 0\neq\langle \theta | \pi | \theta\rangle$.  In this case the rotational invariance of the Hamiltonian is not exhibited in any single ground state: the symmetry is dynamically broken with interactions being responsible for $\langle \theta | \sigma | \theta\rangle \neq 0\neq\langle \theta | \pi | \theta\rangle$.

The connection between dynamics and symmetries is now in plain view.  The elementary excitations of QCD's action are absent from the strong interaction spectrum: neither a quark nor a gluon ever reaches a detector alone.  This is the physics of confinement.  Chirality is the projection of a particle's spin onto its direction of motion.  It is a Lorentz invariant for massless fermions.  To classical QCD interactions, left-handed and right-handed quarks are indistinguishable.  This symmetry has implications for the spectrum that do not appear to be realised.  That is DCSB.  Our challenge is to understand the emergence of confinement and DCSB from the QCD Lagrangian, and therefrom describe their impact on the strong interaction spectrum and hadron dynamics.  These two phenomena need not be separate.  They are both likely to be manifestations of the same mechanism.  That mechanism must be elucidated.  It is certainly nonperturbative.

\section{Nonperturbative Tool in the Continuum}
\index{Dyson-Schwinger equations (DSEs)}
\label{dsea}
\setcounter{equation}{0}
The Dyson-Schwinger equations (DSEs) provide a nonperturbative tool for the study of continuum strong QCD.  At the simplest level they provide a generating tool for perturbation theory.  Since QCD is asymptotically free, this means that any model-dependence in the application of these methods can be restricted to the infrared or, equivalently, to the long-range domain.  In this mode, the DSEs provide a means by which to use nonperturbative strong interaction phenomena to map out, e.g., the behaviour at long range of the interaction between light-quarks.  A nonperturbative solution of the DSEs enables the study of: hadrons as composites of dressed-quarks and -gluons; the phenomena of confinement and DCSB; and therefrom an articulation of any connection between them.  The solutions of the DSEs are Schwinger functions and because all cross-sections can be constructed from such $n$-point functions the DSEs can be used to make predictions for real-world experiments.  One of the merits in this is that any assumptions employed, or guesses made, can be tested, verified and improved, or rejected in favour of more promising alternatives.  The modern application of these methods is described in Refs.\,\cite{cdrsms,ralvs,pmcdr,hrw,cf06} and a pedagogical overview provided in Ref.\,

Let's return to the \index{Quark propagator} dressed-quark propagator, which is given by the solution of QCD's gap equation that is expressed in Euclidean space as:
\begin{eqnarray}
\label{Sqdse} 
\lefteqn{S(p)^{-1} = i\gamma\cdot p \,+ m + \Sigma(p)\,,}\\
\Sigma(p) &= & \int {d^4\ell\over (2\pi)^4} \, g^2\,D_{\mu\nu}(p-\ell)\, \gamma_\mu\frac{\lambda^a}{2} \frac{1}{i\gamma\cdot \ell A(\ell^2) + B(\ell^2)} \, \Gamma_\nu^a(\ell,p). \label{qdse} 
\end{eqnarray}
At zero temperature and chemical potential the most general Poincar\'e covariant solution of this gap equation involves two scalar functions.  There are three common expressions:
\begin{equation}
\label{Sgeneral}
S(p) = \frac{1}{i\gamma\cdot p \, A(p^2) + B(p^2)} = \frac{Z(p^2)}{i \gamma\cdot p + M(p^2)} = -i \gamma\cdot p\, \sigma_V(p^2) + \sigma_S(p^2)\,,
\end{equation}
which are equivalent.  In the second form, $Z(p^2)$ is called the wave-function renormalisation and $M(p^2)$ is the dressed-quark mass function.  (A derivation of Eq.\,(\ref{Sqdse}) and a discussion of its elementary and perturbative properties is presented in Sect.~2.2 of Ref.\,\cite{hrw}.)

A weak coupling expansion of the DSEs produces every diagram in perturbation theory.  The general result in perturbation theory can be summarised
\begin{equation}
B_{\rm pert}(p^2) = m \left( 1 - \frac{\alpha}{\pi} \ln \left[\frac{p^2}{m^2}\right] + \ldots \right),
\end{equation}
where the ellipsis denotes terms of higher order in $\alpha$ that involve $(\ln [p^2/m^2])^2$ and $(\ln\ln [p^2/m^2])$, etc.  However, at arbitrarily large finite order in perturbation theory it is always true that 
\begin{equation}
\lim_{m\to 0} B_{\rm pert}(p^2) \equiv 0.
\end{equation}
This means that if one starts with a chirally symmetric theory, then in perturbation theory one also ends up with a chirally symmetric theory: the fermion DSE cannot generate a gap if there is no bare-mass seed in the first place.  Thus \index{Dynamical chiral symmetry breaking (DCSB)} DCSB is \label{impossible} impossible in perturbation theory.  Our question is whether this conclusion can ever be avoided; namely, are there circumstances under which it is possible to obtain a nonzero dressed-quark mass function in the chiral limit; viz., for $m\to 0$? \index{Chiral limit}

\subsection{Dynamical Mass Generation}
\index{Dynamical mass generation}
\label{dynamicalmass}
To begin the search for an answer, consider Eqs.\,(\ref{Sqdse}), (\ref{qdse}) with the following model forms for the dressed-gluon propagator and quark-gluon vertex:\footnote{The form for the gluon two-point function implements a four-dimensional-cutoff version of the Nambu--Jona-Lasinio model, which has long been used to model QCD at low energies, e.g., Refs.\,\protect\cite{weisenjl,klevanskynjl,ebertnjl}.  This model stands within a class of four-fermion interaction models \protect\cite{tandyrev,cahillrev}.}
\begin{eqnarray}
\label{Dnjl}
g^2 D_{\mu\nu}(p-\ell) & = &  \delta_{\mu\nu}\, 
\frac{1}{m_G^2}\,\theta(\Lambda^2-\ell^2)\,,\\
\label{Gnjl}
\Gamma_\nu^a(k,p) & = & \gamma_\mu \frac{\lambda^a}{2}\,,
\end{eqnarray}
wherein $m_G$ is some gluon mass-scale and $\Lambda$ serves as a cutoff.  The model thus obtained is not renormalisable so that the regularisation scale $\Lambda$, upon which all calculated quantities depend, plays a dynamical role.  In this case the gap equation is
\begin{eqnarray} 
\nonumber \lefteqn{ 
i\gamma\cdot p\, A(p^2) + B(p^2)= i \gamma\cdot p + m_0 }\\ 
 &  & + 
\frac{4}{3}\, \frac{1}{m_G^2}\,\int \frac{d^4\ell}{(2\pi)^4}\, 
\theta(\Lambda^2 -\ell^2)\, \gamma_\mu \,\frac{-i\gamma\cdot \ell A(\ell^2) + 
B(\ell^2)}{\ell^2 A^2(\ell^2) + B^2(\ell^2)}\, \gamma_\mu\,,  \label{njlgap2} 
\end{eqnarray} 
wherein $m_0$ is the mass that explicitly breaks chiral symmetry. 

If one multiplies Eq.\,(\ref{njlgap2}) by $(-i\gamma\cdot p)$ and subsequently evaluates a matrix trace over spinor (Dirac) indices, then one finds
\begin{equation} 
\label{Aeqnnjl}
 p^2 \, A(p^2)= p^2 + \frac{8}{3}\, \frac{1}{m_G^2}\,\int 
\frac{d^4\ell}{(2\pi)^4}\,\theta(\Lambda^2 
-\ell^2)\,p\cdot\ell\,\frac{A(\ell^2)}{\ell^2 A^2(\ell^2) + B^2(\ell^2)} \,.
\end{equation} 
It is straightforward to show that $\int d^4\ell\, p\cdot \ell = 0$; i.e., the angular integral in Eq.\,(\ref{Aeqnnjl}) vanishes, from which it follows that
\begin{equation} 
\label{Aisonenjl} A(p^2) \equiv 1\,. 
\end{equation} 
This owes to the fact that models of the Nambu--Jona-Lasinio type are defined via a four-fermion contact-interaction in configuration space, which entails momentum-independence of the interaction and therefore also of the gap equation's solution in momentum space.  

If, on the other hand, one multiplies Eq.\,(\ref{njlgap2}) by $\mbox{\boldmath $I$}_{\rm D}$, uses Eq.\,(\ref{Aisonenjl}) and subsequently evaluates a trace over Dirac indices, then \begin{equation} 
B(p^2) = m_0 + \frac{16}{3}\, \frac{1}{m_G^2}\,\int 
\frac{d^4\ell}{(2\pi)^4}\,\theta(\Lambda^2 -\ell^2)\,\frac{B(\ell^2)}{\ell^2 + 
B^2(\ell^2)}\,. \label{BMnjl} 
\end{equation} 
Since the integrand here is $p^2$-independent then a solution at one value of $p^2$ must be the solution at all values; viz., any nonzero solution must be of the form \begin{equation}
B(p^2) = {\rm constant}=M\,.
\end{equation}
Using this result, Eq.\,(\ref{BMnjl}) becomes 
\begin{eqnarray} 
M & = &  m_0 + M\,\frac{1}{3\pi^2} \, \frac{1}{m_G^2}\, { C}(M^2,\Lambda^2)\,,\\ 
\label{CMLamda} { C}(M^2,\Lambda^2) & = & \Lambda^2 - M^2 \ln\left[1+\Lambda^2/M^2\right]
\,. 
\end{eqnarray} 
Recall now that $\Lambda$ defines the mass-scale in a nonrenormalisable model.  Hence one can set $\Lambda \equiv 1$ and hereafter merely interpret all other mass-scales as being expressed in units of $\Lambda$, whereupon the gap equation becomes 
\begin{equation} 
\label{pgapnjl} M = m_0 + M\,\frac{1}{3\pi^2} \, \frac{1}{m_G^2}\, { 
C}(M^2,1)\,. 
\end{equation} 

Let us consider Eq.\,(\ref{pgapnjl}) in the chiral limit: $m_0=0$,
\begin{equation}
\label{pgapnjl0}
M = M\,\frac{1}{3\pi^2} \, \frac{1}{m_G^2}\, { 
C}(M^2,1)\,.
\end{equation}
One solution is obviously $M\equiv 0$.  This is the result that connects smoothly with perturbation theory: one starts with no mass, and no mass is generated.  In this instance the theory is said to realise chiral symmetry in the \index{Chiral symmetry: realisation, Wigner-Weyl mode} Wigner-Weyl mode.  

Suppose, however, that $M\neq 0$ in Eq.\,(\ref{pgapnjl0}).  That is possible if, and only if, the following equation has a solution:
\begin{equation}
1= \frac{1}{3\pi^2} \, \frac{1}{m_G^2}\,  { C}(M^2,1)\,.
\end{equation}
It is plain from Eq.\,(\ref{CMLamda}) that $C(M^2,1)$ is a monotonically decreasing function of $M$ whose maximum value occurs at $M=0$: $ {C}(0,1)=1$.  Consequently, $\exists M \neq 0$ solution if, and only if, 
\begin{equation}
\label{critmG1}
\frac{1}{3\pi^2} \, \frac{1}{m_G^2} > 1\,.
\end{equation}
It is thus apparent that there is always a domain of values for the gluon mass-scale, $m_G$, for which a nontrivial solution of the gap equation can be found.  If one supposes $\Lambda \sim 1\,$GeV, which is a scale that is typical of hadron physics, then $\exists M \neq 0$ solution for 
\begin{equation}
\label{critmG2}
m_G^2 < \frac{\Lambda^2}{3\pi^2} \simeq (0.2\,{\rm GeV})^2.
\end{equation}

This result, derived in a straightforward manner, is astonishing!  It reveals the power of a nonperturbative solution to nonlinear equations.  Although we started with a model of massless fermions, the interaction alone has provided the fermions with mass.  This is dynamical chiral symmetry breaking; namely, the generation of mass \textit{from nothing}.  When this happens chiral symmetry is said to be realised in the \index{Chiral symmetry: realisation, Nambu-Goldstone mode} Nambu-Goldstone mode.  It is clear from Eqs.\,(\ref{critmG1}), (\ref{critmG2}) that \index{Dynamical chiral symmetry breaking (DCSB)} DCSB is guaranteed to be possible so long as the interaction exceeds a particular minimal strength.  This behaviour is typical of gap equations.

\begin{figure}[t]
\centerline{%
\includegraphics[clip,width=0.75\textwidth]{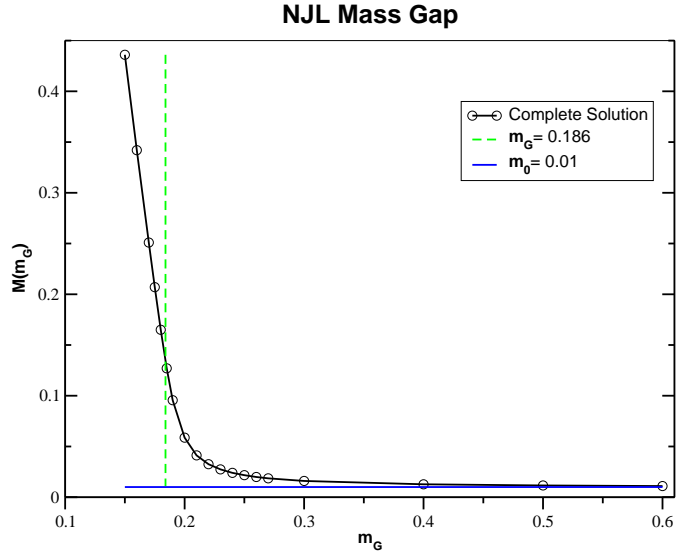}
}

\caption{\label{NJLDCSB} With a bare mass $m=0.01$, the mass gap obtained using the interaction specified by Eqs.\,(\ref{Dnjl}), (\ref{Gnjl}) -- solid curve with circles.  (All dimensioned quantities measured in units of $\Lambda$, the regularisation scale. Adapted from Ref.\,\protect\cite{hrw}.)}
\end{figure}

Figure~\ref{NJLDCSB} depicts the $m_G$-dependence of the nontrivial self-consistent solution of Eq.\,(\ref{pgapnjl}) obtained with a current-quark mass $m = 0.01$ (measured in units of $\Lambda$).  The vertical line marks 
\begin{equation}
\label{mGcr}
m_G^{\rm cr} = \frac{1}{\surd{3}\,\pi}\,;
\end{equation}
namely, the value of $m_G$ above which a DCSB solution of Eq.\,(\ref{pgapnjl0}) is impossible.  Evidently, for $m_G > m_G^{\rm cr}$, $M(m_G) \approx m_0$ and the self-consistent solution is well approximated by the perturbative result.  However, a transition takes place for $m_G \simeq m_G^{\rm cr}$, and for $m_G < m_G^{\rm cr}$ the dynamical mass is much greater than the bare mass, with $M$ increasing rapidly as $m_G$ is reduced and the effective strength of the interaction is thereby increased.

\subsection{Dynamical Mass and Confinement}
\index{Dynamical mass and confinement}
\label{confinement}
One aspect of quark confinement is the absence from the strong interaction spectrum of free-particle-like quarks.  What does the model of Sect.\,\ref{dynamicalmass} have to contribute in this connection?  Well, whether one works within the domain of the model on which DCSB takes place, or not, the \index{Quark propagator} quark propagator always has the form
\begin{equation}
S^{\rm NJL}(p)=\frac{1}{i\gamma\cdot p \, [A(p^2)=1] + [B(p^2)=M]} =  \frac{- i\gamma\cdot p  + M}{p^2 + M^2}\,,
\end{equation}
where $M$ is constant.  This expression has a pole at $p^2 + M^2 = 0$ and thus is always effectively the propagator for a noninteracting fermion with mass $M$.  Hence, while it is generally true that models of the Nambu--Jona-Lasinio type support DCSB, they do not exhibit confinement.  

Consider an alternative \cite{mn83} defined via Eq.\,(\ref{Gnjl}) and 
\begin{equation}
\label{mnprop} g^2 D_{\mu\nu}(k) = (2\pi)^4\, \check{G}\,\delta^4(k) \left[\delta_{\mu\nu} - \frac{k_\mu k_\nu}{k^2}\right].
\end{equation}
Here $\check{G}$ defines the model's mass-scale and the interaction is a $\delta$-function in momentum space, which may be compared with models of the Nambu--Jona-Lasinio type wherein the interaction is instead a $\delta$-function in configuration space.  In this instance the gap equation is 
\begin{equation} 
i\gamma\cdot p\, A(p^2) + B(p^2)  =  i \gamma\cdot p + m_0 + \check{G}\,\gamma_\mu\, 
\frac{-i\gamma\cdot p\, A(p^2) + B(p^2)}{p^2 A^2(p^2) + B^2(p^2)}\, 
\gamma_\mu\,,\label{mngap1} 
\end{equation} 
which yields the following coupled nonlinear algebraic equations: 
\begin{eqnarray} 
A(p^2) & = & 1 + 2 \, \frac{A(p^2)}{p^2 A^2(p^2) + B^2(p^2)} \, ,\label{AMN}\\ 
B(p^2) & = & m_0 + 4\, \frac{B(p^2)}{p^2 A^2(p^2) + B^2(p^2)} \,. \label{BMN}
\end{eqnarray} 
Equation (\ref{mngap1}) yields an ultraviolet finite model and hence there is no regularisation mass-scale.  In this instance one can therefore refer all dimensioned quantities to the model's mass-scale and set $\check{G}=1$. 

Consider the chiral limit of Eq.\,(\ref{BMN}):
\begin{equation} 
B(p^2)  =   4\, \frac{B(p^2)}{p^2 A^2(p^2) + B^2(p^2)}\,. \label{BMN0} 
\end{equation} 
Obviously, like Eq.\,(\ref{pgapnjl0}), this equation admits a trivial solution $B(p^2) \equiv 0$ that is smoothly connected to the perturbative result, but is there another? 
The existence of a $B\not\equiv 0$ solution; i.e., a solution that dynamically 
breaks chiral symmetry, requires (in units of $\check{G}$)
\begin{equation} 
p^2 A^2(p^2) + B^2(p^2) = 4\,.
\end{equation} 
Suppose this identity to be satisfied, then its substitution into Eq.~(\ref{AMN}) gives \begin{equation} 
A(p^2) - 1 = \frac{1}{2}\,A(p^2) \; \Rightarrow \; A(p^2) \equiv 2\,, 
\end{equation} 
which in turn entails 
\begin{equation} 
B(p^2) = 2\,\sqrt{1 - p^2}\,. 
\end{equation} 

A complete chiral-limit solution is composed subject to the physical requirement that the quark self-energy is real on the spacelike momentum domain, and hence 
\begin{eqnarray} 
\label{AMNres}
A(p^2) &= & \left\{ \begin{array}{ll} 
2\,;\; & p^2\leq 1\\ 
\frac{1}{2}\left( 1 + \sqrt{1+8/p^2} \right)\,;\; & p^2>1 \end{array} \right.\\ 
\label{BMNres}
B(p^2) &= & \left\{ \begin{array}{ll} 
2 \sqrt{1-p^2}\,;\; & p^2\leq 1\\ 
0\,; & p^2>1 \,.\end{array} \right. 
\end{eqnarray} 
In this case both scalar functions characterising the \index{Quark propagator} dressed-quark propagator differ significantly from their free-particle forms and are momentum dependent.  (We will see that this is also true in QCD.)  It is noteworthy that the magnitude of the model's mass-scale plays no role in the appearance of this \index{Dynamical chiral symmetry breaking (DCSB)} DCSB solution of the gap equation; viz., reinstating $\check G$, one has in the chiral limit $B(0) = 2 \check G \neq 0$ $\forall \check G \neq 0$.  Thus, in models of the Munczek-Nemirovsky type, the interaction is always strong enough to support the generation of mass from nothing.

The DCSB solution of Eqs.\,(\ref{AMNres}), (\ref{BMNres}) is defined and continuous for all $p^2$, including timelike momenta, $p^2<0$.  It gives a dressed-quark propagator whose denominator
\begin{equation} 
p^2\,A^2(p^2) + B^2(p^2) > 0\,, \; \forall \, p^2\,. 
\end{equation} 
This is a novel and remarkable result, which means that the propagator does not exhibit any free-particle-like poles!  It has long been argued that this feature can be interpreted as a realisation of quark confinement; e.g., \index{Confinement} Refs.\,\cite{stingl,cudell,entire1,entire2,krein,mandula}. 

The Munczek-Nemirovsky interaction has taken a massless quark and turned it into something which at timelike momenta bears little resemblance to the perturbative quark.  It does that for all nonzero values of the model's mass-scale.  In this model one exemplifies an intriguing possibility that all models with quark-confinement necessarily exhibit DCSB.  It is obvious from Sect.\,\ref{dynamicalmass} that the converse is certainly not true.

\begin{figure}[t]
\centerline{%
\includegraphics[clip,width=0.75\textwidth]{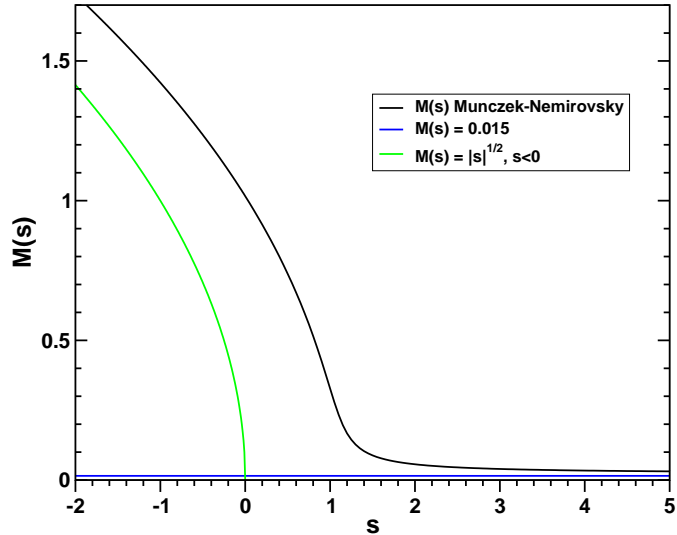}
}

\caption{\label{MNmassfn} With a bare mass $m_0=0.015$, the dressed-quark mass function \mbox{$M(s=p^2)$} obtained using the interaction specified by Eqs.\,(\ref{Gnjl}), (\ref{mnprop}) -- solid curve.  A free-particle-like mass-pole would occur at that value of $s$ for which the solid and green curves intersect.  As the figure suggests, this never happens in models of the Munczek-Nemirovsky type.  (All dimensioned quantities measured in units of $\check{G}$, the model's mass-scale.  Adapted from Ref.\,\protect\cite{hrw}.)}
\end{figure}

In the chirally asymmetric case the gap equation yields 
\begin{eqnarray} 
A(p^2) & = & \frac{2 \,B(p^2)}{ m_0+ B(p^2) }\,,\\ 
B(p^2) & = & m_0 + \frac{4\, [m_0 + B(p^2)]^2}{B(p^2) ([m_0+B(p^2)]^2 + 4 p^2)}\,. 
\end{eqnarray} 
The second of these is a quartic equation for $B(p^2)$.  It can be solved algebraically.  There are four solutions, obtained in closed form, only one of which possesses the physically sensible ultraviolet spacelike behaviour: $B(p^2) \to m_0$ as \mbox{$p^2\to \infty$}.  The physical solution is depicted in Fig.\,\ref{MNmassfn}.  At large spacelike momenta, $M(s=p^2)\to m_0^+$ and a perturbative analysis is reliable.  That is never the case for $s\lsim 1$ (in units of $\check{G}$), on which domain $M(s)\gg m_0$ and the difference $(M(s) - |s|)$ is always nonzero, a feature that is consistent with confinement.

\begin{figure}[t]
\centerline{%
\includegraphics[clip,width=0.8\textwidth]{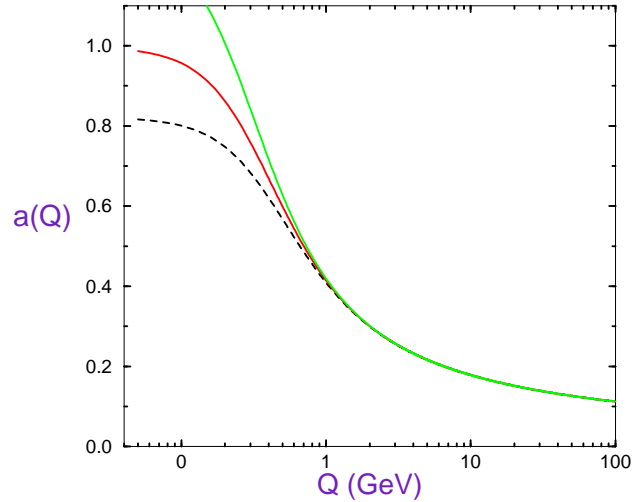}
}\vspace*{-2ex}

\caption{\label{generalinteraction} Two classes of effective interaction in the gap equation.  A theory will exhibit dynamical chiral symmetry breaking if, and only if, $a(Q=0) \geq 1$.  Here the ultraviolet behaviour of the interactions is constrained to be that of QCD.}
\end{figure}

Two variations on the theme of dynamical mass generation via the gap equation have been illustrated.  A general class of models for asymptotically free theories may be discussed in terms of an effective interaction
\begin{equation}
g^2 D(Q^2) = 4\pi \frac{a(Q^2)}{Q^2}\,,
\end{equation}
where $a(Q^2)$ is such that $g^2 D(Q^2)$ evolves according to QCD's renormalisation group in the ultraviolet.  This situation is depicted in \protect\index{Dynamical chiral symmetry breaking} Fig.\,\ref{generalinteraction}.  The types of effective interaction fall into two classes.  In those for which $a(0) < 1$, the only solution of the gap equation in the chiral limit is \mbox{$M(s) \equiv 0$}.  Whereas, when $a(0)\geq 1$, $M(s)\neq 0$ is possible in the chiral limit and, indeed, corresponds to the energetically favoured ground state.  This behaviour was anticipated long ago in the context of attempts to understand QED at strong coupling \cite{bjw,fk}.  That theory unmodified, however, is incomplete \cite{reenders}.

It is the right point to recall the conundrums described in Sect.\,\ref{hadronphysics}. We have seen how confinement and DCSB can emerge in the nonperturbative solution of a theory's DSEs.  Therefore theories whose classical Lagrangian possesses no mass-scale can, in fact, via DCSB, behave like theories with large quark masses and can also exhibit confinement.  This understanding is a step toward a resolution of the riddles posed by strong interaction physics.  

\subsection{Dynamical chiral symmetry breaking and a critical mass}
\index{Critical mass and DCSB}
\label{dcsbmc}
A novel aspect of the interplay between explicit and dynamical chiral symmetry breaking has recently been elucidated \cite{leichang}.  To relay this, suppose DCSB takes place so that $B(p^2) \not \equiv 0$.  Then with the choice $\theta = \pi/2$, Eq.\,(\ref{chiralT}) corresponds to mapping $B(p^2) \to - B(p^2)$.  It follows that if $B(p^2)$ is a solution of the gap equation in the chiral limit, then so is $[- B(p^2]$.  While these two solutions are distinct, the chiral symmetry entails that each yields the same pressure \footnote{The pressure is defined as the negative of the effective-action.  Hence, the effective-action difference is zero between two vacuum configurations of equal pressure.  A system's ground state is that configuration for which the pressure is a global maximum or, equivalently, the effective-action is a global minimum.  An elucidation may be found in Ref.\,\protect\cite{haymaker}.} and hence they correspond to equivalent vacua.  This is an analogue of the chiral-limit equivalence between the $(\sigma =1,\pi=0)$ and $(\sigma =-1,\pi=0)$ vacua in Fig.\,\ref{mexicanhat}, as elucidated in Refs.\,\cite{reggcm,cdrqed}.\footnote{NB.\ More generally, given a solution of the $m_0 > 0$ gap equation characterised by $\{A_{m_0}(p^2),B_{m_0}(p^2)\}$, then $\{A_{-m_0}(p^2),-B_{-m_0}(p^2)\}$ is a solution of the gap equation obtained with $[-m_0]$.}

In common with Sects.\,\ref{dynamicalmass} and \ref{confinement}, studies of DCSB have hitherto focused on a positive definite solution of the gap equation because the introduction of a positive current-quark bare-mass favours this solution; viz., if another solution exists, then it has a lower pressure.  Returning again to Fig.\,\ref{mexicanhat}, such a bare-mass tilts this so-called wine-bottle potential, producing a global minimum at $(\sigma =1,\pi=0)$.  However, whether the massive gap equation admits solutions other than that which is positive definite, the effect of the current-quark mass on such solutions, and their interpretation, are questions little considered.

\subsubsection{Exemplar}
\index{Critical mass and DCSB: NJL model}
\label{Exsecnjl}
In order to begin addressing these questions, one may return to the model introduced in Sect.\,\ref{dynamicalmass}.  The gap equation in this model, Eq.\,(\ref{pgapnjl}), can be rewritten as 
\begin{equation} 
\label{pgapnjlR} G(M):= M - m_0 - M\,\frac{1}{3\pi^2} \, \frac{1}{m_G^2}\, { 
C}(M^2,1) = 0
\end{equation} 
and, as noted above, an $M\neq 0$ solution exists when $m_0=0$ if and only if $m_G$ satisfies Eq.\,(\ref{mGcr}).  Hence, to proceed let us choose
\begin{equation}
\label{usedmG1}
m_G^2 = \frac{3}{4}\, \frac{1}{3\pi^2}\,.
\end{equation}

\begin{figure}[t]
\centerline{
\includegraphics[clip,width=0.75\textwidth]{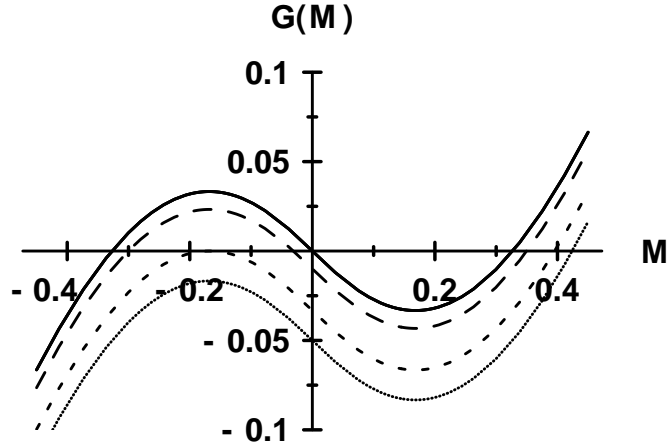}}
\vspace*{2ex}

\caption{\label{figNJL} The zeros of $G(M)$ give the solution of the gap equation defined by Eqs.\,(\protect\ref{pgapnjl}), (\protect\ref{usedmG1}).
\textit{Solid curve}: obtained with $m_0=0$, in which case $G(M)$ is odd under $M\to -M$; \textit{long-dashed curve}: $m_0=0.01$; \textit{short-dashed curve}: $m_0=m_0^{\rm cr} = 0.033$; \textit{dotted curve}: $m_0=0.05$.
(All dimensioned quantities in units of $\tilde\Lambda$.  Adapted from Ref.\,\protect\cite{leichang}.)}
\end{figure}

For this case $G(M)$ is plotted in Fig.\,\ref{figNJL}.  One reads from the figure that in the chiral limit there are three solutions to Eq.\,(\ref{pgapnjlR}):
\begin{equation}
\label{gapsolnjl}
M = \left\{
\begin{array}{l}
M_W = 0\,, \\
M_\pm = \pm M^0 = \pm 0.33\,.
\end{array}\right.
\end{equation}
As observed after Eq.\,(\ref{pgapnjl0}), $M_W$ describes a realisation of chiral symmetry in the Wigner mode.  It corresponds to the vacuum configuration in which the possibility of DCSB is not realised; i.e., the position of the ball in Fig.\,\ref{mexicanhat}.  In the chiral limit this is the only solution accessible in perturbation theory.  The solutions $M_\pm$ are essentially nonperturbative.  Plainly, they represent the realisation of chiral symmetry in the Nambu-Goldstone mode; namely, DCSB. \index{Dynamical chiral symmetry breaking (DCSB)}

$M=M_+>0$ is the solution usually tracked in connection with QCD phenomenology; e.g., it is the solution plotted in Fig.\,\ref{NJLDCSB}.   In models of this type $M=M_+$ is identified as a \index{Constituent-quark mass} constituent-quark mass.  As $m_0$ is increased, $M_+$ also increases. 

As evident in Fig.\,\ref{figNJL}, the other two solutions of the gap equation do not immediately disappear when $m_0$ increases from zero.  Nor do they always persist.  Instead, these solutions exist on a domain 
\begin{equation}
{\cal D}(m_0)= \{m_0\; | \;0 \leq m_0 < m_0^{\rm cr} \}.
\end{equation}
$M_W,M_-$ also evolve smoothly with $m_0$.  Moreover, at the critical current-quark mass, $m_0^{\rm cr}$, these two solutions coalesce.  

To understand the origin of a critical mass one observes from Eq.\,(\protect\ref{pgapnjlR}) and Fig.\,\ref{figNJL} that introducing a current-quark mass merely produces a constant pointwise negative shift in the curve $G(M)$.  Hence, the critical current-quark mass is that value of this mass for which $G(M)=0$ at its local maximum.  The local maximum occurs at 
\begin{equation}
M_{\rm lm} = \{M \;|\; M<0\,,\; G^\prime(M)\} = 0
\end{equation}
and therefore in the present illustration the critical current-quark mass 
\begin{equation}
\label{mbmcr}
m^{\rm cr}_{0} = \{ m_0 \; | \;G(M_{\rm lm}) = 0 \} = 0.033 \,.
\end{equation}
An interpretation of $m^{\rm cr}_{0}$ follows shortly.

\begin{figure}[t]
\centerline{
\includegraphics[width=0.75\textwidth]{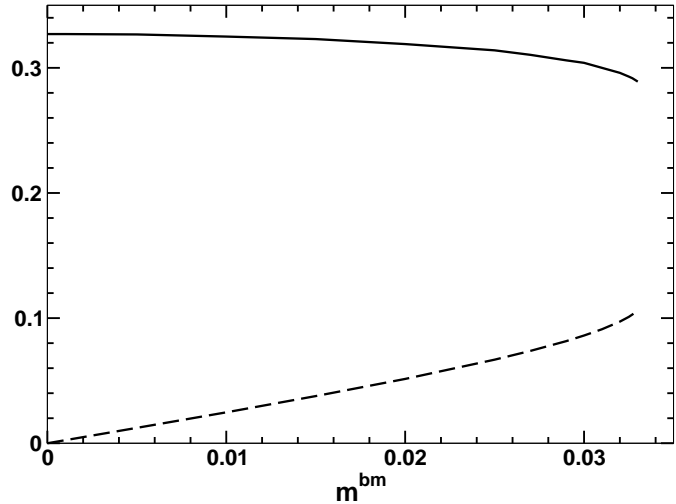}}
\vspace*{2ex}

\caption{\label{figbarM} Evolution with current-quark mass of $\bar M$ (solid curve) and $\check M$ (dashed curve).
(All dimensioned quantities in units of $\tilde\Lambda$.  Adapted from Ref.\,\protect\cite{leichang}.)}
\end{figure}

The two combinations 
\begin{equation}
\label{MbMc}
\bar M := \frac{1}{2} \left(M_+ - M_-\right)\,,\; \check M := \frac{1}{2} \left(M_+ + M_-\right)
\end{equation}
are of interest.  In the chiral limit, $\bar M = M^0$ and $\check M = M_W$.  The evolution of each with current-quark mass is depicted in Fig.\,\ref{figbarM}.  It is apparent that $\bar M$ is continuous on ${\cal D}(m_0)$ and evolves from the DCSB solution with increasing $m_0$: $\bar M(m_0)$ is a monotonically decreasing function.  These features can be understood as illustrating that the essentially dynamical component of chiral symmetry breaking decreases with increasing current-quark mass.  

The alternative combination, $\check M$, is also continuous on ${\cal D}(m_0)$.  With the value of the coupling given in Eq.\,(\ref{usedmG1}), $\check M(m_0)$ evolves from the Wigner solution according to
\begin{equation}
\check M \stackrel{m_0 \sim 0}{=}  m_0\left[1 + \frac{2}{3}\, \frac{3 (M^0 )^2 - 1 }{(M^0 )^2+ 1 }\right]+ \ldots\ \;.
\end{equation}
The development of $\check M(m_0)$ might be viewed as a gauge of the destabilising effect that DCSB has on this model's Wigner mode.

What now can one make of the critical current-quark mass, Eq.\,(\ref{mbmcr})? $M_W(m_0)$ can be calculated perturbatively in the neighbourhood of $m_0=0$:
\begin{equation}
 M_W(m_0) = - 3 \, m_0 + \ldots\ \;.
\end{equation}
However, with increasing current-quark mass $M_W$ decreases steadily toward $M_-$, which is an essentially nonperturbative quantity, until at $m_0^{\rm cr}$, $M_W =M_-$.  At this point a solution which began perturbatively has melded with a solution that is inaccessible in perturbation theory and actually characteristic of DCSB.  A related view sees $M_-$ as a DCSB solution whose modification by a current-quark mass can no longer be evaluated perturbatively when that mass exceeds $m_{0}^{\rm cr}$.  These observations suggest that $m_{0}^{\rm cr}$ specifies an upper bound on the domain within which, for physically relevant quantities, a perturbative expansion in the current-quark mass can be valid; i.e., it is a (possibly weak) upper bound on the radius of convergence for such an expansion.  This view and the numerical result in Eq.\,(\ref{mbmcr}) coincide with Ref.\,\cite{hatsuda}.  It is natural to ask whether analogous behaviour exists in QCD, and this is a question that will be addressed.  First, however, additional concepts must be introduced.

\section{Gap equation in QCD}
\index{Gap equation in QCD}
\label{gapQCD}
\setcounter{equation}{0}
The gap equation in QCD may \emph{properly} be written
\begin{equation}
\label{gendse} S(p)^{-1} = Z_2 \,(i\gamma\cdot p + m^{\rm bare}) +\, Z_1
\int^\Lambda_q \, g^2 D_{\mu\nu}(p-q) \frac{\lambda^a}{2}\gamma_\mu S(q)
\Gamma^a_\nu(q;p) \,.
\end{equation}
This is the renormalised DSE for the \index{Quark propagator} dressed-quark propagator wherein $D_{\mu\nu}(k)$ is the renormalised dressed-gluon propagator, $\Gamma^a_\nu(q;p)$ is the renormalised dressed-quark-gluon vertex, $m^{\rm bare}$ is the $\Lambda$-dependent current-quark bare mass that appears in the QCD Lagrangian and \mbox{$\int^\Lambda_q := \int^\Lambda d^4 q/(2\pi)^4$} represents a \textit{Poincar\'e-invariant} regularisation of the integral, with $\Lambda$ the regularisation mass-scale.  In addition, $Z_1(\zeta^2,\Lambda^2)$ and $Z_2(\zeta^2,\Lambda^2)$ are the quark-gluon-vertex and quark wave function renormalisation constants, which depend on the renormalisation point, $\zeta$, and the regularisation mass-scale.  The solution of Eq.\,(\ref{gendse}) is obtained subject to the renormalisation condition
\begin{equation}
\label{renormS} \left.S(p)^{-1}\right|_{p^2=\zeta^2} = i\gamma\cdot p +
m(\zeta)\,,
\end{equation}
where $m(\zeta)$ is the renormalised mass: 
\begin{equation}
Z_2(\zeta^2,\Lambda^2) \, m^{\rm bare}(\Lambda) = Z_4(\zeta^2,\Lambda^2) \, m(\zeta)\,,
\end{equation}
with $Z_4$ the Lagrangian mass renormalisation constant.  In QCD the chiral limit is unambiguously defined by \index{Chiral limit: QCD} \index{Current-quark mass}
\begin{equation}
\label{limchiral}
Z_2(\zeta^2,\Lambda^2) \, m^{\rm bare}(\Lambda) \equiv 0 \,, \forall \Lambda \gg \zeta \,,
\end{equation}
which is equivalent to stating that the renormalisation-point-invariant current-quark mass vanishes; i.e., $\hat m = 0$.

The discussion in Sect.\,\ref{dsea} leads to the key question: what is the behaviour of the kernel of QCD's gap equation, Eq.\,(\ref{gendse}).  This kernel is constituted by the contraction of the dressed-gluon propagator and the dressed-quark-gluon vertex:
\begin{equation}
D_{\mu\nu}(p-q) \, \Gamma_\nu(q)\,.
\end{equation}

\begin{figure}[t]
\centerline{%
\includegraphics[clip,width=0.75\textwidth]{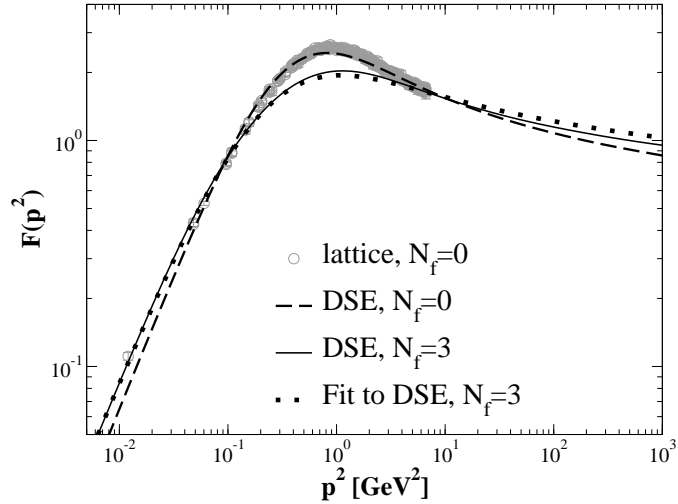}
}

\caption{\label{gluoncf} Solid curve -- $F(p^2)$ in Eq.\,(\protect\ref{gluonZ}) obtained through the solution of one particular truncation of the coupled ghost-gluon-quark DSEs with $N_f=3$ flavours of massless quark.  Dashed curve -- DSE solution with the same truncation but omitting coupling to the quark DSE.  Open circles -- quenched lattice-QCD results.  NB.\ There is sufficient lattice data at intermediate spacelike $p^2$ for the open circles to appear as a grey band.  Within the lattice errors, the quenched DSE and lattice results are indistinguishable.  (Adapted from Ref.\,\protect\cite{alkoferdetmold}.)}
\end{figure}

In Landau gauge the two-point gluon Schwinger function can be expressed \index{Gluon propagator}
\begin{equation}
\label{gluonZ}
D_{\mu\nu}(p)= \left(\delta_{\mu\nu} - \frac{p_\mu p_\nu}{p^2}\right) \frac{F(p^2)}{p^2}\,,
\end{equation}
where $F(p^2)$ involves the gluon vacuum polarisation discussed, e.g., in Sect.\,2.1 of Ref.\,\cite{hrw}.  The modern DSE perspective on $F(p^2)$ is reviewed in Refs.\,\cite{ralvs,cf06} and the predictions described therein were verified in contemporary simulations of lattice-regularised QCD \cite{latticegluon}.  The agreement is illustrated in Fig.\,\ref{gluoncf}.  

The DSE result depicted in Fig.\,\ref{gluoncf} describes a gluon two-point function that is suppressed at small $p^2$; i.e., in the infrared.  This deviation from expectations based on perturbation theory becomes apparent at $p^2 \simeq 1\,$GeV$^2$.  A mass-scale of this magnitude has long been anticipated as characteristic of nonperturbative gauge-sector dynamics.  Its origin is fundamentally the same as that of $\Lambda_{\rm QCD}$, which appears in perturbation theory.  This phenomenon whereby the value of a dimensionless quantity becomes essentially linked to a dynamically generated mass-scale is sometimes called \emph{dimensional transmutation}.  

\begin{figure}[t]
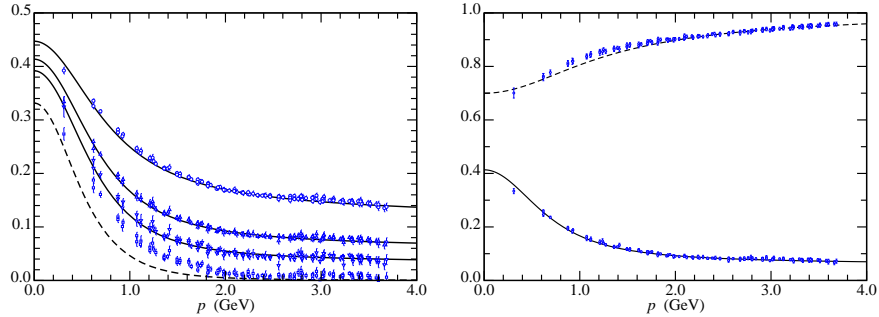

\centerline{%
\resizebox{0.47\textwidth}{!}{\includegraphics{CDRFigs/fig1mb.eps}}\hspace*{1em}
\resizebox{0.47\textwidth}{!}{\includegraphics{CDRFigs/fig3mb.eps}}
}

\caption{\label{quarkcf} \textit{Left Panel} -- Dashed-curve: gap 
equation's solution in the chiral limit; solid curves: solutions for $M(p^2)$ 
obtained using the current-quark masses in Eq.\ (\protect\ref{amvalues}).  Data, upper three sets: lattice results for $M(p^2)$ in GeV at $am$ values in Eq.\,(\protect\ref{amvalues}); lower points (boxes): linear extrapolation of these results to $a m =0$ \protect\cite{bowman2}. \textit{Right Panel} -- Dashed curve, $Z(p^2)$, and solid curve, $M(p^2)$ calculated from the gap equation with $m(\zeta)=55\,$MeV.  Data, quenched lattice-QCD results for $M(p^2)$ and $Z(p^2)$ obtained with $am = 0.036$ \protect\cite{bowman2}.  \index{Quark propagator} (Adapted from Ref.\,\protect\cite{bhagwat}.) \index{Quark propagator cf.\ lattice simulation}}
\end{figure}

The dressed-quark two-point function has the form presented in Eq.\,(\ref{Sgeneral}).  One knows that for a free particle $Z(p^2)=1$ and $M(p^2) = m_{\rm current}$.  Asymptotic freedom \index{Asymptotic freedom} in QCD entails that at some large spacelike $p^2= \zeta^2$ the quark propagator is exactly the free propagator \textit{except} that the bare mass is replaced by the renormalised mass.  At one-loop order, the vector part of the dressed self-energy is proportional to the running gauge parameter.  In Landau gauge, that parameter is zero.  Hence, in this gauge the vector part of the renormalised dressed-quark self-energy vanishes at one-loop order in perturbation theory.  The same is true for a charged fermion in QED.   

On the other hand, the behaviour of these functions on the nonperturbative domain in QCD is a longstanding prediction of DSE studies \cite{cdragw} and could have been anticipated from Refs.\,\cite{lane,politzer}.  These DSE predictions, too, are confirmed in numerical simulations of lattice-QCD \cite{bowman2,bowman}, and the conditions have been explored under which pointwise agreement between DSE results and lattice simulations may be obtained \cite{alkoferdetmold,bhagwat,bhagwat2}.  This agreement is illustrated in Fig.\,\ref{quarkcf}, wherein the nonzero current-quark masses are 
\begin{equation} 
\label{amvalues} 
\begin{array}{l|lll} 
a\,m_{\rm lattice} ~&~ 0.018 & 0.036 & 0.072 \\
\hline m(\zeta) ~({\rm GeV}) &~
0.030 & 0.055 & 0.110 
\end{array}\,.
\end{equation} 
The top row here lists the values used in the quenched-QCD lattice simulations, with $a$ the lattice spacing so that $ a m$ is dimensionless \cite{bowman2} and the second row provides the matched current-quark masses used in the DSE study, with the renormalisation scale $\zeta = \zeta_{19} = 19\,$GeV \cite{bhagwat}.

Figure \ref{quarkcf} confirms that \index{Dynamical chiral symmetry breaking (DCSB)} DCSB is a reality in QCD.  At ultraviolet momenta the magnitude of the mass function is determined by the current-quark mass.  In the infrared, however, for light-quarks, $M(p^2)$ is orders-of-magnitude larger.  The DSE analysis alone, and its correlation of lattice data, indicates that the mass function is nonzero and retains its magnitude in the chiral limit.  In Fig.\,\ref{quarkcf} and subsequently Fig.\ref{Mpsqplot} the evolution in QCD from current-quark to constituent-quark is in plain view.

What \label{confpage} about confinement?  \index{Confinement} It is notable that confinement between static quarks is a quantum mechanical archetype.  Numerical simulations of lattice-regularised QCD exhibit a linear potential at long range between such infinitely heavy sources/sinks of colour charge.  The physics of confinement in the light-quark domain, however, is not obviously amenable to such a description.  For example, a veracious realisation of dynamical chiral symmetry breaking is impossible in quantum mechanical potential models.  Moreover, in the presence of light- (dynamical-) quarks, the breaking of the string that characterises the potential between static quarks appears to be an instantaneous process, with delocalised light-quark-antiquark pair production \cite{gunnarbali}. Thus either no information is to be had about a string between light-quarks, or such a string cannot exist. 

Another perspective on confinement has already been mentioned; namely, that this phenomenon might be expressed in the analyticity properties of the dressed propagators.  In fact it is sufficient for confinement that dressed propagators for coloured excitations not possess a Lehmann representation, since this is associated with a violation of reflection positivity.  \index{Reflection positivity} An excitation connected with a propagator that violates reflection positivity cannot appear in the Hilbert space of physical states; viz., it won't propagate to a detector.  These notions may be traced from Refs.\, \cite{stingl,cudell,entire1,entire2,krein,mandula} and are described in Refs.\,\cite{cdragw,cdrsms,ralvs}.  (See also, e.g., the discussion of the reconstruction theorem in Ref.\,\cite{glimm}.)  Any Schwinger function that exhibits an inflexion point cannot be expressed through a Lehmann representation.  An inspection of the DSE and lattice results for $F(p^2)/p^2$ and $\sigma_S(p^2)$ suggests strongly that the dressed-gluon and -quark propagators display an inflexion point.  Moreover, in QCD there are DSE studies which suggest that quark confinement and DCSB both owe to the same dynamical mechanism \cite{dsenonzeroT,dsenonzeromu}, and therefore that one does not appear without the other.

In connection with Fig.\,\ref{generalinteraction} in Sect.\,\ref{confinement} a description was provided of a class of effective interactions in the gap equation that can generate DCSB.  The dressed-gluon propagator obtained from $F(p^2)$ in Fig.\,\ref{gluoncf}, combined with Eq.\,(\ref{Gnjl}), does not yield an interaction that is a member of that class \cite{papavassiliou,cdrwien,hawes}.  How then is it possible that Ref.\,\cite{bhagwat} unified the gluon and quark two-point functions?  The problem was anticipated in Ref.\,\cite{hawes} and an answer suggested; namely, the \index{Quark-gluon vertex} dressed-quark-gluon vertex must exhibit an enhancement in the infrared.  This is precisely the means employed in Ref.\,\cite{bhagwat}.   The exact nature of the enhancement and its origin in QCD is the subject of contemporary research \cite{bhagwat2,jisvertex,bhagwatvertex}.

\section{Bound States}
\index{Bound states: mesons}
\label{BoundStates}
\setcounter{equation}{0}

\subsection{Bethe-Salpeter Equation}
\index{Bethe-Salpeter equation}
\label{BSEQCD}
It will now be apparent that a semi-quantitatively reliable picture of QCD's key propagators and vertices is established.  What about bound states?  Without them, of course, a direct comparison with experiment is impossible.  Bound states appear as pole contributions to colour-singlet Schwinger functions and this observation may be viewed as the origin of the Bethe-Salpeter equation.  The Bethe-Salpeter equation (BSE) has an history that predates QCD, which can be traced, e.g., from Ref.\,\cite{iz80}, Chap.\,10.  

The axial-vector vertex is of primary interest to hadron physics.  It may be obtained as the solution of the inhomogeneous Bethe-Salpeter equation
\begin{equation}
\label{avbse}
\left[\Gamma_{5\mu}(k;P)\right]_{tu}
 =  Z_2 \left[\gamma_5\gamma_\mu\right]_{tu} + \int^\Lambda_q
[S(q_+) \Gamma_{5\mu}(q;P) S(q_-)]_{sr} K_{tu}^{rs}(q,k;P)\,,
\end{equation}
where $q_\pm = q \pm P/2$ and the colour-, Dirac- and flavour-matrix structure of the elements in the equation is denoted by the indices $r,s,t,u$.  In Eq.\,(\ref{avbse}), $K(q,k;P)$ is the fully-amputated quark-antiquark scattering kernel.  It is one-particle-irreducible and hence, by definition, does not contain quark-antiquark to single gauge-boson annihilation diagrams, such as would describe the leptonic decay of the pion, nor diagrams that become disconnected by cutting one quark and one antiquark line.  If one knows the form of $K$ then one completely understands the nature of the interaction between quarks in QCD.

\subsection{Model-independent results}
\label{modelindependent}
\subsubsection{Pseudoscalar meson mass formula}
Much has been made of chiral symmetry in the preceding discussion.  In quantum field theory, chiral symmetry and the pattern by which it is broken are expressed via the chiral Ward-Takahashi identity: \index{Ward-Takahashi identity: axial-vector}
\begin{equation}
\label{avwtim}
P_\mu \Gamma_{5\mu}^H(k;P)  = \check{S}(k_+)^{-1} i \gamma_5\frac{T^H}{2}
+  i \gamma_5\frac{T^H}{2} \check{S}(k_-)^{-1}
- i\,\{ {M}^\zeta ,\Gamma_5^H(k;P) \} ,
\end{equation}
where the pseudoscalar vertex is given by
\begin{equation}
\label{genpve}
\left[\Gamma_{5}^H(k;P)\right]_{tu} =
Z_4\,\left[\gamma_5 \frac{T^H}{2}\right]_{tu} \,+
\int^\Lambda_q \,
\left[ \chi_5^H(q;P)\right]_{sr}
K^{rs}_{tu}(q,k;P)\,,
\end{equation}
with $\chi_5^H(q;P) = \check{S}(q_+) \Gamma_{5}^H(q;P) \check{S}(q_-)$, $\check{S}= {\rm diag}[S_u,S_d,S_s,\ldots]$ and $M ^\zeta = {\rm diag}[m_u(\zeta),m_d(\zeta),m_s(\zeta),\ldots]$.  

Equations~(\ref{avwtim}), (\ref{genpve}) are written for the case of a flavour-nonsinglet vertex in a theory with $N_f$ quark flavours.  The matrices $T^H$ are constructed from the generators of $SU(N_f)$ with, e.g., \mbox{$T^{\pi^+}=\mbox{\small $\frac{1}{2}$} (\lambda^1+i\lambda^2)$} providing for the flavour content of a positively charged pion.  Writing the equations in this manner is straightforward.  However, a unified description of light- and heavy-quark systems is not.  Truncations and approximations that are reliable in one sector need not be valid in the other.

The axial-vector Ward-Takahashi identity relates the solution of a BSE to that of the gap equation.  If the identity is always to be satisfied and in a model-independent manner, as it must be in order to preserve an essential symmetry of the strong interaction and its breaking pattern, then the kernels of the gap and Bethe-Salpeter equations must be intimately related.  Any truncation or approximation of these equations must preserve that relation.  This is an extremely tight constraint.  Perturbation theory is one systematic truncation that, order by order, guarantees Eq.\,(\ref{avwtim}).  However, as has been emphasised, perturbation theory is inadequate in the face of QCD's emergent phenomena.  Something else is needed. 

Fortunately, at least one systematic, nonperturbative and symmetry preserving truncation of the DSEs exists \cite{bhagwatvertex,munczek,truncscheme}.  \index{Truncation of DSEs}  This makes it possible to prove \index{Goldstone theorem} Goldstone's theorem in QCD \cite{mrt98}.  Namely, when chiral symmetry is dynamically broken: the axial-vector vertex, Eq.\,(\ref{avbse}), is dominated by the pion pole for $P^2\sim 0$ and the homogeneous, isovector, pseudoscalar BSE has a massless ($P^2 = 0$) solution.  The converse is also true, so that DCSB \index{Dynamical chiral symmetry breaking (DCSB)} is a sufficient and necessary condition for the appearance of a massless pseudoscalar bound state of dynamically-massive constituents which dominates the axial-vector vertex for infrared total momenta.  

Furthermore, from the axial-vector Ward-Takahashi identity and the existence of a systematic, nonperturbative symmetry-preserving truncation, one can prove the following identity involving the mass-squared of a pseudoscalar meson \cite{mrt98}: \index{Pseudoscalar meson mass formula}
\begin{equation}
\label{gengmor}
f_H \, m_H^2 = 
\rho_H(\zeta) {M}_H^\zeta,
\end{equation}
where ${M}_H^\zeta = m_{q_1}(\zeta) + m_{q_2}(\zeta)$ is the sum of the
current-quark masses of the meson's constituents;
\begin{equation}
\label{fH}
f_H \, P_\mu = Z_2 {\rm tr} \int_q^\Lambda \! \sfrac{1}{2} (T^H)^T \gamma_5
\gamma_\mu \check{S}(q_+)\, \Gamma^H(q;P)\, \check{S}(q_-)\,,
\end{equation}
where $(\cdot)^{\rm T}$ indicates matrix transpose, and
\begin{equation}
\label{qbqH}
\rho_H(\zeta) =   Z_4\, {\rm tr}\int_q^\Lambda \!
\sfrac{1}{2}(T^H)^T \gamma_5 \check{S}(q_+) \,\Gamma^H(q;P)\, \check{S}(q_-) \,.
\end{equation}
The renormalisation constants in Eqs.\,(\ref{fH}), (\ref{qbqH}) play a pivotal role.  Indeed, the expressions would be meaningless without them.  They serve to guarantee that the quantities described are gauge invariant and finite as the regularisation scale is removed to infinity, which is the final step in any calculation.  Moreover, $Z_2$ in Eq.\,(\ref{fH}) and $Z_4$ in Eq.\,(\ref{qbqH}) ensure that both $f_H$ and the product $\rho_H(\zeta) M_H^\zeta$ are renormalisation point independent, which is an absolute necessity for any observable quantity.

Taking note that in a Poincar\'e invariant theory a pseudoscalar meson Bethe-Salpeter amplitude \index{Bethe-Salpeter amplitude: pion} assumes the form
\begin{eqnarray}
\nonumber
\lefteqn{i \Gamma_{H}^j(k;P) = T^H \gamma_5
\left[ i E_H(k;P)  + \gamma\cdot P \, F_H(k;P)\right.} \\
 & &
\left.+ \, \gamma\cdot k \,k\cdot P\, G_H(k;P)+
 \sigma_{\mu\nu}\,k_\mu P_\nu \,H_H(k;P) \right],
\label{genpvv}
\end{eqnarray}
then, in the chiral limit, one can also prove that \index{Goldberger-Treiman relations: pseudoscalar mesons}
\begin{eqnarray}
\label{bwti}
f_H E_H(k;0)  &= &  B(k^2)\,, \\
 F_R(k;0) +  2 \, f_H F_H(k;0)  & = & A(k^2)\,,
 \label{fwti}\\
G_R(k;0) +  2 \,f_H G_H(k;0)    & = & 2 A^\prime(k^2)\,,
\label{gwti}\\
\label{hwti}
H_R(k;0) +  2 \,f_H H_H(k;0)    & = & 0\,.
\end{eqnarray}
The functions $F_R$, $G_R$, $H_R$ are associated with terms in the axial-vector vertex that are regular in the neighbourhood of $P^2 +m_H^2 = 0$ and do not vanish at $P_\mu = 0$.  These four identities are quark-level Goldberger-Treiman relations for pseudoscalar mesons.  They are exact in QCD and are a pointwise expression of Goldstone's theorem.  These identities relate the pseudoscalar meson Bethe-Salpeter amplitude directly to the \index{Quark propagator} dressed-quark propagator.  Equation (\ref{bwti}) explains why DCSB and the appearance of a Goldstone mode are so intimately connected, and Eqs.\,(\ref{fwti}) -- (\ref{hwti}) entail that in general a pseudoscalar meson Bethe-Salpeter amplitude has what might be called pseudovector components; namely: $F_H$, $G_H$, $H_H$.  The latter are connected with the  presence of quark orbital angular momentum \index{Orbital angular momentum} in the pion, which is a necessary feature of a Poincar\'e covariant treatment, and guarantee that the electromagnetic pion form factor behaves as $1/Q^2$ at large spacelike momentum transfer \cite{farrar,mrpion}.

Equation (\ref{gengmor}) and its corollaries are of fundamental importance in QCD.  To exemplify let's focus first on the chiral limit behaviour of Eq.\,(\ref{qbqH})
whereat, using Eqs.\,(\ref{genpvv}), (\ref{bwti})-(\ref{hwti}), one finds readily
\begin{equation}
f_H^0\,\rho^0_H(\zeta) = 
Z_4(\zeta,\Lambda)\, N_c\, {\rm tr}_{\rm D} \int^\Lambda_q  S_{\hat m =0}(q) = - \langle \bar q q \rangle_\zeta^0  \,, \label{qbq0}
\end{equation}
where $f_H^0$ is the chiral limit value from Eq.\,(\ref{fH}), which is nonzero when chiral symmetry is dynamically broken.  Equation (\ref{qbq0}) is unique as the expression for the chiral limit \textit{vacuum quark condensate}, \index{Quark condensate: vacuum} and is the true definition of the order parameter first described in Eq.\,(\ref{condensatesimple}).  It thus follows from Eqs.\,(\ref{gengmor}), (\ref{qbq0}) that in the neighbourhood of the chiral limit \index{Gell-Mann--Oakes--Renner relation} 
\begin{equation}
\label{gmor}
(f_H^0)^2 \, m_H^2 = - \, M_H^\zeta\, \langle \bar q q \rangle_\zeta^0 +
{\rm O}(\hat M^2)\,.
\end{equation}
Hence what is commonly known as the Gell-Mann--Oakes--Renner relation is a \textit{corollary} of Eq.\,(\ref{gengmor}).

Let's now consider another extreme; viz., when one of the constituents is a heavy quark, a domain on which Eq.\,(\ref{gengmor}) is equally valid.  \index{Heavy-light pseudoscalar mesons} In this case Eq.\,(\ref{fH}) yields the model-independent result \cite{marismisha} 
\begin{equation}
\label{fHheavy}
f_H \propto \frac{1}{\sqrt{M_H}}\,;
\end{equation}
i.e., it reproduces a well-known consequence of \index{Heavy-quark symmetry} heavy-quark symmetry \cite{neubert93}.  A similar analysis of Eq.\,(\ref{qbqH}) gives a new result \cite{marisAdelaide,mishaSVY}
\begin{equation}
\label{qbqHheavy}
- \langle \bar q q \rangle^H_\zeta = \mbox{constant} +
  O\left(\frac{1}{m_H}\right) \mbox{~for~} \frac{1}{m_H} \sim 0\,.
\end{equation}
Combining Eqs.\,(\ref{fHheavy}), (\ref{qbqHheavy}), one finds \cite{marisAdelaide,mishaSVY}
\begin{equation}
m_H \propto \hat m_Q \;\; \mbox{for} \;\; \frac{1}{\hat m_Q} \sim 0\,,
\end{equation}
where $ \hat m_Q$ is the renormalisation-group-invariant current-quark mass of the flavour-nonsinglet pseudoscalar meson's heaviest constituent.  This is the result one would have anticipated from constituent-quark models but here is indicated a direct proof in QCD. 

\subsubsection{Pseudoscalar meson excited states}
\index{Pseudoscalar meson excited states}
Pseudoscalar mesons hold a special place in QCD and there are three states, composed of $u$,$d$ quarks, in the hadron spectrum with masses below $2\,$GeV \cite{pdg}: $\pi(140)$; $\pi(1300)$; and $\pi(1800)$.  Of these, the pion [$\pi(140)$] is naturally well known and much studied.  The other two are observed, e.g., as resonances in the coherent production of three pion final states via pion-nucleus collisions \cite{experiment}.  In the context of a model constituent-quark Hamiltonian, these mesons are often viewed as the first three members of a $Q\bar Q$ $n\, ^1\!S_0$ trajectory, where $n$ is the principal quantum number; i.e., the $\pi(140)$ is viewed as the $S$-wave ground state and the others are its first two radial excitations.  By this reasoning the properties of the $\pi(1300)$ and $\pi(1800)$ are likely to be sensitive to details of the long-range part of the quark-quark interaction because the constituent-quark wave functions will possess material support at large interquark separation.\footnote{One might note that, in comparison with $\pi(1300)$, the $\pi(1800)$ is narrow, with a width of $207\pm 13\,$MeV \protect\cite{pdg}, and has a decay pattern that may be consistent with its interpretation as a \emph{hybrid} meson in constituent-quark models \cite{page}.  This picture has the constituent-quarks' spins aligned to produce $S_{Q\bar Q} = 1$ with $J=0$ obtained by coupling $S_{Q\bar Q}$ to a spin-$1$ excitation of the confinement potential.  From this perspective, too, the properties of $\pi(1800)$ are sensitive to the interaction responsible for light-quark confinement.}
Hence the development of an understanding of their properties may provide information about \index{Confinement} light-quark confinement, which complements that obtained via angular momentum excitations\,\cite{a1b1,watson}.  As has been emphasised, the development of an understanding of confinement is one of the most important goals in physics.

We have already seen that Eq.\,(\ref{gengmor}) is a powerful result.  That is further emphasised by the fact that it is applicable here, too \cite{krassnigg1,krassnigg2}.  The result holds at each pole common to the pseudoscalar and axial-vector vertices and therefore it also impacts upon the properties of non ground state pseudoscalar mesons.  Let's work with a label $n\geq 0$ for the pseudoscalar mesons: $\pi_n$, with $n=0$ denoting the ground state, $n=1$ the state with the next lowest mass, and so on.  Plainly, by assumption, $m_{\pi_{n\neq 0}}>m_{\pi_0}$ , and hence $m_{\pi_{n\neq 0}} > 0$ in the chiral limit.  Moreover, the ultraviolet behaviour of the quark-antiquark scattering kernel in QCD guarantees that Eq.\,(\ref{qbqH}) is cutoff independent.  Thus
\begin{equation}
0 < \rho_{\pi_{n}}^0(\zeta):= \lim_{\hat m\to 0} \rho_{\pi_{n}}(\zeta) < \infty \,, \; \forall \, n\,.
\end{equation}
Hence, it is a necessary consequence of chiral symmetry and its dynamical breaking in QCD; viz., Eq.\,(\ref{gengmor}), that \index{Leptonic decay constant: pseudoscalar excited states}
\begin{equation}
\label{fpinzero}
f_{\pi_n}^0 \equiv 0\,, \forall \, n\geq 1\,.
\end{equation}
This result is consistent with Refs.\,\cite{dominguez}, as appreciated in Ref.\,\cite{volkov}. It means that in the presence of \index{Dynamical chiral symmetry breaking (DCSB)} DCSB all pseudoscalar mesons except the ground state decouple from the weak interaction.  

This argument is legitimate in any theory that has a valid chiral limit.  It is logically possible that such a theory does not exhibit DCSB; i.e., realises chiral symmetry in the Wigner-Weyl mode, as was illustrated in Sect.\,\ref{dynamicalmass} and is the case in QCD above the critical temperature for chiral symmetry restoration \cite{cdrsms}.  Equation (\ref{gengmor}) is still valid in the Wigner phase.  However, its implications are different; namely, in the Wigner phase, one has 
\begin{equation}
\label{Bm0}
B^W(0,\zeta^2)\propto m(\zeta) \propto\hat m\,;
\end{equation}
i.e., the mass function and constituent-quark mass vanish in the chiral limit.  This result is accessible in perturbation theory.  Equation (\ref{bwti}) applies if there is a massless bound state in the chiral limit.  Suppose such a bound state persists in the absence of DCSB.  It then follows from Eqs.\,(\ref{bwti}) and (\ref{Bm0}) that \begin{equation}
f^W_{\pi_0} \propto \hat m\,.
\end{equation}
In this case the leptonic decay constant of the ground state pseudoscalar meson also vanishes in the chiral limit, and hence all pseudoscalar mesons are blind to the weak interaction.

It is always true that 
\begin{equation}
f_{\pi_0}\,\rho_{\pi_{0}}(\zeta) \stackrel{\hat m\approx 0}{\propto} - \langle \bar q q \rangle_\zeta^0 \,.
\end{equation}
In the Wigner phase\,\cite{kurtcondensate}, $ \langle \bar q q \rangle^{0\,W}_\zeta \propto \hat m^3$.  Hence, via Eq.\,(\ref{gengmor}), if a rigorously chirally symmetric theory possesses a massless pseudoscalar bound state then\footnote{cf.\ The case of DCSB; i.e., the Nambu phase, wherein $m^N_{\pi_o} \stackrel{\hat m\approx 0}{\propto} \surd \hat m$, Eq.\,(\ref{gmor}).}
\begin{equation}
m^W_{\pi_0} \stackrel{\hat m\approx 0}{\propto} \hat m \,.
\end{equation}
In this case there is also a degenerate scalar meson partner whose mass behaves in the same manner.

\subsubsection{Pseudoscalar meson two photon decays}
\index{Pseudoscalar meson two photon decays}
In the presence of DCSB the ground state neutral pseudoscalar meson decays predominantly into two photons, a process connected with the \index{Abelian anomaly} Abelian anomaly.  Given that $f_{\pi_n \neq 0} \equiv 0$ in the chiral limit, it is natural to ask whether $\pi_{n\neq 0} \to \gamma\gamma$ is similarly affected.  Since rainbow-ladder is the leading order in a symmetry preserving truncation, it can be used to provide a model-independent analysis of this process.  In this instance, the two-photon coupling for all $u,d$ pseudoscalar mesons, including $n=0$, is described by the renormalised triangle diagrams
\begin{eqnarray}
\nonumber T^3_{5\mu\nu\rho}(k_1,k_2) &=& {\rm tr}\int_\ell^M \check{S}(\ell_{0+}) \, \Gamma^3_{5\rho}(\ell_{0+},\ell_{-0}) \, \check{S}(\ell_{-0}) \\%
 & \times&  \, i \check{Q}\Gamma_\mu(\ell_{-0},\ell) \, \check{S}(\ell) \, i \check{Q}\Gamma_\nu(\ell,\ell_{0+})\,,\label{Tmnr}\\
\nonumber T^3_{5\mu\nu}(k_1,k_2) &=& {\rm tr}\int_\ell^M \check{S}(\ell_{0+}) \, \Gamma^3_{5}(\ell_{0+},\ell_{-0}) \, \check{S}(\ell_{-0}) \\
 &\times&  \, i \check{Q}\Gamma_\mu(\ell_{-0},\ell) \, \check{S}(\ell) \, i \check{Q}\Gamma_\nu(\ell,\ell_{0+})\,,\label{Pmnr}
\end{eqnarray}
where $\ell_{\alpha\beta}=\ell+\alpha k_1+\beta k_2$, the electric charge matrix $\check{Q}={\rm diag}[e_u,e_d]=e\,{\rm diag}[2/3,-1/3]$, $\check{S}= {\rm diag}[S_u,S_d]$ and 
\begin{equation}
\label{photonvertex}
\left[\Gamma_{\mu}(k;P)\right]_{tu} = Z_2 \left[\gamma_\mu  \right]_{tu}\\
+ \int^\Lambda_q
[\check{S}(q_+) \Gamma_{\mu}(q;P) \check{S}(q_-)]_{sr} K_{tu}^{rs}(q,k;P) 
\end{equation}
is the renormalised dressed-quark-photon vertex.\footnote{This is an inhomogeneous BSE for the dressed-quark-photon vertex.  Its lowest mass pole-contribution is the ground-state $\rho$-meson \cite{marisphotonvertex}.  That fact underlies the success of vector meson dominance phenomenology.}  Is is subsequently assumed that $m_u(\zeta)=m(\zeta)=m_d(\zeta)$ so that $S_u =S_d$.

The dressed-quark propagators in Eqs.\,(\ref{Tmnr}) -- (\ref{photonvertex}) are understood to be calculated using a rainbow-truncation gap equation, and $K$ is a ladder-like quark-antiquark scattering kernel that ensures Eq.\,(\ref{avwtim}) is satisfied.  At this point, nothing more need be said of these elements.  The results described are independent of the details discussed in Sect.\,(\ref{practical}).

\begin{figure}[t]
\centerline{%
\includegraphics[width=0.95\textwidth]{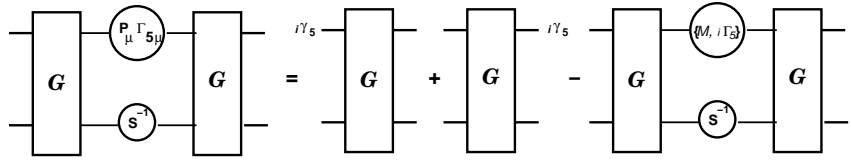}}
\caption{\label{bicudoavwti} This axial-vector Ward-Takahashi identity is an analogue of Eq.\,(\protect\ref{avwtim}).  It is valid if, and only if: the dressed-quark propagator, $S$, is obtained from Eq.\,(\ref{rainbowdse}); the axial-vector vertex, $\Gamma_{5\mu}$, is obtained from Eq.\,(\protect\ref{avbse}) with a ladder-like kernel, $K$, that ensures Eq.\,(\ref{avwtim}); the pseudoscalar vertex is constructed analogously; and the unamputated renormalised quark-antiquark scattering matrix: $G= (SS) + (SS)K(SS) + (SS)K(SS)K(SS)+ [\ldots]$, is constructed from the elements just described.  (Adapted from Ref.\,\protect\cite{krassnigg2}.)}
\end{figure}

Under the conditions just described one may derive\,\cite{bicudo} the Ward-Takahashi identity for the quark-antiquark scattering matrix, $G$, depicted in Fig.\,\ref{bicudoavwti}.  Using this identity it can be shown \cite{lc03}
\begin{equation}
\label{anomaly}
P_\rho T^3_{5\mu\nu\rho}(k_1,k_2) + 2 i m(\zeta) \, T^3_{5\mu\nu}(k_1,k_2) = \frac{\alpha}{2 \pi}\, \varepsilon_{\mu\nu\rho\sigma} k_{1\rho} k_{2\sigma}\,,
\end{equation}
where $\alpha= e^2/(4\pi)$.  This is an explicit demonstration that the triangle-diagram representation of the axial-vector--two-photon coupling calculated in the rainbow-ladder truncation is a necessary and sufficient pairing to preserve the \index{Abelian anomaly} Abelian anomaly.

In general the coupling of an axial-vector current to two photons is described by a six-point Schwinger function, to which Eq.\,(\ref{Tmnr}) is an approximation.  The same is true of the pseudoscalar--two-photon coupling and its connection with Eq.\,(\ref{Pmnr}).  Equation~(\ref{anomaly}) is valid for any and all values of $P^2=(k_1+k_2)^2$.  It is an exact statement of a divergence relation between these two six-point Schwinger functions, which is preserved by the rainbow-ladder truncation, and any systematic improvement thereof.  It is now possible to report corollaries of Eq.\,(\ref{anomaly}) that have important implications for the properties of pseudoscalar bound states.

It follows from the discussion presented above that, unless there is a reason for the residue to vanish, every isovector pseudoscalar meson appears as a pole contribution to the axial-vector and pseudoscalar vertices \cite{mrt98}:
\begin{eqnarray}
\left. \Gamma_{5 \mu}^j(k;P)\right|_{P^2+m_{\pi_n}^2 \approx 0}&=&   \frac{f_{\pi_n} \, P_\mu}{P^2 + 
m_{\pi_n}^2} \Gamma_{\pi_n}^j(k;P) +\; \Gamma_{5 \mu}^{j\,{\rm reg}}(k;P) \,, \label{genavv} \\
\left. i\Gamma_{5 }^j(k;P)\right|_{P^2+m_{\pi_n}^2 \approx 0}
&=&   \frac{\rho_{\pi_n}(\zeta) }{P^2 + 
m_{\pi_n}^2} \Gamma_{\pi_n}^j(k;P) + \; i\Gamma_{5 }^{j\,{\rm reg}}(k;P) \,, \label{genpv} 
\end{eqnarray}
where the residues are given in Eqs.\,(\ref{fH}), (\ref{qbqH}); viz., each vertex may be expressed as a simple pole plus terms regular in the neighbourhood of this pole, with $\Gamma_{\pi_n}^j(k;P)$ being the bound state's canonically normalised Bethe-Salpeter amplitude.\footnote{Canonical normalisation of the bound-state Bethe-Salpeter amplitude is discussed, e.g., in Sect.\,10-2-1 of Ref.\,\protect\cite{iz80} and Ref.\,\protect\cite{llewellyn}.  The normalisation condition assumes a simple form when the two-body scattering amplitude is independent of the total momentum, as illustrated in Eq.\,(\protect\ref{BSEnorm}). \label{fnnorm}}

If one inserts Eqs.\,(\ref{genavv}) and (\ref{genpv}) into Eq.\,(\ref{anomaly}), and uses Eq.\,(\ref{gengmor}), it follows that in the neighbourhood of each electric-charge-neutral pseudoscalar-meson bound-state pole  
\begin{equation}
P_\rho T_{5\mu\nu\rho}^{3\,{\rm reg}}(k_1,k_2) + 2 i m(\zeta) \, T_{5\mu\nu}^{3\,{\rm reg}}(k_1,k_2)
  + f_{\pi_n} \,T^{\pi_n^0}_{\mu\nu}(k_1,k_2) = \frac{\alpha}{2 \pi} i\varepsilon_{\mu\nu\rho\sigma} k_{1\rho} k_{2\sigma}\,.
\label{reganomaly}
\end{equation}
In this equation, $T^{3\,{\rm reg}}(k_1,k_2)$ are nonresonant or \emph{continuum} contributions to the relevant Schwinger functions, whose form is concretely illustrated upon substitution of $\Gamma_{5 \mu}^{j\,{\rm reg}}(k;P)$ and $\Gamma_{5}^{j\,{\rm reg}}(k;P)$ into Eqs.\,(\ref{Tmnr}) and (\ref{Pmnr}), respectively.  Moreover, $T^{\pi_n^0}_{\mu\nu}$ is the six-point Schwinger function describing the bound state contribution, which in rainbow-ladder truncation is realised as   
\begin{eqnarray}
\nonumber T^{\pi_n^0}_{\mu\nu}(k_1,k_2)  &=& {\rm tr}\int_\ell^{M\to\infty} \!\! \check{ S}(\ell_{0+}) \, \Gamma_{\pi_n^0}(\ell_{-\frac{1}{2}\frac{1}{2}};P) \, \check{ S}(\ell_{-0}) \\%
&\times&  \, i\check{Q}\Gamma_\mu(\ell_{-0},\ell) \, \check{S}(\ell) \, i \check{ Q}\Gamma_\nu(\ell,\ell_{0+}).
\label{Tpingg}
\end{eqnarray}
This Schwinger function describes the direct coupling of a pseudoscalar meson to two photons.  The support properties of the bound state Bethe-Salpeter amplitude guarantee that the renormalised Schwinger function is finite so that the regularising parameter can be removed; i.e., $M\to \infty$.

Owing to the $O(4)$ (Euclidean Lorentz) transformation properties of each term on the l.h.s.\ in Eq.\,(\ref{anomaly}), one may write
\begin{eqnarray}
P_\rho T_{5\mu\nu\rho}^{3\,{\rm reg}}(k_1,k_2) & = & \frac{\alpha}{\pi} i\varepsilon_{\mu\nu\rho\sigma} k_{1\rho} k_{2\sigma}\,A^{3\,{\rm reg}}(k_1,k_2) \,,\; \\
T_{5\mu\nu}^{3\,{\rm reg}}(k_1,k_2) & = & \frac{\alpha}{\pi} i\varepsilon_{\mu\nu\rho\sigma} k_{1\rho} k_{2\sigma}\,P^{3\,{\rm reg}}(k_1,k_2)\,,\; \\
T^{\pi_n^0}_{\mu\nu}(k_1,k_2) & = & \frac{\alpha}{\pi} i\varepsilon_{\mu\nu\rho\sigma} k_{1\rho} k_{2\sigma}\,G^{\pi_n^0}(k_1,k_2)\,, \; \label{TGdef}
\end{eqnarray}
so that Eq.\,(\ref{anomaly}) can be compactly expressed as
\begin{equation}
\label{reganomaly0}
A^{3\,{\rm reg}}(k_1,k_2) + 2 i m(\zeta) P^{3\,{\rm reg}}(k_1,k_2) + f_{\pi_n} G^{\pi_n^0}(k_1,k_2) = \frac{1}{2}.
\end{equation}
It follows from Eqs.\,(\ref{fpinzero}), (\ref{reganomaly}) that in the chiral limit all pseudoscalar mesons, \emph{except} the Goldstone mode, decouple from the divergence of the axial-vector--two-photon vertex.  

In the chiral limit the pole associated with the ground state pion appears at $P^2=0$ and thus  
\begin{equation}
\left. P_\rho T_{5\mu\nu\rho}^{3}(k_1,k_2)\right|_{P^2\neq 0}
  = \left.P_\rho T_{5\mu\nu\rho}^{3\,{\rm reg}}(k_1,k_2)\right|_{P^2\neq 0} = \frac{\alpha}{2 \pi}\, i\varepsilon_{\mu\nu\rho\sigma} k_{1\rho} k_{2\sigma}\,;
\end{equation}
namely, outside the neighbourhood of the ground state pole the regular (or continuum) part of the divergence of the axial-vector vertex saturates the anomaly in the divergence of the axial-vector--two-photon coupling.  On the other hand, in the neighbourhood of $P^2=0$ 
\begin{eqnarray} 
\left. A^{3\,{\rm reg}}(k_1,k_2) \right|_{ P^2\simeq 0} + f_{\pi_0} \,G^{\pi_0}(k_1,k_2)
& =& \frac{1}{2}\,; \label{anomalypion}
\end{eqnarray}
i.e., on this domain the contribution to the axial-vector--two-photon coupling from the regular part of the divergence of the axial-vector vertex combines with the direct $\pi_0^0 \gamma \gamma$ vertex to fulfill the anomaly.  This fact was illustrated in Ref.\,\cite{mrpion} by direct calculation: Eqs.\,(\ref{bwti}) -- (\ref{hwti}) are an essential part of that demonstration.

If one defines 
\begin{equation}
\label{TpiG}
\check{T}_{\pi_n^0}(P^2,Q^2) = \left. G^{\pi_n^0}(k_1,k_2) \right|_{k_1^2=Q^2=k_2^2},
\end{equation}
in which case $ P^2= 2(k_1\cdot k_2+Q^2)$, then the physical width of the neutral ground state pion is determined by
\begin{equation}
g_{\pi_0^0 \gamma\gamma}:= \check{T}_{\pi_0^0}(-m_{\pi_0^0}^2,0) ;
\end{equation}
viz., the second term on the l.h.s.\ of Eq.\,(\ref{anomalypion}) evaluated at the on-shell points.  This result is not useful unless one has a means of estimating the contribution from the first term; viz., $A^{3\,{\rm reg}}(k_1,k_2)$.  However, that is readily done.  A consideration\,\cite{mrt98} of the structure of the regular piece in Eq.\,(\ref{genavv}) indicates that the impact of this continuum term on the $\pi_0^0 \gamma\gamma$ coupling is modulated by the magnitude of the pion's mass, which is small for realistic $u$ and $d$ current-quark masses and vanishes in the chiral limit.  One therefore expects this term to contribute very little and anticipates from \index{Abelian anomaly} Eq.\,(\ref{anomalypion}) that 
\begin{equation}
\label{anomalycouple}
g_{\pi_0^0 \gamma\gamma} = \frac{1}{2} \frac{1}{f_{\pi_0}}
\end{equation}
is a good approximation.  This is verified in explicit calculations; e.g., in Ref.\,\cite{maristandy4}, which evaluates the triangle diagrams described herein, the first term on the l.h.s.\ modifies the result in Eq.\,(\ref{anomalycouple}) by less than 2\%.  

There is no reason to expect an analogous result for pseudoscalar mesons other than the $\pi(140)$; i.e., the states which are denoted by $n\geq 1$.  Indeed, as all known such pseudoscalar mesons have experimentally determined masses that are greater than $1\,$GeV, the reasoning used above suggests that the presence of the continuum terms, $A^{3\,{\rm reg}}(k_1,k_2)$ and $P^{3\,{\rm reg}}(k_1,k_2)$, must materially impact upon the value of $g_{\pi_1^0 \gamma\gamma}$.  This was shown to be true in Ref.\,\cite{krassnigg2}; e.g., $g_{\pi_1^0 \gamma\gamma} = - 0.13\,g_{\pi_0^0 \gamma\gamma}$. 


It was stated that the rainbow-ladder truncation preserves the one-loop renormalisation group properties of QCD.  It follows that Eq.\,(\ref{Tpingg}) should reproduce the leading large-$Q^2$ behaviour of the $\gamma^\ast(Q) \pi_n(P) \gamma^\ast(Q)$ transition form factor inferred from perturbative QCD.  That is true \cite{kekez}.

Reference~\cite{krassnigg2} considered the general case; i.e., arbitrary $n$, and proved \begin{equation}
\check{T}_{\pi_n^0}(-m_{\pi_n}^2,Q^2) \stackrel{Q^2\gg \Lambda_{\rm QCD}^2}{=} \frac{4\pi^2}{3}
\left[ \frac{f_{\pi_n}}{Q^2} + F_n^{(2)}(-m_{\pi_n}^2) 
\frac{\ln^{\gamma} Q^2/\omega_{\pi_n}^2}{Q^4}
\right] ,
\label{UVnot0}
\end{equation}
where: $\gamma$ is an anomalous dimension, which cannot be determined accurately in rainbow-ladder truncation; $\omega_{\pi_n}$ is a mass-scale that gauges the momentum space width of the pseudoscalar meson; and $F_n^{(2)}(-m_{\pi_n}^2)$ is a structure-dependent constant, similar but unrelated to $f_{\pi_n}$.\footnote{With the interaction of Eq.\,(\ref{gk2}), $F_1^{(2)}(-m_{\pi_n}^2) \simeq -\langle\bar q q\rangle^0$, and it is generally nonzero in the chiral limit.}   It is now plain that $\forall\, n\geq 1$ 
\begin{equation}
\lim_{\hat m\to 0} \check{T}_{\pi_n^0}(-m_{\pi_n}^2,Q^2) 
 \stackrel{Q^2\gg \Lambda_{\rm QCD}^2}{=} \frac{4\pi^2}{3}\left. F^{(2)}_{n }(-m_{\pi_n}^2)\frac{\ln^{\gamma} Q^2/\omega_{\pi_n}^2}{Q^4}\right|_{\hat m=0} ;
\label{UVchiralnot0}
\end{equation}
namely, in the chiral limit the leading-order power-law in the transition form factor for excited state pseudoscalar mesons is O$(1/Q^4)$.  This is a novel, nonperturbative and model-independent result.

Reference~\cite{krassnigg2} also provides quantitative illustrations of these and other results relating to the electromagnetic properties of ground- and excited-state pseudoscalar mesons.  For example, the two photon width of the $\pi_1$-meson is estimated
\begin{equation}
\label{Gpiggbest}
\Gamma_{\pi_1\gamma\gamma} \simeq 240\,{\rm eV} \simeq 30\,\Gamma_{\pi_0\gamma\gamma}\,,
\end{equation} 
as is the electromagnetic radius of the charge states
\begin{equation}
r_{\pi_1} \simeq 1.4 \,r_{\pi_0}\,.
\end{equation}

\begin{figure}[t]
\begin{center}
\includegraphics[clip,width=0.75\textwidth]{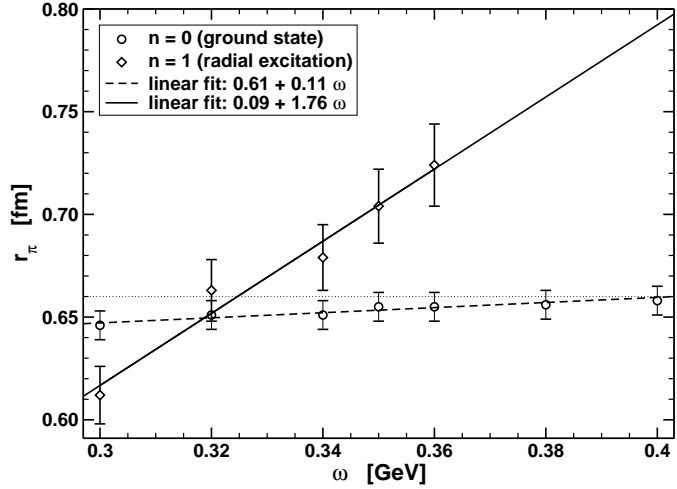}
\caption{\label{fig:emradii} Evolution of ground and first excited state pseudoscalar mesons' electromagnetic charge radii with the scale parameter $\omega$ in Eq.\,(\protect\ref{gk2}): $r_a=1/\omega$ gauges the range of the confining interaction between light quarks.  \textit{Dotted line}: $r_\pi=0.66\,$fm, which indicates the experimental value of the ground state's radius.  The radius is evaluated numerically from the electromagnetic form factor.  That is the origin of the theory error depicted in the figure, which corresponds to a relative error  $\lsim 1$\% for $n=0$ and $\lsim 3$\% for $n=1$.  (Adapted from Ref.\,\protect\cite{krassnigg2}.)}
\end{center}
\end{figure}

\index{Charge radii: pseudoscalar mesons} These results were obtained with the model rainbow-ladder interaction described below, in Sect.\,\ref{practical}.  However, the calculation of the latter was used to verify that the properties of excited states are indeed a sensitive probe of the interaction between light-quarks at long-range.  This is shown in Fig.\,\ref{fig:emradii}.  As with so many properties of the ground state Goldstone mode, its radius is protected against rapid evolution by \index{Dynamical chiral symmetry breaking (DCSB)} DCSB.  However, the charge radius of the first excited state changes swiftly with increasing $\omega$, with the ratio $r_{\pi_1}/r_{\pi_0}$ varying from $0.9$ -- $1.2$.  This outcome can readily be interpreted.  The length-scale $r_a := 1/\omega$ is a measure of the range of strong attraction: magnifying $r_a$ increases the active range of the confining piece of the interaction.  This effectively strengthens the \index{Confinement} confinement force and that compresses the bound state, as one observes in Fig.\,\ref{fig:emradii}: $r_{\pi_1}$ decreases quickly with increasing $r_a=1/\omega$.  While it is natural to suppose $r_{\pi_1}>r_{\pi_0}$; namely, that a radial excitation is larger than the associated ground state, the calculations of Ref.\,\cite{krassnigg2} illustrated that with the ground state pseudoscalar meson's properties constrained by \index{Goldstone theorem} Goldstone's theorem and its pointwise consequences, Eqs.\,(\ref{bwti}) -- (\ref{hwti}), it is possible in quantum field theory for a confining interaction to compress the excited state with the consequence that $r_{\pi_1}<r_{\pi_0}$.  

A related result for the evolution of the mass was observed in Ref.\,\cite{krassnigg1}; namely, the mass of the first excited state dropped rapidly with increasing $r_a$.  (However, in this case DCSB guarantees $m_{\pi_1}>m_{\pi_0}$.)  On the $\omega$-domain illustrated in Fig.\,\ref{fig:emradii}, the mass of the ground state obtained with nonzero current-quark mass varied by only 3\% while that of the first excited state changed by 14\%.  It is natural to expect that an increase in the strength of the confinement force should increase the magnitude of the binding energy and hence reduce the mass.  That is precisely what occurs.  

\subsection{Practical, predictive tool}
\index{Truncation of DSEs: rainbow-ladder}
\label{practical}
It was stated that there is a systematic, nonperturbative symmetry-preserving truncation of the DSEs.  The leading-order is provided by the renormalisation-group-improved rainbow-ladder truncation, which has been used widely; e.g., Refs.\,\cite{jainmunczek,mr97,klabucar}, and references thereto.  A practical renormalisation-group-improved rainbow-ladder truncation preserves the one-loop ultraviolet behaviour of perturbative QCD.  However, a model assumption is required for the behaviour of the kernel in the infrared; viz., on the domain $Q^2 \lsim 1\,$GeV$^2$, which corresponds to length-scales $\gsim 0.2\,$fm.  This is the confinement domain \index{Confinement} whereupon little is truly known about the interaction between light-quarks.  That information is, after all, what we seek!  

The systematic application of a single model to an extensive range of JLab-related phenomena is pursued in Refs.\,\cite{marisphotonvertex,maristandy4,maristandy1,maristandy3,marisji,%
marisbicudo,maristandy5,mariscotanch,wright05}.  The model interaction is expressed via
\begin{equation}
\frac{\check{G}(Q^2)}{Q^2} = \frac{4\pi^2}{\omega^6} D\, Q^2 {\rm
e}^{-Q^2/\omega^2} + \, \frac{ 8\pi^2\, \gamma_m } { \ln\left[\tau + \left(1 +
Q^2/\Lambda_{\rm QCD}^2\right)^2\right]} \, {\check F}(Q^2) \,, \label{gk2}
\end{equation}
wherein ${\check F}(Q^2)= [1 - \exp(-Q^2/[4 m_t^2])]/Q^2$, $m_t$ $=$ $0.5\,$GeV; $\tau={\rm e}^2-1$; $\gamma_m = 12/25$; and $\Lambda_{\rm QCD} = \Lambda^{(4)}_{\overline{\rm MS}}=0.234\,$GeV. (NB.\ Eq.\,(\protect\ref{gk2}) gives $\alpha(m_Z^2)= 0.126$.)  This simple form represents the interaction strength as a sum of two terms.  The second ensures that \index{Asymptotic freedom} perturbative behaviour is preserved at short-range.  The first, on the other hand, provides for the possibility of enhancement in $K$ and the gap equation's kernel at long-range \cite{bhagwat}.  The model assumption explicit here is restricted to the infrared; viz., to the interval $Q^2 \lsim 1\,$GeV$^2$, which corresponds to length-scales $\gsim 0.2\,$fm. 

The true parameters in Eq.\,(\ref{gk2}) are $D$ and $\omega$, which together determine the integrated infrared strength of the rainbow-ladder kernel; i.e., the so-called interaction tension  $\sigma^\Delta$ \cite{cdrwien}.  However, it is worth emphasising that they are not independent \cite{maristandy1}: in fitting to a selection of observables, a change in one is compensated by altering the other; e.g., on the domain $\omega\in[0.3,0.5]\,$GeV, the fitted observables are approximately constant along the trajectory \cite{raya3}
\begin{equation}
\label{omegaD}
\omega \,D = (0.72\,{\rm GeV})^3 =: m_g^3.
\end{equation}
This correlation: a reduction in $D$ compensating an increase in $\omega$, acts to keep a fixed value of the interaction tension.  Equation (\ref{gk2}) is thus a one-parameter model.  NB.\ As we saw in connection with Fig.\,\ref{gluoncf}, this value of $m_g$ is typical of the mass-scale associated with nonperturbative gluon dynamics.

\begin{figure}[t]
\centerline{%
\includegraphics[clip,width=0.75\textwidth]{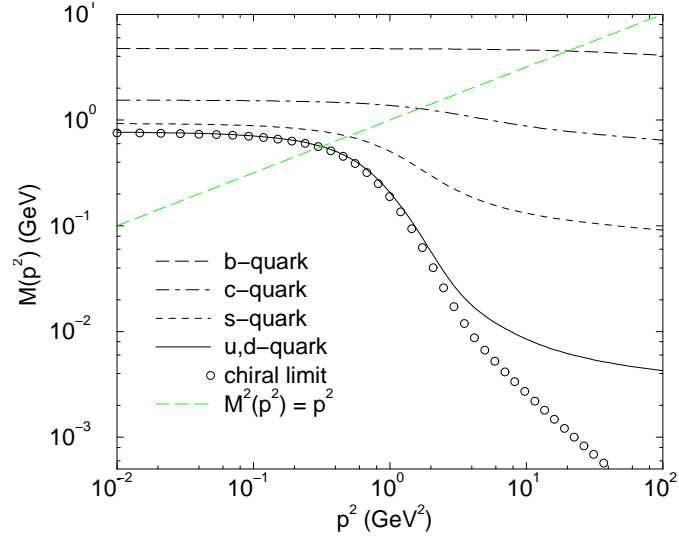}
}
\caption{\label{Mpsqplot}  Quark mass function obtained by solving the complex formed by Eqs.\,(\protect\ref{renormS}), (\protect\ref{gk2}), (\protect\ref{rainbowdse}), (\protect\ref{rainbowmasses}). NB.\ A nonzero solution is obtained in the chiral limit.  This solution is power law suppressed in the ultraviolet\,\protect\cite{lane,politzer} and essentially nonperturbative.  Thus when the $u,d$-quark solution melds with the chiral limit solution one has entered the domain on which all effects are primarily nonperturbative.  The transition takes place for $m_g^2 \lsim p^2 \lsim 6\,m_g^2$. (Adapted from Ref.\,\protect\cite{mishaSVY}.)}
\end{figure}

With the interaction in Eq.\,(\ref{gk2}), one obtains the rainbow-truncation gap equation:
\begin{equation}
S(p)^{-1} = Z_2 \,(i\gamma\cdot p + m^{\rm bare}) + \int^\Lambda_q\! {\check G}((p-q)^2) D_{\mu\nu}^{\rm free}(p-q) \frac{\lambda^a}{2}\gamma_\mu S(q) \frac{\lambda^a}{2}\gamma_\nu , \label{rainbowdse} 
\end{equation} 
wherein $D_{\mu\nu}^{\rm free}(\ell)$ is the free gauge boson propagator; namely, Eq.\,(\ref{gluonZ}) with $F(p^2)\equiv 1$.  The self-consistent solution obtained with current-quark masses (in GeV):
\begin{equation}
\label{rainbowmasses}
\begin{array}{l|c cccc}
f ~&~ \mbox{chiral} & u,d & s & c & b \\\hline
m_f(\zeta_{19}) ~&~ 0.0 & 0.0037 & 0.082 & 0.97 & 4.1\\
\end{array}
\end{equation}
is depicted in Fig.\,\ref{Mpsqplot}.   It is important to bear in mind and evident in the figure that, because the truncation preserves the one-loop renormalisation group properties of QCD, the ultraviolet behaviour of the solutions of Eqs.\,(\ref{gendse}) and (\ref{rainbowdse}) is precisely that of QCD; viz., 
\begin{equation}
\label{Mp2uv}
M(p^2) \stackrel{p^2\gg \Lambda^2_{\rm QCD}}{=} \frac{\hat m}{(\mbox{\small $\frac{1}{2}$}\ln p^2/\Lambda^2_{\rm QCD})^{\gamma_m}},
\end{equation}
where $\hat m$ is the renormalisation-group-invariant current-quark mass.  \index{Current-quark mass}

The mass function is a renormalisation group invariant and can be used to define a Euclidean constituent-quark mass; \index{Constituent-quark mass}
viz.,\footnote{The model produces dressed-quarks, which are confined in the sense that their propagator does not exhibit a true Minkowski space mass pole; i.e., there is no solution of \mbox{$s + M(s)^2 = 0$}.}
\begin{equation}
\label{CQM}
(M^E)^2 := s \, |\, s = M(s)^2 .
\end{equation}
For the solutions depicted in Fig.\,\ref{Mpsqplot}, one finds (in GeV)
\begin{equation}
\label{rainbowCQmasses}
\begin{array}{l|ccccc}
f ~&~ \mbox{chiral} & u,d & s & c & b \\\hline
\rule{0em}{2.5ex} M^E_f ~&~ 0.42 & 0.42 & 0.56 & 1.57 & 4.68
\end{array}
\end{equation}
The ratio $\check L_f:=M^E_f/m_f(\zeta)$ is a measure of the impact of the \index{Dynamical chiral symmetry breaking (DCSB)} DCSB mechanism on a particular flavour of quark.  A comparison between Eqs.\,(\ref{rainbowmasses}) and (\ref{rainbowCQmasses}) shows that for quark flavours with $\hat m_f \ll m_g$ the effect of \index{Dynamical chiral symmetry breaking (DCSB)} DCSB on their propagation characteristics is very great $\forall \,s\lsim 6 m_g^2$.  The domain on which the impact is large diminishes rapidly as the current-quark mass increases.

\begin{figure}[t]
\centerline{\includegraphics[clip,width=0.75\textwidth]{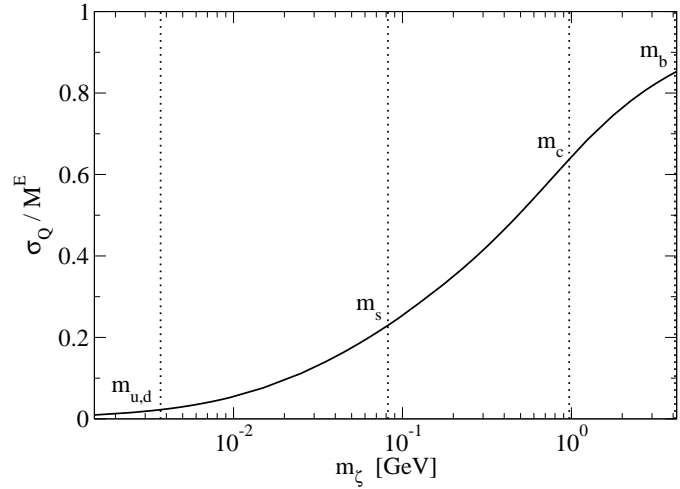}}
\caption{\label{fig:quark_ratio} Ratio $\sigma_Q/M^E_Q$ [Eqs.\,(\protect\ref{CQM}) \& (\protect\ref{sMEQ})] as a function of current-quark mass.  It is a measure of the current-quark-mass-dependence of dynamical chiral symmetry breaking: a zero value indicates complete dominance of dynamical over explicit chiral symmetry breaking; and a value of one, the opposite.  The vertical dotted lines correspond to the $u=d$, $s$, $c$ and $b$ current-quark masses listed in Eq.\,(\protect\ref{rainbowmasses}). (Adapted from Ref.\,\protect\cite{wright05}.)}
\end{figure}

Another way of looking at this is via a constituent-quark $\sigma$-term \cite{wright05,sigmaterms}:
\begin{equation}
\label{sMEQ}
\sigma_f := m_f(\zeta) \, \frac{\partial  M_f^E}{\partial m_f(\zeta)}\,.
\end{equation}
It is a renormalisation-group-invariant that can be determined from solutions of the gap equation, Eq.\,(\ref{rainbowdse}), and is a keen probe of the impact of explicit chiral symmetry breaking on the mass function.  The ratio $\sigma_{f}/M^E_f$ plotted in Fig.\,\ref{fig:quark_ratio} therefore measures the effect of explicit chiral symmetry breaking on the dressed-quark mass-function compared with the sum of the effects of explicit and dynamical chiral symmetry breaking.  It is evident in the figure that this ratio vanishes for light-quarks because the magnitude of their constituent-mass owes primarily to dynamical chiral symmetry breaking, while for heavy-quarks it approaches one.  

An illustration of the chiral limit behaviour can easily be obtained through the model in Sect.\,\ref{confinement}, which yields algebraic results in the neighbourhood of the chiral limit \cite{changlei}: $M^E_0=\mbox{\small $\frac{1}{\surd 2}$}\, \check G$, $\sigma_0 = \mbox{\small $\frac{3}{2}$}\,m$, and hence $\mbox{\small $\frac{\sigma_{0}}{M^E_0}$} = \mbox{\small $\frac{3}{\surd 2}$} \mbox{\small $\frac{m}{\check G}$}$.  With the parameter values in Ref.\,\cite{mn83}, this gives an estimate $\mbox{\small $\frac{\sigma_{0}}{M^E_0}$} \approx 0.04$ for $u=d$ quarks.  

For the purpose of a more realistic illustration one may also consider the rainbow-truncation defined with Eq.\,(\ref{gk2}) and unequal light-quark masses.  For example, setting $m_u(\zeta_1)=4\,$MeV and $m_d(\zeta_1)=7\,$MeV through $\zeta_1=1\,$GeV in Eq.\,(\ref{Mp2uv}), one obtains $M^E_u=0.411\,$GeV and $M^E_d=0.416\,$GeV, and $\sigma_u=0.015$ and $\sigma_d=0.025$.  It is interesting to note that the \index{Dynamical chiral symmetry breaking (DCSB)} DCSB mechanism amplifies the \index{Light-quark mass difference} current-quark mass difference in the separation between light-quark constituent-quark masses: 
\begin{equation}
M^E_d-M^E_u= 1.5\,[m_d(\zeta_1)-m_u(\zeta_1)]\,.
\end{equation}

A value for the vacuum quark condensate, Eq.\,(\ref{qbq0}), may be obtained from the chiral limit solution depicted in Fig.\,\ref{Mpsqplot}
\begin{equation}
 -\langle \bar q q \rangle_\zeta^0 =
\lim_{\Lambda\to \infty} Z_4(\zeta,\Lambda)\, N_c\, {\rm tr}_{\rm D} \int^\Lambda_q  S_{\hat m =0}(q) 
=(0.275\,{\rm GeV})^3\,.
\end{equation}
\index{Quark condensate: vacuum} Since QCD is multiplicatively renormalisable, then
\begin{equation}
\langle \bar q q \rangle_{\zeta^\prime}^0 = Z_m(\zeta^\prime,\zeta) \, \langle \bar q q \rangle_\zeta^0\,, 
\end{equation}
where the mass renormalisation constant $Z_m:= Z_4/Z_2$.  Applying one-loop evolution, Eq.\,(\ref{Mp2uv}), to define the vacuum condensate at a typical hadronic scale, one therefore obtains \cite{mr97}
\begin{equation}
\label{qbq1}
- \langle \bar q q \rangle_{\zeta_1}^0 = (0.24\,{\rm GeV})^3.
\end{equation}
This condensate might be interpreted as measuring the density of quark-antiquark pairs in the vacuum of chiral QCD, in which connection the result in Eq.\,(\ref{qbq1}) corresponds to $\rho_{\bar q q} = 1.8\,{\rm fm}^{-3}$.  Were one to assume that this vacuum could be viewed as a medium of close-packed spheres, then each would occupy a volume $V_{\bar q q} = 0.55\,{\rm fm}^3$, which corresponds to a radius $r_{\bar q q} = 0.51\,$fm.  For comparison, the measured electromagnetic charge radius of a pion can be written $r_\pi= 0.77 r_{\bar q q}$ and that of a proton, $r_p = 0.58\, r_{\bar q q}$.  It is therefore clear that a veracious understanding of the length-scale defined by the chiral limit vacuum quark condensate is a keystone for the bridge between theory and experiment. 

The renormalisation-group-improved rainbow-ladder truncation obtained with Eq.\,(\ref{gk2}) has been employed very successfully.  For example, it predicted\,\cite{marisphotonvertex,maristandy3} the \index{Electromagnetic form factor: pion} electromagnetic pion form factor measured\,\cite{Volmer2000} at JLab,\footnote{The result is perhaps too good given that pion-loop contributions were omitted.  Such effects, which appear beyond rainbow-ladder truncation, are important at small momentum transfer.  For example, they add $\lsim 15$\% to the rainbow-ladder result for $r_\pi$ \protect\cite{benderpion}.} and over an illustrative basket of thirty-one calculated quantities, tabulated in Ref.\,\cite{pmcdr}, its agreement with data has an average relative error of $2$\% and standard-deviation of $15$\%.

\subsection{Critical mass, DCSB and QCD}
\index{Critical mass and DCSB: QCD}
\label{mcrQCD}
At this point one can return to the question posed at the end of Sect.\,\ref{dcsbmc}; namely, whether a critical mass is also associated with DCSB in QCD.  This has been explored \cite{leichang} with the renormalisation-group-improved (RGI) rainbow truncation of the gap equation's kernel described by Eqs.\,(\ref{gk2}), (\ref{rainbowdse}).

Equations~(\ref{gendse}) and (\ref{rainbowdse}) obviously admit the \mbox{$M(p^2)\equiv 0$} solution in the chiral limit, Eq.\,(\ref{limchiral}).  This is the perturbative solution and is analogous to $M_W$ in Eq.\,(\ref{gapsolnjl}).  It is moreover plain in Fig.\,\ref{Mpsqplot} that the chiral limit rainbow gap equation also yields a DCSB solution.  This capacity is the basis for much of the phenomenological success of the RGI rainbow-ladder truncation.  The truncation preserves the feature that if $M_+(p^2)=M(p^2)>0$, $\forall p^2>0$, is a solution of the chiral-limit gap equation, then so is $M_-(p^2):= [-M(p^2)]$.  These solutions are the analogues of $M_\pm$ in Eq.\,(\ref{gapsolnjl}).  

\begin{figure}[t]
\centerline{\includegraphics[width=0.80\textwidth]{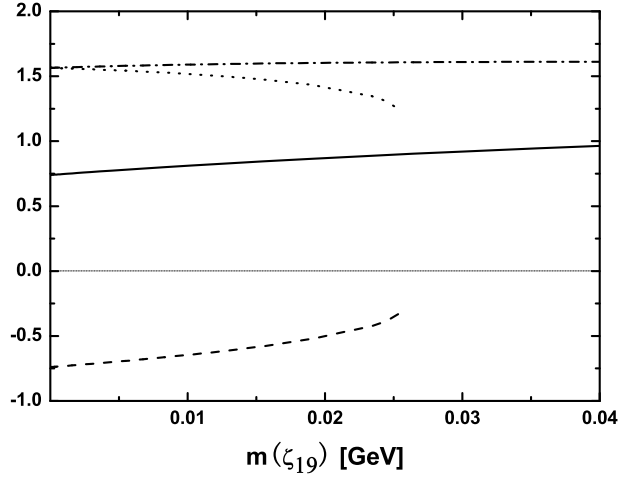}}

\caption{\label{MTAB} Evolution with current-quark mass, $m(\zeta_{19})$, of \mbox{$A(p^2=0,\zeta^2)$} (dimensionless), $B(p^2=0,\zeta^2)$ (GeV) as calculated with $\omega=0.4\,$GeV in Eq.\,(\protect\ref{gk2}): \textit{solid curve} -- $B_+(0)$; \textit{dash-dot curve} -- $A_+(0)$; \textit{dashed curve} -- $B_-(0)$; \textit{dotted curve} -- $A_-(0)$.  The light dotted curve simply distinguishes the ordinate zero.}
\end{figure}

It is apparent in Fig.\,\ref{MTAB} that this interaction model, too, exhibits a bounded domain of current-quark mass on which $M_-(p^2)$ and $M_+(p^2)$ exist simultaneously:
\begin{equation}
{\cal D}(\hat m)= \{\hat m \; | \;0 \leq \hat m < \hat m_{\rm cr} \}\,.
\end{equation}
The critical current-quark mass depends on $\omega$:
\begin{equation}
\label{mcrarray}
\begin{array}{l|ccc}
\omega \, [{\rm GeV}] & 0.3 & 0.4 & 0.5 \\\hline
\rule{0ex}{2.5ex}\hat m_{\rm cr} \, [{\rm MeV}] & 71 & 63 & 31\\\hline
\rule{0ex}{2.5ex} m_{\rm cr}(\zeta_{19}) \, [{\rm MeV}] & 35 & 31 & 15\\
\rule{0ex}{2.5ex} m_{\rm cr}(\zeta_1) \, [{\rm MeV}] & 60 & 53 & 26
\end{array}\,,
\end{equation}
where the third and fourth rows report the critical mass at the renormalisation scales described above.  NB.\ In most phenomenological applications $0.3 < \omega < 0.4$. 

It is noteworthy that when applying Eq.\,(\ref{renormS}) only the $M_\pm(p^2)$ solutions appear; i.e., in this case, for nonzero current-quark mass an analogue of $M_W$ in Eq.\,(\ref{gapsolnjl}) was not found.

In considering alternative rainbow interaction models, it was determined that $m_{\rm cr}$ assumes similar values in most models that provide a reasonable description of the same low-energy observables.  An exception is the model of Ref.\,\cite{mn83}, in which a solution with $B(s=0)<0$, and $A(s)$ and $B(s)$ continuous on $s\in [0,\infty)$ exists only in the chiral limit.  It is also natural to enquire after the effect of dressing the quark-gluon vertex.  As noted above, symmetry ensures that in the chiral limit the gap equation simultaneously admits $M_-(p^2)$ and $M_+(p^2)$ solutions in this instance, too.  The extent of the domain of current-quark mass on which the $M_-(p^2)$ solution persists will probably depend on the structure of the vertex.  However, that is also true of the parameters required to describe a given set of low-energy observables.  It is thus likely that in physical terms the domain ${\cal D}(\hat m)$ will not be much altered by realistic vertex dressing.  These issues are currently being explored \cite{leichang2}.

Equation~(\ref{Mp2uv}) is valid for both $M_+(p^2)$ and $M_-(p^2)$.  In fact, one can make a stronger statement, while both $M_+(p^2)$, $M_-(p^2)$ exist:
\begin{equation}
\label{MPMeq}
M_+(p^2)\stackrel{p^2\gg \Lambda_{\rm QCD}^2}{=} M_-(p^2)\,.
\end{equation}
This follows from \index{Asymptotic freedom} asymptotic freedom, which is a feature of the RGI model and QCD.  One may argue for this result as follows.  On the perturbative domain 
\begin{equation}
A_\pm(p^2,\zeta^2) \approx 1\,,\; \frac{M_\pm(p^2)}{p^2+M^2_\pm(p^2)} \approx \frac{M_\pm(p^2)}{p^2}\,.
\end{equation}
Hence, the gap equation becomes a single linear integral equation for $M_\pm(p^2)$.  Within the domain on which the preceding steps are valid, that equation can be approximated by a linear second-order ordinary differential equation (d.e.) (e.g., Ref.\,\cite{atkinson}).  The d.e.\ is the same for both $M_\pm(p^2)$, as is the ultraviolet boundary condition, which is determined by the current-quark mass.  Thus follows Eq.\,(\ref{MPMeq}).  

\begin{figure}[t]
\centerline{\includegraphics[width=0.88\textwidth]{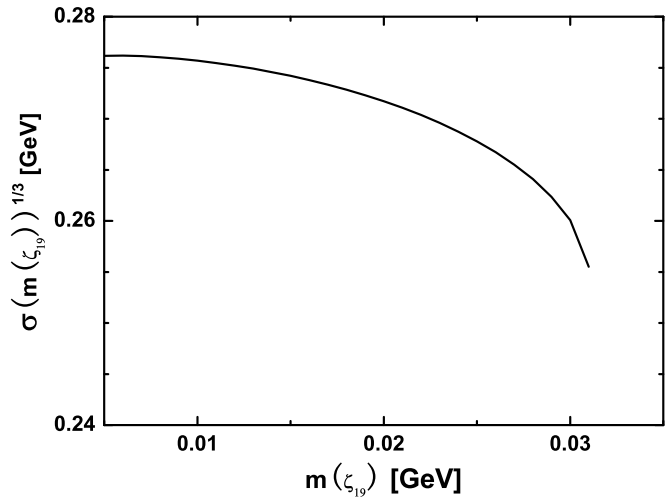}}

\caption{\label{figsm} Evolution with current-quark mass, $m(\zeta_{19})$, of the massive-quark condensate defined in Eq.\,(\protect\ref{barsigma}), calculated with $\omega=0.4\,$GeV and at the renormalisation point $\zeta=19\,$GeV.}
\end{figure}

It is true in general that the model gap equation may usefully be approximated by a second-order d.e., which is nonlinear in $M(p^2)$ for infrared momenta but linear in the ultraviolet \cite{atkinson,tang,mckellar,munczekde}.  The ultraviolet boundary condition for both $M_\pm(p^2)$ solutions is still fixed by the current-quark mass.  However, they have different infrared boundary conditions.  The d.e.\ is solved in terms of two independent solutions, which may be described as the irregular and regular solutions.  The regular solution is power-law suppressed cf.\ the irregular form.  The irregular solution of the d.e.\ is the functional form accessible in perturbation theory.  It dominates in the ultraviolet, simultaneously fixing both $M_\pm(p^2)$ to a value determined by the current-quark mass.  Any difference between $M_+(p^2)$ and $M_-(p^2)$ appears through the regular component of the solution, which is essentially nonperturbative and has a magnitude that is primarily determined by the infrared boundary condition.

From Eq.\,(\ref{MPMeq}) and the subsequent discussion it follows that \index{Quark condensate: massive}
\begin{eqnarray}
\label{barsigma}
\bar\sigma(m(\zeta)) &:=&  \lim_{\Lambda\to \infty} 
Z_4(\zeta^2,\Lambda^2)\, N_c \, {\rm tr}_{\rm D}\int^\Lambda_q\! \bar S^{m(\zeta)}(q,\zeta)\,,\\
\bar S^{m(\zeta)}(q,\zeta)  &= &\frac{1}{2} \left[S_+^{m(\zeta)}(q,\zeta)-S_-^{m(\zeta)}(q,\zeta)\right],
\end{eqnarray}
provides a gauge invariant, current-quark-mass-dependent quark condensate in QCD, which is finite, well-defined on a bounded interval, and is truly the vacuum quark condensate in the chiral limit.  The behaviour of $\bar\sigma(m(\zeta))$ obtained with the RGI model employed herein is depicted in Fig.\,\ref{figsm}.  As we saw in connection with Fig.\,\ref{figbarM}, here, too, the essentially dynamical component of chiral symmetry breaking decreases with increasing current-quark mass. 

It must be stressed that the straightforward definition of a massive-quark condensate via the trace of a $\hat m \neq 0$ dressed-quark propagator is not useful because it gives a quantity that is quadratically divergent and therefore very difficult to define unambiguously \cite{kurtcondensate,furnstahl}.  

It is now clear that on a bounded interval of current-quark mass, ${\cal D}(\hat m)=\{\hat m \; | \;0 \leq \hat m < \hat m_{\rm cr} \}$ with Eq.\,(\ref{mcrarray}), realistic models of QCD's gap equation can simultaneously admit two inequivalent dynamical chiral symmetry breaking solutions for the dressed-quark mass-function, $M_\pm(p^2)$.  These solutions are distinguished by their value at the origin: $M_+(p^2)>0$ and $M_-(p^2)<0$.  In the ultraviolet they both coincide with the running current-quark mass.  

The pointwise values of both solutions evolve continuously with current-quark mass within ${\cal D}(\hat m)$.  Beyond the upper boundary, however, only the positive solution exists.  It is fundamental that $M_-(p^2)$ is an essentially nonperturbative quantity whose modification by a current-quark mass can be evaluated perturbatively within ${\cal D}(\hat m)$.  However, outside that domain such an analysis fails because the current-quark mass is large enough to completely destabilise this solution.  Thus $\hat m_{\rm cr}$ specifies an upper bound on the domain within which a perturbative expansion in the current-quark mass can uniformly be valid.  \index{Convergence of expansion in pion mass} In the renormalisation-group-improved rainbow-ladder truncation of QCD's gap equation described herein the critical current-quark mass corresponds to a pseudoscalar meson mass \mbox{$m_{0^-} \sim 0.45\,$GeV} (\mbox{$m_{0^-}^2 \sim 0.20\,$GeV$^2$}) \cite{andreaspi}.  

A value of similar magnitude was deduced in Refs.\,\cite{awtconverge,youngA} as the scale below which accuracy may be expected from the approximation of observables through a perturbative expansion in pseudoscalar meson mass.  This scale also marks the boundary below which observables should exhibit that curvature as a function of pseudoscalar meson mass which is characteristic of chiral effective theories.  

%

\begin{figure}[t]
\centerline{%
\includegraphics[clip,width=0.80\textwidth]{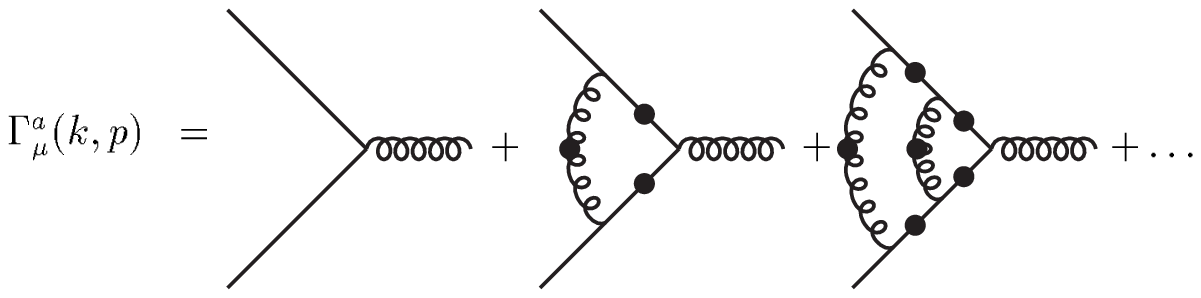}}
\centerline{%
\includegraphics[width=0.80\textwidth]{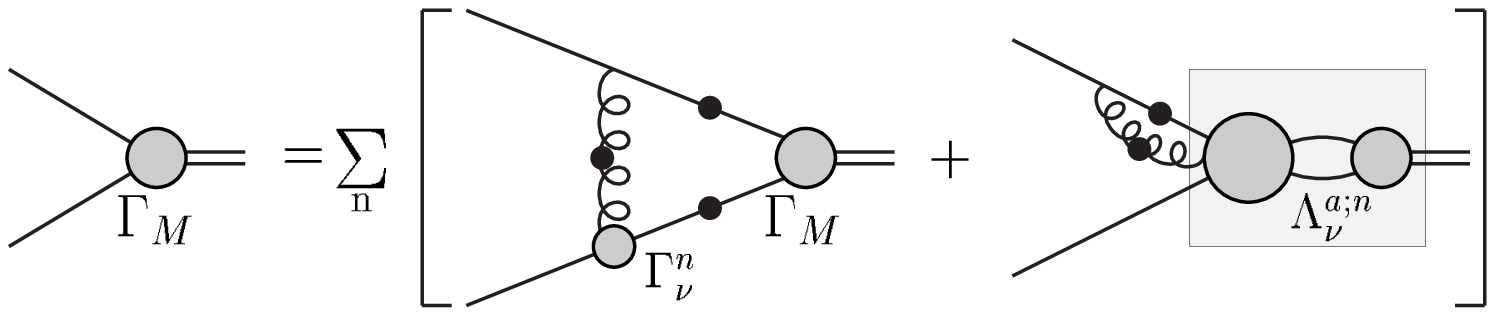}}
\centerline{%
\includegraphics[width=0.80\textwidth]{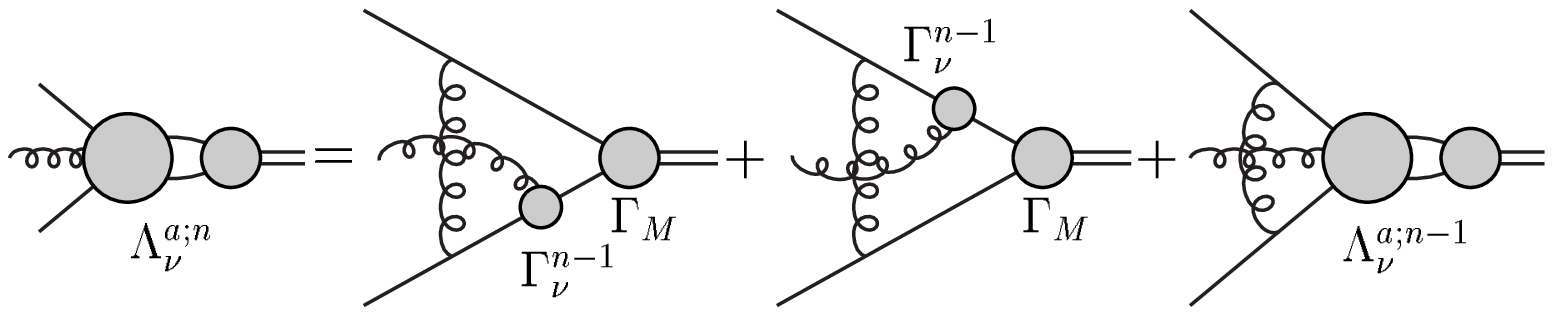}}
\caption{\label{detmoldvertex}  
\textit{Top line} --  Integral equation for a planar dressed-quark-gluon vertex.   ``Springs'' between dressed-quark lines indicate the dressed-gluon propagator.  First diagram depicts the rainbow-truncation; second adds a single gluon rung; etc.  While this series apparently neglects the three-gluon vertex, its effect has satisfactorily been modelled via the \textit{Ansatz} $g^2 \to -\,\check{C} g^2$ \protect\cite{bhagwatvertex}.  The change of sign is an important consequence of the non-Abelian nature of QCD.  
%
\textit{Lower two lines} -- Integral equation for a symmetry preserving Bethe-Salpeter kernel that is consistent with the vertex depicted.  Beyond rainbow-truncation, symmetry preservation requires that this kernel be nonplanar even though the vertex in the gap equation is planar.  (Adapted from Ref.\,\protect\cite{detmold}.)}
\end{figure}

\subsection{Beyond rainbow-ladder truncation}
\index{Truncation of DSEs: beyond rainbow-ladder}
\label{beyond}
Rainbow-ladder is the leading order in a nonperturbative, systematic and symmetry preserving truncation of the DSEs.  This is explained and illustrated in Refs.\,\cite{bhagwatvertex,truncscheme,detmold}.  Figure \ref{detmoldvertex} depicts a natural extension of the vertex \textit{Ansatz} in Eq.\,(\ref{Gnjl}).  It modifies the gap equation's kernel.  In order to preserve the axial-vector Ward-Takahashi identity, Eq.\,(\ref{avwtim}), the Bethe-Salpeter kernel, $K$, must also be modified.  That is systematically accomplished for this vertex via the procedure indicated in the lower two lines of Fig.\,\ref{detmoldvertex}, as detailed in Refs.\,\cite{bhagwatvertex,detmold}. 

\begin{table}[t]
\caption{Calculated $\pi$- and $\rho$-meson masses, in GeV, obtained with $\check{G} \to -\,\check{C} \,\check{G}$ in Eq.\,(\ref{mnprop}), $\check{C} =0.35$, and quoted with $\check{G}=0.65\,$GeV in which case $m=0.016 \check{G} = 10\,$MeV.  $n$ is the number of interaction rungs retained in dressing the quark-gluon vertex, see Fig.\,(\protect\ref{detmoldvertex}), and hence the order of the vertex-consistent
Bethe-Salpeter kernel.  NB.\ $n=0$ corresponds to the rainbow-ladder truncation, in which case $m_\rho = \sqrt{2}\, \check{G}$ for $m=0$.}
{\normalsize
\begin{tabular*}
{\hsize} {l@{\extracolsep{0ptplus1fil}}
|c@{\extracolsep{0ptplus1fil}}c@{\extracolsep{0ptplus1fil}}
c@{\extracolsep{0ptplus1fil}}c@{\extracolsep{0ptplus1fil}}}
%
 & $M_H^{n=0}$ & $M_H^{n=1}$ & $M_H^{n=2}$ & $M_H^{n=\infty}$\\\hline
$\pi$, $m=0$ & 0 & 0 & 0 & 0\\
$\pi$, $m=0.011$ & 0.147 & 0.135 & 0.139 & 0.138\\\hline
$\rho$, $m=0$ & 0.920 & 0.648 & 0.782 & 0.754\\
$\rho$, $m=0.011$ & 0.936 & 0.667 & 0.798 &0.770
\end{tabular*}\label{pirhores}}
\end{table}

The model interaction of Eq.\,({\ref{mnprop}) provides an excellent tool with which to illustrate the procedure.  When used for the ``spring'' in the diagrams of Fig.\,\ref{detmoldvertex}, it gives the results in Table~\ref{pirhores}.  The first striking observation is that $m_\pi=0$ in the chiral limit for every value of $n$.  Plainly, the truncation procedure preserves the Ward-Takahashi identity, Eq.\,(\ref{avwtim}).  In addition, all of the $\rho$-$\pi$ mass splitting is present in the chiral limit.  \index{Vector-meson ($\rho$) mass and DCSB} This answers the questions raised on page~\pageref{eightfold} and in connection with Table~\ref{tableSpectrum}: the remarkably large difference between the $\pi$ and $\rho$ masses owes to \index{Dynamical chiral symmetry breaking (DCSB)} DCSB, which forces $m_\pi$ to be unnaturally small.  It is also apparent that $m_\pi$ is not very sensitive to the order of the truncation.  This is another corollary of DCSB for the Goldstone mode; namely, the cancellations which ensure $m_\pi = 0$ in the chiral limit are still quite effective for small nonzero current-quark masses.  Finally, $m_\rho$ does exhibit some sensitivity to the order of the truncation.  With a value of $\check{C} =0.35$ the rainbow-ladder truncation is accurate to $21$\%; the one-loop correction, to $13$\%; and the two-loop result, to $4$\%.\footnote{With $\check{C} =0.51$ the rainbow-ladder truncation overestimates by $29$\% the complete result obtained with the vertex in Fig.\,\protect\ref{detmoldvertex}; with $\check{C} =0.25$, by $16$\%; and with $\check{C} =0.15$, by $10$\%.  The discrepancy obviously vanishes as $\check{C} \to 0$ because this limit defines the rainbow-ladder truncation.  A comparison with these ratios serves as a true gauge against which to judge the importance of modifications of the vertex in Ref.\,\protect\cite{bhagwatvertex}.}  Hence, at the two-loop level one is certainly very sensitive to details of the model and is thus in a position to make a quantitatively accurate map of the interaction between light-quarks at long-range via comparison with experiment.  These results explain the level of accuracy attained with a rainbow-ladder truncation based on Eq.\,(\ref{gk2}).

\begin{figure}[t] 
\centerline{\includegraphics[clip,width=0.75\textwidth]{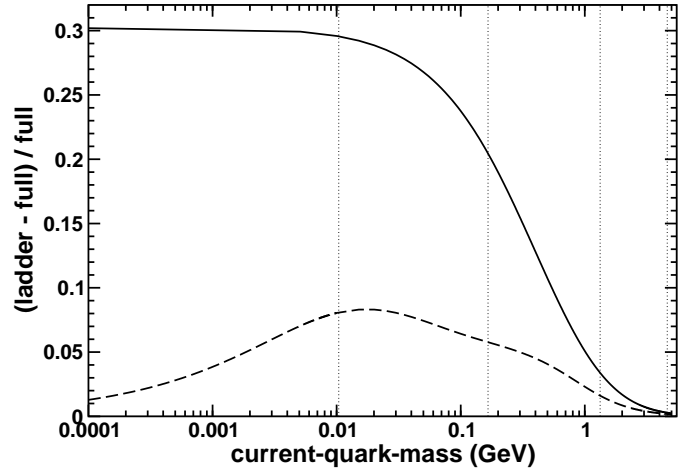}}
\caption{\label{ladderfull} Evolution with current-quark mass of the relative difference between the $\bar q q$ meson mass calculated in the rainbow-ladder truncation and the exact value obtained with $\check C=0.51$.  Solid lines: vector meson trajectories; and dashed-lines; pseudoscalar meson trajectories.  The dotted vertical lines mark the $u=d$, $s$, $c$ and $b$ current-quark masses. (Adapted from Ref.\,\protect\cite{bhagwatvertex}).}
\end{figure} 

Reference~\protect\cite{bhagwatvertex} provides a discussion of the transition between the light-quark and heavy-quark sectors, and the relative strength of the corrections to rainbow-ladder truncation as this transition is made.  Herein the results are illustrated with Fig.\,\ref{ladderfull}, from which it is apparent that with increasing current-quark mass the contributions from nonplanar diagrams and vertex corrections are suppressed.   Naturally, they must still be included in precision spectroscopic calculations.  As usual, for small current-quark masses, owing to the effects of DCSB, the pseudoscalar channel is a little different.  However, the trend in this channel becomes the same as that in the vector channel for current-quark masses above $\sim 2\,m_s$.  A contemporary challenge is to generalise the procedure described in this section to mesons composed of constituents with different current-quark masses.  

\section{Baryon Properties}
\index{Bound states: baryons}
\label{baryon}
\setcounter{equation}{0}
Significant progress has been made with the study of mesons.  While that is good, it does not directly impact on the important challenge of baryons.  Mesons fall within the class of two-body problems.  They are the simplest bound states for theory.  However, the absence of meson targets poses significant difficulties for the experimental verification of predictions such as those reported above.  On the other hand, it is relatively straightforward to construct a proton target but, as a three-body problem in relativistic quantum field theory, here the difficulty is for theory.  With this problem the current expertise is approximately at the level it was for mesons ten years ago; namely, model building and phenomenology, making as much use as possible of the results and constraints outlined above.  

A natural primary aim is to provide a true picture of the proton's electromagnetic form factors and therefrom a reliable determination of the ratio depicted in Fig.\,\ref{gepgmpdata}.  On the domain of momentum transfer for which there is apparently a discrepancy between the experimental results; namely, $Q^2 > M^2$, where $M$ is the nucleon's mass, a veracious understanding of these and other contemporary data require a Poincar\'e covariant description of the nucleon.  This is apparent in applications of relativistic quantum mechanics; e.g., Refs.~\cite{millerfrank,boffi,fuda,stan,bruno}.  A different tack follows the formulation of a Poincar\'e covariant Faddeev equation\,\cite{regfe,hugofe}.  \index{Faddeev equation} Its foundation is understood through the observation that the same interaction which describes colour-singlet mesons also generates quark-quark \index{Diquark correlations} (diquark) correlations in the colour-$\bar 3$ (antitriplet) channel\,\cite{regdq}.  While diquarks do not survive as asymptotic states \cite{bhagwatvertex,truncscheme,detmold,njldiquark}; i.e., they do not appear in the strong interaction spectrum, the attraction between quarks in this channel grounds a picture of baryons in which two quarks are always correlated as a colour-$\bar 3$ diquark pseudoparticle, and binding is effected by the iterated exchange of roles between the bystander and diquark-participant quarks. 

A first numerical study of this Faddeev equation for the nucleon was reported 
in Ref.\,\cite{cjbfe}, and following that there have been numerous more 
extensive analyses; e.g., Refs.~\cite{bentzfe,oettelfe,hechtfe,birsefe}.  It has become apparent that the dominant correlations for ground state octet and decuplet baryons are scalar and axial-vector diquarks, primarily because the associated mass-scales are smaller than the masses of these baryons\,\cite{cjbsep,marisdq} and the positive parity of the correlations matches that of the baryons.  Both scalar and axial-vector diquarks provide attraction in the Faddeev equation; e.g., a scalar diquark alone provides for a bound octet baryon and including axial-vector correlations reduces that baryon's mass.

With the retention of axial-vector diquark correlations a quantitative description of baryon properties is attainable.  Indeed, the formulation of Ref.\,\cite{oettelfe} employs confined quarks, and confined pointlike-scalar and -axial-vector diquark correlations, to obtain a spectrum of octet and decuplet baryons in which the rms-deviation between the calculated mass and experiment is only $2$\%.  The model also reproduces nucleon form factors over a large range of momentum transfer\,\cite{oettel2}, and its descriptive success in that application is typical of such Poincar\'e covariant 
treatments; e.g., Refs.\,\cite{jacquesA,jacquesmyriad,cdrqciv,nedm}. 

However, these successes might themselves indicate a flaw in the application of the Faddeev equation to the nucleon.  For example, in the context of spectroscopy, studies using the Cloudy Bag Model (CBM)\,\cite{tonyCBM1,tonyCBM2,jerryCBM} indicate that the dressed-nucleon's mass receives a negative contribution of as much as $300$-$400\,$MeV from pion self-energy corrections; i.e.,\,\cite{tonyANU,bruceCBM} $\delta M_+ = -300\,$ to $-400\,$MeV.  Furthermore, a perturbative study, using the Faddeev equation, of the mass shift induced by pointlike-$\pi$ exchange between the quark and diquark constituents of the nucleon obtains\,\cite{ishii} $\delta M_+ = -150$ to $-300\,$MeV.  Unameliorated these mutually consistent results would much diminish the value of the $2$\% spectroscopic accuracy obtained using only quark and diquark degrees of freedom. 

In addition to masses, pseudoscalar meson loops make important contributions to many other baryon properties; e.g., to charge and magnetic radii, and magnetic moments \cite{radiiCh,young}.  These effects must not be overlooked because the size and qualitative impact of meson contributions provide material constraints on the development of a realistic quark-diquark picture of the nucleon, and its interpretation and application.  

\subsection{Faddeev Equation}
\index{Faddeev equation}
\label{faddeev}
For quarks in the fundamental representation of colour-$SU(3)$:
\begin{equation} 
3_c \otimes 3_c \otimes 3_c = (\bar 3_c \oplus 6_c) \otimes 3_c = 1_c \oplus 
8_c^\prime \oplus 8_c \oplus 10_c\,,
\end{equation} 
and hence any two quarks in a colour-singlet three-quark bound state must constitute a relative colour-antitriplet.  This fact enables the derivation of a Faddeev equation for the bound state contribution to the three quark scattering kernel\,\cite{regfe} because the same kernel that describes mesons so well\,\cite{pmcdr} is also attractive for quark-quark scattering in the colour-$\bar 3$ channel.  

In this truncation of the three-body problem the interactions between two selected quarks are added to yield a quark-quark scattering matrix, which is then approximated as a sum over all possible diquark pseudoparticle terms \,\cite{cjbfe}: Dirac-scalar $+$ -axial-vector $+ [\ldots]$.  The Faddeev equation thus obtained describes the baryon as a composite of a dressed-quark and nonpointlike diquark with an iterated exchange of roles between the bystander and diquark-participant quarks.  The baryon is consequently represented by a Faddeev amplitude: 
\begin{equation} 
\Psi = \Psi_1 + \Psi_2 + \Psi_3 \,, 
\end{equation} 
where the subscript identifies the bystander quark and, e.g., $\Psi_{1,2}$ are obtained from $\Psi_3$ by a correlated, cyclic permutation of all the quark 
labels.  

The Faddeev equation is simplified further by retaining only the lightest diquark correlations in the representation of the quark-quark scattering matrix.  A simple Goldstone-theorem-preserving, rainbow-ladder DSE-model\,\cite{cjbsep} yields the following diquark pseudoparticle masses (isospin symmetry is assumed): 
\begin{equation} 
\label{dqmass} 
\begin{array}{l|cccc} 
(qq)_{J^P}           & (ud)_{0^+} & (us)_{0^+}  & (uu)_{1^+}& (us)_{1^+}\\ 
 m_{qq}\,({\rm GeV}) & 0.74       & 0.88        & 0.95      & 1.05 \\\hline 
(qq)_{J^P}           & (ss)_{1^+} & (uu)_{1^-} & (us)_{1^-} & (ss)_{1^-}  \\ 
 m_{qq}\,({\rm GeV}) & 1.13 & 1.47 & 1.53 & 1.64 
\end{array} 
\end{equation} 
The \index{Diquark masses} mass ordering is characteristic and model-independent (cf.\ Refs.\,\cite{marisdq,gunner}, lattice-QCD estimates\,\cite{hess} and 
studies of the spin-flavour dependence of parton distributions\,\cite{Close:br,Schreiber,Alberg}), and indicates that a study of the $N$ and $\Delta$ must retain at least the scalar $[ud]$ correlations and pseudovector $(uu)$, $(ud)$ and $(dd)$ correlations if it is to be accurate.  NB.\ The spin-$3/2$ $\Delta$ is inaccessible unless pseudovector correlations are retained.

The simplest realistic representation of the Faddeev amplitude for the spin- and isospin-$1/2$ nucleon is therefore \index{Faddeev amplitude: nucleon}
\begin{equation} 
\label{Psi} \Psi_3(p_i,\alpha_i,\tau_i) = \check{N}_3^{0^+} + \check{N}_3^{1^+};
\end{equation} 
namely, a sum of scalar and axial-vector diquark correlations, with $(p_i,\alpha_i,\tau_i)$ the momentum, spin and isospin labels of the quarks constituting the bound state, and $P=p_1+p_2+p_3$ the system's total momentum.\footnote{NB.\ Hereafter isospin symmetry of the strong interaction is assumed; i.e., the $u$- and $d$-quarks are indistinguishable but for their electric charge.  This simplifies the form of the Faddeev amplitudes.}  For the $\Delta$, since it is not possible to combine an isospin-$0$ diquark with an isospin-$1/2$ quark to obtain isospin-$3/2$, the spin- and isospin-$3/2$ $\Delta$ contains only an axial-vector diquark component \index{Faddeev amplitude: $\Delta$}
\begin{equation}
\label{PsiD} \Psi^\Delta_3(p_i,\alpha_i,\tau_i) = \check{D}_3^{1^+}.
\end{equation} 

The scalar diquark piece in Eq.\,(\ref{Psi}) is \index{Diquark: scalar}
\begin{eqnarray} 
\check{N}_3^{0^+}(p_i,\alpha_i,\tau_i)&=& [\Gamma^{0^+}(\sfrac{1}{2}p_{[12]};K)]_{\alpha_1 
\alpha_2}^{\tau_1 \tau_2}\, \Delta^{0^+}(K) \,[\mathbf{S}(\ell;P) u(P)]_{\alpha_3}^{\tau_3}\,,%
\label{calS} 
\end{eqnarray} 
where: the spinor satisfies Eq.\,(\ref{DiracN}), and it is also a spinor in isospin space with $\varphi_+= {\rm col}(1,0)$ for the proton and $\varphi_-= {\rm col}(0,1)$ for the neutron; $K= p_1+p_2=: p_{\{12\}}$, $p_{[12]}= p_1 - p_2$, $\ell := (-p_{\{12\}} + 2 p_3)/3$; $\Delta^{0^+}$ is a pseudoparticle propagator for the scalar diquark formed from quarks $1$ and $2$, and $\Gamma^{0^+}\!$ is a Bethe-Salpeter-like amplitude describing their relative momentum correlation; and ${\mathbf S}$, a $4\times 4$ Dirac matrix, describes the relative quark-diquark momentum correlation.  (${\mathbf
S}$, $\Gamma^{0^+}$ and $\Delta^{0^+}$ are discussed in Sect.\,\ref{completing}.)  The necessary colour antisymmetry of $\Psi_3$ is implicit in $\Gamma^{J^P}\!\!$, with the 
Levi-Civita tensor, $\epsilon_{c_1 c_2 c_3}$, expressed via the antisymmetric Gell-Mann matrices; viz., defining 
\begin{equation} 
\{H^1=i\lambda^7,H^2=-i\lambda^5,H^3=i\lambda^2\}\,, 
\end{equation} 
then $\epsilon_{c_1 c_2 c_3}= (H^{c_3})_{c_1 c_2}$.  [See Eqs.\,(\ref{Gamma0p}), (\ref{Gamma1p}).]

The axial-vector component in Eq.\,(\ref{Psi}) is \index{Diquark: pseudovector}
\begin{equation} 
\check{N}^{1^+}(p_i,\alpha_i,\tau_i) =  [{\tt t}^i\,\Gamma_\mu^{1^+}(\sfrac{1}{2}p_{[12]};K)]_{\alpha_1 
\alpha_2}^{\tau_1 \tau_2}\,\Delta_{\mu\nu}^{1^+}(K)\, 
[{\mathbf A}^{i}_\nu(\ell;P) u(P)]_{\alpha_3}^{\tau_3}\,,
\label{calA} 
\end{equation} 
where the symmetric isospin-triplet matrices are 
\begin{equation} 
{\tt t}^+ = \frac{1}{\surd 2}(\tau^0+\tau^3) \,,\; 
{\tt t}^0 = \tau^1\,,\; 
{\tt t}^- = \frac{1}{\surd 2}(\tau^0-\tau^3)\,, 
\end{equation} 
and the other elements in Eq.\,(\ref{calA}) are straightforward generalisations of those in Eq.\,(\ref{calS}). 

The general form of the Faddeev amplitude for the spin- and isospin-$3/2$ $\Delta$ is complicated.  However, isospin symmetry means one can focus on the $\Delta^{++}$ with it's simple flavour structure, because all the charge states are degenerate, and consider 
\begin{equation}
{\mathbf D}_3^{1^+}= [{\tt t}^+ \Gamma^{1^+}_\mu(\sfrac{1}{2}p_{[12]};K)]_{\alpha_1 \alpha_2}^{\tau_1 \tau_2}
\, \Delta_{\mu\nu}^{1^+}(K) \, [{\mathbf D}_{\nu\rho}(\ell;P)u_\rho(P)\, \varphi_+]_{\alpha_3}^{\tau_3}\,, \label{DeltaAmpA} 
\end{equation} 
where $u_\rho(P)$ is a Rarita-Schwinger spinor, Eq.\,(\ref{rarita}).

The general forms of the matrices ${\mathbf S}(\ell;P)$, ${\mathbf A}^i_\nu(\ell;P)$ and ${\mathbf D}_{\nu\rho}(\ell;P)$, which describe the momentum space correlation between the quark and diquark in the nucleon and the $\Delta$, respectively, are described in Ref.\,\cite{oettelfe}.  The requirement that ${\mathbf S}(\ell;P)$ represent a positive energy nucleon; namely, that it be an eigenfunction of $\Lambda_+(P)$, Eq.\,(\ref{Lplus}), entails
\begin{equation}
\label{Sexp} 
{\mathbf S}(\ell;P) = s_1(\ell;P)\,I_{\rm D} + \left(i\gamma\cdot \hat\ell - \hat\ell \cdot \hat P\, I_{\rm D}\right)\,s_2(\ell;P)\,, 
\end{equation} 
where $\hat \ell^2=1$, $\hat P^2= - 1$.  Placing the same constraint on the axial-vector component, one has
\begin{equation}
\label{Aexp}
 {\mathbf A}^i_\nu(\ell;P) = \sum_{n=1}^6 \, p_n^i(\ell;P)\,\gamma_5\,A^n_{\nu}(\ell;P)\,,\; i=+,0,-\,,
\end{equation}
where ($ \hat \ell^\perp_\nu = \hat \ell_\nu + \hat \ell\cdot\hat P\, \hat P_\nu$, $ \gamma^\perp_\nu = \gamma_\nu + \gamma\cdot\hat P\, \hat P_\nu$)
\begin{equation}
\begin{array}{lll}
A^1_\nu= \gamma\cdot \hat \ell^\perp\, \hat P_\nu \,,\; &
A^2_\nu= -i \hat P_\nu \,,\; &
A^3_\nu= \gamma\cdot\hat \ell^\perp\,\hat \ell^\perp\,,\\
A^4_\nu= i \,\hat \ell_\mu^\perp\,,\; &
A^5_\nu= \gamma^\perp_\nu - A^3_\nu \,,\; &
A^6_\nu= i \gamma^\perp_\nu \gamma\cdot\hat \ell^\perp - A^4_\nu\,.
\end{array}
\end{equation}
Finally, requiring also that ${\mathbf D}_{\nu\rho}(\ell;P)$ be an eigenfunction of $\Lambda_+(P)$, one obtains
\begin{equation}
{\mathbf D}_{\nu\rho}(\ell;P) = {\mathbf S}^\Delta(\ell;P) \, \delta_{\nu\rho} + \gamma_5{\mathbf A}_\nu^\Delta(\ell;P) \,\ell^\perp_\rho \,,
\end{equation}
with ${\mathbf S}^\Delta$ and ${\mathbf A}^\Delta_\nu$ given by obvious analogues of Eqs.\,(\ref{Sexp}) and (\ref{Aexp}), respectively.

It is now possible to write the Faddeev equation satisfied by $\Psi_3$: \index{Faddeev equation: nucleon}
\begin{equation} 
 \left[ \begin{array}{r} 
{\mathbf S}(k;P)\, u(P)\\ 
{\mathbf A}^i_\mu(k;P)\, u(P) 
\end{array}\right]\\ 
 = -\,4\,\int\frac{d^4\ell}{(2\pi)^4}\,{\mathbf M}(k,\ell;P) 
\left[ 
\begin{array}{r} 
{\mathbf S}(\ell;P)\, u(P)\\ 
{\mathbf A}^j_\nu(\ell;P)\, u(P) 
\end{array}\right] .
\label{FEone} 
\end{equation} 
The kernel in Eq.~(\ref{FEone}) is 
\begin{equation} 
\label{calM} {\mathbf M}(k,\ell;P) = \left[\begin{array}{cc} 
{\mathbf M}_{00} & ({\mathbf M}_{01})^j_\nu \\ 
({\mathbf M}_{10})^i_\mu & ({\mathbf M}_{11})^{ij}_{\mu\nu}\rule{0mm}{3ex} 
\end{array} 
\right] 
\end{equation} 
with 
\begin{equation} 
\label{calM00}
 {\mathbf M}_{00} = \Gamma^{0^+}\!(k_q-\ell_{qq}/2;\ell_{qq})\, 
S^{\rm T}(\ell_{qq}-k_q) \,\bar\Gamma^{0^+}\!(\ell_q-k_{qq}/2;-k_{qq})\, 
S(\ell_q)\,\Delta^{0^+}(\ell_{qq}) \,, 
\end{equation} 
where: $\ell_q=\ell+P/3$, $k_q=k+P/3$, $\ell_{qq}=-\ell+ 2P/3$, 
$k_{qq}=-k+2P/3$, $\bar\Gamma$ is defined in Eq.\,(\ref{chargec}) and the superscript ``T'' denotes matrix transpose; and
\begin{eqnarray}
\nonumber
\lefteqn{({\mathbf M}_{01})^j_\nu= {\tt t}^j \,
\Gamma_\mu^{1^+}\!(k_q-\ell_{qq}/2;\ell_{qq})} \\
&& \times 
S^{\rm T}(\ell_{qq}-k_q)\,\bar\Gamma^{0^+}\!(\ell_q-k_{qq}/2;-k_{qq})\, 
S(\ell_q)\,\Delta^{1^+}_{\mu\nu}(\ell_{qq}) \,, \label{calM01} \\ 
\nonumber \lefteqn{({\mathbf M}_{10})^i_\mu = 
\Gamma^{0^+}\!(k_q-\ell_{qq}/2;\ell_{qq})\, 
}\\ 
&&\times S^{\rm T}(\ell_{qq}-k_q)\,{\tt t}^i\, \bar\Gamma_\mu^{1^+}\!(\ell_q-k_{qq}/2;-k_{qq})\, 
S(\ell_q)\,\Delta^{0^+}(\ell_{qq}) \,,\\ 
\nonumber \lefteqn{({\mathbf M}_{11})^{ij}_{\mu\nu} = {\tt t}^j\, 
\Gamma_\rho^{1^+}\!(k_q-\ell_{qq}/2;\ell_{qq})}\\ 
&&\times \, S^{\rm T}(\ell_{qq}-k_q)\,{\tt t}^i\, \bar\Gamma^{1^+}_\mu\!(\ell_q-k_{qq}/2;-k_{qq})\, 
S(\ell_q)\,\Delta^{1^+}_{\rho\nu}(\ell_{qq}) \,. \label{calM11} 
\end{eqnarray} 

The $\Delta$'s Faddeev equation is \index{Faddeev equation: $\Delta$}
\begin{eqnarray} 
{\mathbf D}_{\lambda\rho}(k;P)\,u_\rho(P) & = & 4\int\frac{d^4\ell}{(2\pi)^4}\,{\mathbf
M}^\Delta_{\lambda\mu}(k,\ell;P) \,{\mathbf D}_{\mu\sigma}(\ell;P)\,u_\sigma(P)\,, \label{FEDelta} 
\end{eqnarray} 
with
\begin{eqnarray}
\nonumber\lefteqn{ {\mathbf M}^\Delta_{\lambda\mu} = {\tt t}^+ 
\Gamma_\sigma^{1^+}\!(k_q-\ell_{qq}/2;\ell_{qq})}\\
&& \times\, 
 S^{\rm T}\!(\ell_{qq}-k_q)\, {\tt t}^+\bar\Gamma^{1^+}_\lambda\!(\ell_q-k_{qq}/2;-k_{qq})\, 
S(\ell_q)\,\Delta^{1^+}_{\sigma\mu}\!(\ell_{qq}). \label{MDelta}
\end{eqnarray}

The Faddeev equation is illustrated in Fig.\,\ref{faddeevpic}.  It is a linear, homogeneous matrix equation whose solution yields the Poincar\'e covariant Faddeev amplitude, which describes the relative quark-diquark isospin-, linear- and angular-momentum-distribution within the baryon.  An unamputated Faddeev \textit{wave function} is constructed from the amplitude by attaching the quark and diquark propagators; namely, 
\begin{equation}
\hat\Psi_3 = \left(\begin{array}{c} 
\hat {\mathbf S}(\ell;P)\, u(P)\\ 
\hat {\mathbf A}^j_\rho(\ell;P)\, u(P) 
\end{array}\right) := S(\ell_q)
\left(
\begin{array}{c} 
\Delta^{0^+}(\ell_{qq}) \, {\mathbf S}(\ell;P)\, u(P)\\ 
\Delta^{1^+}_{\rho\nu}(\ell_{qq}) {\mathbf A}^j_\nu(\ell;P)\, u(P) 
\end{array}\right)\,.
\end{equation}
It is this quantity that is most closely related to a wave function in quantum mechanics.  One can express $\hat {\mathbf S}(\ell;P) u(P)$ in the form of Eq.\,(\ref{Sexp}).  If one labels the two functions that appear as $\hat s_{1,2}$ then, in the nucleon rest frame, $\hat s_{1,2}$ describe, respectively, the upper, lower component of the bound-state nucleon's spinor.  The lower component is an essentially relativistic effect.  It is noteworthy that $\hat s_1$ has angular momentum zero while $\hat s_1$ has angular momentum one.  The lower component is sizeable even in the Bag Model of the nucleon.  It follows that the quarks possess a material amount of orbital angular momentum \index{Orbital angular momentum} even in that simple model.  Orbital angular momentum is not a Poincar\'e invariant.  However, if absent in a particular frame, it will almost inevitably appear in another frame related via a Poincar\'e transformation.  Nonzero quark orbital angular momentum is the necessary outcome of a Poincar\'e covariant description.  This is why the covariant Faddeev amplitude is a matrix-valued function with a rich structure that, in the baryons' rest frame, corresponds to a relativistic wave function with $s$-wave, $p$-wave and even $d$-wave components. (Details can be found in Ref.\,\cite{oettelthesis}, Sect.\,2.4.) 

\begin{figure}[t]
\centerline{%
\includegraphics[clip,width=0.70\textwidth]{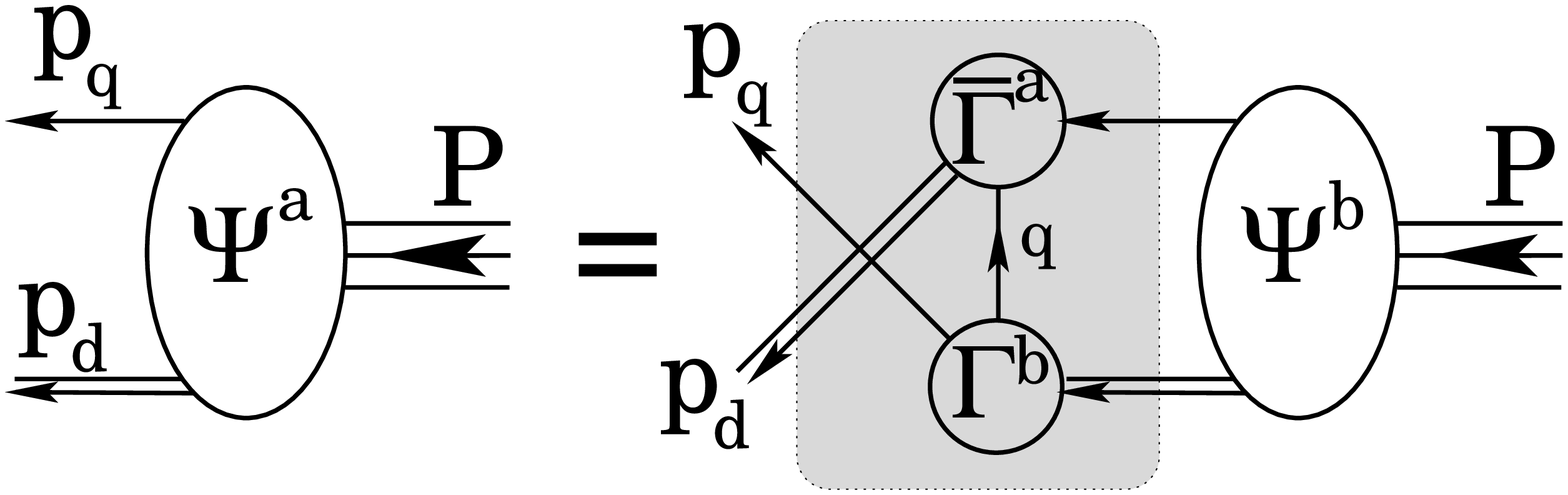}}
\caption{\label{faddeevpic} Pictorial representation of the Faddeev equation, Eq.\,(\ref{FEone}).  A nucleon of four-momentum $P$ is constituted from a dressed-quark (single line, momentum $p_q=k_q$) and dressed-diquark (double line, momentum $p_d=k_{qq}$).  Binding is effected by an iterated exchange of roles between the bystander and diquark-participant quarks, which is described by the kernels $\mathbf M$ in Eqs.\,(\protect\ref{calM00}) -- (\protect\ref{calM11}), (\protect\ref{MDelta}).  The exchange takes place within the shaded region.}
\end{figure}

\subsubsection{Faddeev equation kernels}
\index{Faddeev equation kernels}
\label{completing} 
To complete the Faddeev equations, Eqs.\,(\ref{FEone}) \& (\ref{FEDelta}), one must specify the dressed-quark propagator, the diquark Bethe-Salpeter amplitudes and the diquark propagators that appear in the kernels.

\paragraph{Dressed-quark propagator}
This propagator has the general form given in Eq.\,(\ref{Sgeneral}) and herein an overview of its properties has already been provided.  In solving the Faddeev equation described above, one is required to repeatedly evaluate some eight-dimensional integrals.  Thus, while it is straightforward to obtain a numerical solution of a truncation of QCD's gap equation, the utility of an algebraic form for $S(p)$ is self-evident.  An efficacious parametrisation, which exhibits the features described above and has been used extensively in studies of hadron properties, is \index{Quark propagator: algebraic model} expressed via
\begin{eqnarray} 
\bar\sigma_S(x) & =&  2\,\bar m \,{\check F}(2 (x+\bar m^2)) + {\check
F}(b_1 x) \,{\check F}(b_3 x) \,  
\left[b_0 + b_2 {\check F}(\epsilon x)\right]\,,\label{ssm} \\ 
\label{svm} \bar\sigma_V(x) & = & \frac{1}{x+\bar m^2}\, \left[ 1 - {\check F}(2 (x+\bar m^2))\right]\,, 
\end{eqnarray}
with $x=p^2/\lambda^2$, $\bar m$ = $m/\lambda$, ${\check F}(x)= (1-\mbox{\rm e}^{-x})/x$, $\bar\sigma_S(x) = \lambda\,\sigma_S(p^2)$ and $\bar\sigma_V(x) =
\lambda^2\,\sigma_V(p^2)$.  The mass-scale, $\lambda=0.566\,$GeV, and
parameter values\footnote{$\epsilon=10^{-4}$ in Eq.\ (\ref{ssm}) serves only to
decouple the large- and intermediate-$p^2$ domains.}
\begin{equation} 
\label{tableA} 
\begin{array}{ccccc} 
   \bar m& b_0 & b_1 & b_2 & b_3 \\\hline 
   0.00897 & 0.131 & 2.90 & 0.603 & 0.185 
\end{array}\;, 
\end{equation} 
were fixed in a least-squares fit to light-meson observables\,\cite{mark,valencedistn}.  The dimensionless $u=d$ current-quark mass in Eq.~(\ref{tableA}) corresponds to
\begin{equation} 
\label{mcq}
m=5.1\,{\rm MeV}\,,
\end{equation} 
and the parametrisation yields a Euclidean \index{Constituent-quark mass} constituent-quark mass, Eq.\,(\ref{CQM}),
\begin{equation} 
\label{MEq} M_{u,d}^E = 0.33\,{\rm GeV},
\end{equation}
which agrees semiquantitatively with that produced by a renormalisation-group-improved rainbow-ladder truncation based on Eq.\,(\ref{gk2}) [see Eq.\,(\ref{rainbowCQmasses})].

\paragraph{Diquark Bethe-Salpeter amplitudes}
\index{Diquark Bethe-Salpeter amplitudes}
The rainbow-ladder DSE truncation yields asymptotic diquark states in the strong interaction spectrum.  Such states are not observed and their appearance is an artefact of the truncation.  Higher order terms in the quark-quark scattering kernel, whose analogue in the quark-antiquark channel do not much affect the properties of vector and flavour non-singlet pseudoscalar mesons, ensure that QCD's quark-quark scattering matrix does not exhibit singularities which correspond to asymptotic diquark states \cite{bhagwatvertex,truncscheme,detmold,njldiquark}.  Nevertheless, studies with kernels that do not produce diquark bound states, do support a physical interpretation of the masses, $m_{(qq)_{J^P}}$, obtained using the rainbow-ladder truncation: the quantity $l_{(qq)_{J^P}}=1/m_{(qq)_{J^P}}$ may be interpreted as a range over which the diquark correlation can persist inside a baryon.  These observations motivate the {\it Ansatz} for the quark-quark scattering matrix that is employed in deriving the Faddeev equation: \begin{equation} 
[G_{qq}(k,q;K)]_{rs}^{tu} = \sum_{J^P=0^+,1^+,\ldots} \bar\Gamma^{J^P}\!(k;-K)\, \Delta^{J^P}\!(K) \, \Gamma^{J^P}\!(q;K)\,. \label{AnsatzMqq} 
\end{equation}  

One practical means of specifying the $\Gamma^{J^P}\!\!$ in Eq.\,(\ref{AnsatzMqq}) is to employ the solutions of a rainbow-ladder quark-quark BSE.  Using the properties of the Gell-Mann matrices one finds easily that $\Gamma^{J^P}_C:= \Gamma^{J^P}C^\dagger$ satisfies exactly the same equation as the $J^{-P}$ colour-singlet meson {\it but} for a halving of the coupling strength\,\cite{regdq}.  This makes clear that the interaction in the ${\bar 3_c}$ $(qq)$ channel is strong and attractive.\footnote{The same analysis shows the interaction to be strong and repulsive in the ${6_c}$ (colour-sextet) $(qq)$ channel.}  Moreover, it follows as a feature of the rainbow-ladder truncation that, independent of the specific form of a model's interaction, the calculated masses satisfy \begin{equation}
m_{(qq)_{J^P}} > m_{(\bar q q)_{J^{-P}}}\,.
\end{equation}
This is a useful guide for all but scalar diquark correlations because the partnered mesons in that case are pseudoscalars whose ground state masses, as has been shown herein, are constrained to be small by Goldstone's theorem and which therefore provide a weak lower bound.  For the correlations relevant herein, models typically give masses (in GeV, recall Eq.\,(\ref{dqmass}) and the associated discussion):
\begin{equation}
\label{diquarkmass}
m_{(ud)_{0^+}} = 0.74 - 0.82 \,,\; m_{(uu)_{1^+}}=m_{(ud)_{1^+}}=m_{(dd)_{1^+}}=0.95 - 1.02\,.
\end{equation}

A solution of the BSE equation requires a simultaneous solution of the quark-DSE \cite{marisdq}.  However, since in the model under discussion the calculations are simplified by parametrising $S(p)$, that expedient should also be employed with $\Gamma^{J^P}\!$:
\begin{eqnarray} 
\label{Gamma0p} \Gamma^{0^+}(k;K) &=& \frac{1}{{\mathbf N}^{0^+}} \, 
H^a\,C i\gamma_5\, i\tau_2\, {\check F}(k^2/\omega_{0^+}^2) \,, \\ 
\label{Gamma1p} {\tt t}^i \Gamma^{1^+}_\mu (k;K) &=& \frac{1}{{\mathbf N}^{1^+}}\, 
H^a\,i\gamma_\mu C\,{\tt t}^i\, {\check F}(k^2/\omega_{1^+}^2)\,, 
\end{eqnarray} 
with the normalisation, ${\mathbf N}^{J^P}\!$, fixed by\footnote{This is the normalisation condition when the two-body scattering kernel is independent of the system's total momentum.  Rainbow-ladder is one such truncation.  For quark-antiquark bound states the condition has the same form except for the replacement $S^{\rm T}(-q+Q/2) \to S(q-Q/2)$ in Eq.\,(\protect\ref{PiKQ}). \label{fnnorm2}}
\begin{eqnarray}
\label{BSEnorm} 
2 \,K_\mu & = & 
\left[ \frac{\partial}{\partial Q_\mu} \Pi(K,Q) \right]_{Q=K}^{{K^2=-m_{J^P}^2}},
\end{eqnarray}
where
\begin{equation}
\label{PiKQ}
\Pi(K,Q) = {\rm tr}\!\! \int\!\! 
\frac{d^4 q}{(2\pi)^4}\, \bar\Gamma(q;-K) \, S(q+Q/2) \, \Gamma(q;K) \, S^{\rm T}(-q+Q/2) .
\end{equation}
These {\it Ans\"atze} retain only that single Dirac-amplitude which would represent a point particle with the given quantum numbers in a local Lagrangian density: they are usually the dominant amplitudes in a solution of the rainbow-ladder BSE for the lowest mass $J^P$ mesons\,\cite{marisphotonvertex,mr97,maristandy1} and diquarks \cite{cjbsep,marisdq}.

\paragraph{Diquark propagators}
\index{Diquark propagators}
Solving for the quark-quark scattering matrix using the rainbow-ladder truncation yields free particle propagators for $\Delta^{J^P}$ in Eq.\,(\ref{AnsatzMqq}).  Higher order contributions remedy that defect, \index{Confinement} eliminating asymptotic diquark states from the spectrum.  The attendant modification of $\Delta^{J^P}$ can be modelled efficiently by simple functions that are free-particle-like at spacelike momenta but pole-free on the timelike axis (see the earlier discussion of confinement, page~\pageref{confpage}); namely, 
\begin{eqnarray} 
\Delta^{0^+}(K) & = & \frac{1}{m_{0^+}^2}\,{\check F}(K^2/\omega_{0^+}^2)\,,\\ 
\Delta^{1^+}_{\mu\nu}(K) & = & 
\left(\delta_{\mu\nu} + \frac{K_\mu K_\nu}{m_{1^+}^2}\right) \, \frac{1}{m_{1^+}^2}\, {\check F}(K^2/\omega_{1^+}^2) \,,
\end{eqnarray} 
where the two parameters $m_{J^P}$ are diquark pseudoparticle masses and 
$\omega_{J^P}$ are widths characterising $\Gamma^{J^P}\!$.  It is useful to require additionally that
\begin{equation}
\label{DQPropConstr}
\left. \frac{d}{d K^2}\,\left(\frac{1}{m_{J^P}^2}\,{\check F}(K^2/\omega_{J^P}^2)\right)^{-1} \right|_{K^2=0}\! = 1 \; \Rightarrow \; \omega_{J^P}^2 = \sfrac{1}{2}\,m_{J^P}^2\,,
\end{equation} 
which is a normalisation that accentuates the free-particle-like propagation characteristics of the diquarks {\it within} the hadron. 

\subsection{Nucleon and \mbox{\boldmath $\Delta$} Masses}
\index{Masses of nucleon and \mbox{\boldmath $\Delta$}}
\label{NDmasses}
The Faddeev equations, Eqs.\,(\ref{FEone}) \& (\ref{FEDelta}), are now completely specified.  The three-body problem is intrinsically complex and thus it may appear to have been a complicated process.  However, one has merely combined existing information about the one- and two-body sectors of QCD. 

\subsubsection{Meson loops and baryon masses}
\index{Mass of nucleon and chiral loops}
\index{Chiral loops and baryon masses}
\label{loopmass}
Before reporting the solution of the Faddeev equation it is worthwhile to explicate the effect of pseudoscalar meson loops on baryon masses.  A thorough discussion is provided in Ref.\,\cite{hechtfe}.  One can begin with the leading term in a pion-nucleon chiral Lagrangian:\footnote{That part which describes the pseudoscalar field alone has been neglected. NB.\ This subsection, Eqs.\,(\protect\ref{LpiNL}) -- (\protect\ref{shiftF}),  employs Minkowski metric for ease of comparison with textbook material.}
\begin{equation} 
\label{LpiNL} \bar N(x)\left[ i \not \!\partial - M  +\frac{g}{2 M} \gamma_5 
\gamma^\mu \,\vec{\tau} \cdot \partial_\mu\vec{\pi}(x) + \ldots \right] N(x)\,. 
\end{equation} 
The rainbow truncation DSE for the nucleon in this theory is
\begin{equation} 
\Sigma(P) = 3i \frac{g^2}{4 M^2} \int \frac{d^4 k}{(2\pi)^4} \, 
 \Delta(k^2,m_\pi^2) \! \not\! k\,\gamma_5\, G(P-k)  \! \not\! k\,\gamma_ 5\,,
\label{AVDSE} 
\end{equation} 
wherein the nucleon propagator 
\begin{eqnarray} 
\lefteqn{G(P) = \frac{1}{\not\! P - M -\Sigma(P)} =:  G^+(P) + G^-(P)}\\
\nonumber & & = \frac{M}{\omega_N(\vec{P})} \left[ \Lambda_+(\vec{P}) 
\frac{1}{P_0 - \omega_N(\vec{P}) + i \varepsilon} + \, \Lambda_-(\vec{P}) \frac{1}{P_0 + \omega_N(\vec{P}) - i 
\varepsilon} \right]\,, \\
\end{eqnarray} 
with $\omega_N^2(\vec{P}) = \vec{P}^2 + M^2$, and where $\Lambda_\pm(\vec{P}) = (\not\!\! \tilde{P} \pm M)/(2M)$, $\tilde P=(\omega(\vec{P}),\vec{P})$, are the Minkowski space positive and negative energy projection operators, respectively; and the pion propagator ($\omega_\pi^2(\vec{k}) = \vec{k}^2 + m_\pi^2$)
\begin{eqnarray} 
\lefteqn{ \Delta(k^2,m_\pi^2)  = \frac{1}{k^2 - m_\pi^2 + i \varepsilon}}\\ 
& = & \frac{1}{2\,\omega_\pi(\vec{k})}\left[ 
\frac{1}{k_0 - \omega_\pi(\vec{k}) + i\varepsilon} - 
\frac{1}{k_0 + \omega_\pi(\vec{k}) - i\varepsilon} \right] .\\ 
\end{eqnarray} 

As written, the integral in Eq.\,(\ref{AVDSE}) is divergent.  It must be regularised to give it meaning.  In this case the Poincar\'e invariant Paul-Villars procedure is useful.  It may be effected by modifying the $\pi$-propagator: 
\begin{equation} 
\label{DpiPV} \Delta(k^2,m_\pi^2) \to \bar\Delta_\pi(k^2) 
= \Delta(k^2,m_\pi^2) + \sum_{i=1,2} c_i\, \Delta(k^2,\lambda_i^2)\,, 
\end{equation} 
and then, with 
\begin{equation} 
c_1= - \,\frac{\lambda_2^2-m_\pi^2}{\lambda^2_2-\lambda^2_1}\,,\; 
c_2= \frac{\lambda_1^2-m_\pi^2}{\lambda^2_2-\lambda^2_1} \,,
\end{equation} 
Eq.~(\ref{DpiPV}) yields 
\begin{eqnarray} 
\nonumber \bar\Delta_\pi(k^2) & = & \Delta(k^2,m_\pi^2)\, 
\prod_{i=1,2} \, (\lambda_i^2-m_\pi^2) \, \Delta(k^2,\lambda_i^2)\,,\\ 
&& 
\end{eqnarray} 
in which case the integrals are convergent for any fixed $\lambda_{1,2}$.  Furthermore, for $m_\pi \ll \lambda_1\to \lambda_2 = \lambda$ 
\begin{eqnarray} 
\label{PVtovertex} 
\bar\Delta_\pi(k^2)& = & \Delta(k^2,m_\pi^2) \, \Delta^2(k^2/\lambda^2,1)
\end{eqnarray} 
i.e., the Pauli-Villars regularisation is equivalent to employing a monopole form factor at each $\pi N N$ vertex: $g \to g\, \Delta(k^2/\lambda^2,1)$, where $k$ is the pion's momentum.  Since this procedure modifies the pion propagator it may be interpreted as expressing compositeness of the pion and regularising its off-shell contribution (a related effect is identified in Refs.\,\cite{reglaws,rhopipipeter}) but that interpretation is not unique. 

The contribution to the nucleon's mass from a positive-energy nucleon, $G^+(P)$, in the loop described by Eq.\,(\ref{AVDSE}) is
\begin{equation} 
\delta^A M_+^+  =  -\,\frac{3 g^2}{16 M^2} \int\frac{d^3 k}{(2\pi)^3} \frac{1}{\omega_N} \sum_{i=0,1,2} c_i\, \frac{\lambda_i^2 (\omega_N - M) + 2 \vec{k}^2 
(\omega_{\lambda_i} + \omega_N)} {\omega_{\lambda_i} [ \omega_{\lambda_i} + 
\omega_N - M ]}\, , 
\label{deltaMApp} 
\end{equation} 
with $\omega_N=\omega_N(\vec{k}^2)$, etc.  The connection between this and other mass-shift calculations can be made transparent by writing Eq.\,(\ref{deltaMApp}) in the form 
\begin{equation} 
\delta^A M_+^+ =  -\,6\pi\, \frac{f^2_{NN\pi}}{m_\pi^2} \int\frac{d^3 
k}{(2\pi)^3} \, \frac{\vec{k}^2\,u^2(\vec{k}^2)}{\omega_\pi(\vec{k}^2) [ 
\omega_\pi(\vec{k}^2) + \omega_N(\vec{k}^2) - M]} \,,
\label{deltaMppCBM} 
\end{equation} 
where $f_{NN\pi}^2 = g^2 m_\pi^2/(16 \pi M^2)$ and
\begin{equation} 
\vec{k}^2\, u^2(\vec{k}^2) :=
\frac{\omega_{\lambda_0}}{2\,\omega_N} \,  
[ \omega_{\lambda_0} + \omega_N - M ] 
\sum_{i=0,1,2} c_i\, \frac{\lambda_i^2 (\omega_N - M) + 2 \vec{k}^2 
(\omega_{\lambda_i} + \omega_N)} {\omega_{\lambda_i} [ \omega_{\lambda_i} + 
\omega_N - M ]}\,.  \label{ukdef} 
\end{equation} 
This is useful because for $m_\pi \ll  \lambda_1 \to \lambda_2= \lambda$; i.e., on the domain in which Eq.\,(\ref{PVtovertex}) is valid, one finds algebraically that 
\begin{equation} 
\label{uklimit} 
u(\vec{k}^2) = 1/(1+\vec{k}^2/\lambda^2)\,, 
\end{equation} 
which firmly establishes the qualitative equivalence between Eq.\,(\ref{deltaMApp}) and the calculation, e.g., in Refs.\,\cite{tonyCBM1,tonyCBM2,jerryCBM,tonyANU}. 

It is instructive to consider Eq.\,(\ref{deltaMApp}) further.  Suppose that $M$ 
is very much greater than the other scales, then on the domain in which the 
integrand has significant support 
\begin{equation} 
\omega_N(\vec{k}^2) - M \approx \frac{\vec{k}^2}{2 M} 
\end{equation} 
and Eq.\,(\ref{deltaMApp}) yields
\begin{equation} 
\nonumber \delta_A M_+^+ \approx  - \frac{3 g^2}{8 M^2} \int \frac{d^3 k}{(2\pi)^3} \, 
\vec{k}^2  \sum_{i=0,1,2} 
\frac{c_i}{\omega_{\lambda_i}^2\!(\vec{k}^2)} \,. \label{deltaMAppNR} 
\end{equation} 
It follows that
\begin{equation} 
\frac{d^2 \,\delta_A M_+^+}{(d m_\pi^2)^2} \approx -\,\frac{3 g^2}{4 M^2} \, 
\int \frac{d^3 k}{(2\pi)^3} \frac{\vec{k}^2}{\omega_\pi^6(\vec{k}^2)}
=-\, \frac{9}{128 \pi} \frac{g^2}{M^2} \frac{1}{m_\pi}\,.
\end{equation} 
Hence on the domain considered, 
\begin{equation} 
\label{LNA} \delta_A M_+^+ = -\,\frac{3}{32\pi} \frac{g^2}{M^2} m_\pi^3 + 
f^+_{(1)}(\lambda_1,\lambda_2)\,m_\pi^2 + f^+_{(0)}(\lambda_1,\lambda_2)\,, 
\end{equation} 
where, as the derivation makes transparent, $f_{(0,1)}$ are scheme-dependent functions of (only) the regularisation parameters but the first term is regularisation-scheme-independent.  Given that $m_\pi^2 \propto \hat m$, Eq.\,(\ref{gmor}), this first term is nonanalytic in the current-quark mass.  It is the leading nonanalytic contribution, a much touted feature of effective field theory, and its coefficient is fixed by chiral symmetry and the pattern by which that symmetry is dynamically broken.  NB.\ The contribution from $G^-(P)$, which produces the so-called $Z$-diagram, is suppressed by $1/M$ \cite{hechtfe}.

While the leading nonanalytic contribution is model-independent, and thus provides a constraint on models that purport to represent QCD, it is not of particular quantitative use in this or related studies.  The pion and nucleon are both of finite size and hence the regularisation parameters $\lambda_{1,2}$, which set a compositeness scale for the $\pi N N$ vertex, must assume soft values; e.g., $\lsim 600\,$MeV \cite{oettel2,jacquesmyriad,tonysoft}.  Therefore the actual value of the pion-loop contribution to the nucleon's mass is completely determined by the regularisation-scheme-dependent terms.\footnote{From a chiral perturbation theory perspective it might be argued that such $\lambda$-dependent contributions correspond to a short distance effect ($d \sim 1/\lambda \sim 1/3\,$fm) and are therefore absorbed into the \textit{a priori} unknown counter-term essential within that framework.  In that approach their effect is thus unobservable.  However, this is not so in a framework that resolves the internal structure of hadrons.  In that case the $\lambda$-dependent contribution from these loops adds unambiguously and quantifiably to the quark core defined via the dressed-quark degrees of freedom.  The sum predicts a value for the counter-term. \label{fnchiral}}

This is thoroughly explored in Ref.\,\cite{hechtfe}, wherein a self-consistent solution of the nucleon's gap equation shows that the one-loop result is accurate to within $95$\%.  A full consideration leads to the conclusion that the shift in the nucleon's mass owing to the $\pi N$-loop is (in GeV, for $g_A=1.26$): 
\begin{equation} 
\label{shiftF} - \delta M_+ \simeq  ( 0.039 - 0.063 ) \, g_A^2 = (0.061 - 
0.099)\,. 
\end{equation} 
Thus one arrives at a robust result: the $\pi N$-loop reduces the nucleon's mass by $100\,$-$200\,$MeV.  Extant calculations; e.g., Refs.\,\cite{tonyCBM1,tonyCBM2,jerryCBM,tonyANU}, show that the contribution from the analogous $\pi \Delta$-loop is of the same sign and no greater in magnitude so that the likely total reduction is $200$-$400\,$MeV.  Based on these same calculations one anticipates that the $\Delta$ mass is also reduced by $\pi$ loops but by a smaller amount ($\sim 50\,$-$\,100\,$MeV less).

How is that effect to be incorporated into the quark-diquark picture of baryons?  Reference \cite{hechtfe} argued that it may be included by solving the Faddeev equation with target nucleon and $\Delta$ masses that are inflated to allow for the loop corrections.  Namely, that a physically sensible picture of the quark-diquark piece of the nucleon can be obtained if its parameters are chosen not so as to give the experimental masses, but higher values; e.g., $M_N=0.94 + 0.2=1.14\,$GeV, $M_\Delta= 1.232+0.1=1.332\,$GeV. 

\subsubsection{\mbox{\boldmath $N$} and \mbox{\boldmath $\Delta$} masses from the Faddeev equation}
\index{Masses of nucleon and \mbox{\boldmath $\Delta$} from the Faddeev equation}
\label{faddeevsolution}
The method described in Ref.\,\cite{oettelcomp} is effective for solving Eqs.\,(\ref{FEone}) \& (\ref{FEDelta}).  Owing to Eq.\,(\ref{DQPropConstr}), the masses of the scalar and axial-vector diquarks are the only variable parameters in the kernels of the Faddeev equations.  It is natural to choose the axial-vector mass so as to obtain a desired mass for the $\Delta$, and fix the scalar mass subsequently by requiring a particular nucleon mass.  Two primary parameter sets are presented in Table~\ref{ParaFix}.  Set~A is obtained by requiring a precise fit to the experimental nucleon and $\Delta$ masses, while Set~B was obtained by fitting to nucleon and $\Delta$ masses that are inflated so as to allow for the additional attractive contribution from the pion cloud, as described in Sect.\,\ref{loopmass}.

\begin{table}[t]
\caption{Mass-scale parameters \index{Diquark masses} (in GeV) for the scalar and axial-vector diquark correlations, fixed by fitting nucleon and $\Delta$ masses: Set~A provides a fit to the actual masses; whereas Set~B provides masses that are offset to allow for ``pion cloud'' contributions, Sect.\,\protect\ref{loopmass}.  Also listed is $\omega_{J^{P}}=m_{J^{P}}/\surd 2$, which is the width-parameter in the $(qq)_{J^P}$ Bethe-Salpeter amplitude, Eqs.\,(\protect\ref{Gamma0p}) \& (\protect\ref{Gamma1p}):  its inverse is an indication of the diquark's matter radius.  Sets A$^\ast$ and B$^\ast$ illustrate effects of omitting the axial-vector diquark correlation: the $\Delta$ cannot be formed and $M_N$ is significantly increased.  It is thus plain that the axial-vector diquark provides significant attraction in the Faddeev equation's kernel. (Adapted from Ref.\,\protect\cite{sigmaterms}.)}
{\normalsize
\begin{tabular*}{1.0\textwidth}{
l@{\extracolsep{0ptplus1fil}}c@{\extracolsep{0ptplus1fil}}c@{\extracolsep{0ptplus1fil}}
c@{\extracolsep{0ptplus1fil}} c@{\extracolsep{0ptplus1fil}}c@{\extracolsep{0ptplus1fil}}c@{\extracolsep{0ptplus1fil}}}
\hline
set & $M_N$ & $M_{\Delta}$~ & $m_{0^{+}}$ & $m_{1^{+}}$~ &
$\omega_{0^{+}} $ & $\omega_{1^{+}}$ \\
\hline
A & 0.94 & 1.23~ & 0.63 & 0.84~ & 0.44=1/(0.45\,{\rm fm}) & 0.59=1/(0.33\,{\rm fm}) \\
B & 1.18 & 1.33~ & 0.79 & 0.89~ & 0.56=1/(0.35\,{\rm fm}) & 0.63=1/(0.31\,{\rm fm}) \\\hline
A$^\ast$ & 1.15 &  & 0.63 &  & 0.44=1/(0.45\,{\rm fm}) &  \\
B$^\ast$ & 1.46 &  & 0.79 &  & 0.56=1/(0.35\,{\rm fm}) &  \\
\hline
\end{tabular*}\label{ParaFix} }
\end{table}

It is apparent in Table~\ref{ParaFix} that a baryon's mass increases with increasing diquark mass, and the fitted diquark mass-scales are commensurate with the anticipated values, cf.\ Eq.\,(\ref{diquarkmass}), with Set~B in better accord.  If coupling to the axial-vector diquark channel is omitted from Eq.\,(\ref{FEone}), then $M_N^{\rm Set\,A} = 1.15\,$GeV and $M_N^{\rm Set\,B} = 1.46\,$GeV, rows labelled A$^\ast$, B$^\ast$, respectively.  It is thus clear that axial-vector diquark correlations provide significant attraction in the nucleon.  Of course, using the Faddeev equation approach, the $\Delta$ does not exist without axial-vector correlations.  In Set~B the amount of attraction provided by axial-vector correlations must be matched by that provided by the pion cloud.  This highlights the constructive interference between the contribution of these two effects to a baryons' mass.  It is related and noteworthy that $m_{1^+}-m_{0^+}$ is only a reasonable approximation to $M_\Delta - M_N=0.29\,$GeV when pion cloud effects are ignored: Set~A, $m_{1^+}-m_{0^+}=0.21\,$GeV cf.\ Set~B, $m_{1^+}-m_{0^+}=0.10\,$GeV.  Plainly, understanding the $N$-$\Delta$ mass splitting requires more than merely reckoning the mass-scales of constituent degrees of freedom.  It is curious that for Set~B the matter radius, $1/\omega_{J^P}$, of the diquarks is smaller than for Set~A.  One might view this as a contraction in the hadron's quark-core in response to the presence of a pion cloud that contributes at longer range.  

\subsection{Nucleon Electromagnetic Form Factors}
\index{Electromagnetic form factors: nucleon}
\label{nucleonform}
\subsubsection{Nucleon-photon vertex}
\index{Photon-nucleon vertex}
\label{Ncurrent}
The nucleon's electromagnetic current is given in Eq.\,(\ref{jmuN}), which introduces the Dirac and Pauli form factors, and also the electric and magnetic form factors, which are related to the electric-charge-density distribution and the magnetic-current-density distribution.  In Eq.\,(\ref{jmuN}), $\Lambda_\mu$ is the nucleon-photon vertex.  It may be constructed following the systematic procedure of Ref.\,\cite{oettelpichowsky}.  That approach has the merit of automatically providing a conserved current for on-shell nucleons described by the Faddeev amplitudes which are obtained simultaneously with the mass.  Moreover, the canonical normalisation condition for the nucleons' Faddeev amplitude is equivalent to requiring $F_1(Q^2=0)=1$ for the proton.  The vertex has six terms, which are depicted in Fig.~\ref{vertex}.  Hereafter they are briefly described.  A full explanation is provided in Ref.\,\cite{arneJ}.
\smallskip

\begin{figure}[t]
\begin{minipage}[t]{\textwidth}
\begin{minipage}[t]{0.45\textwidth}
\leftline{\includegraphics[width=0.90\textwidth]{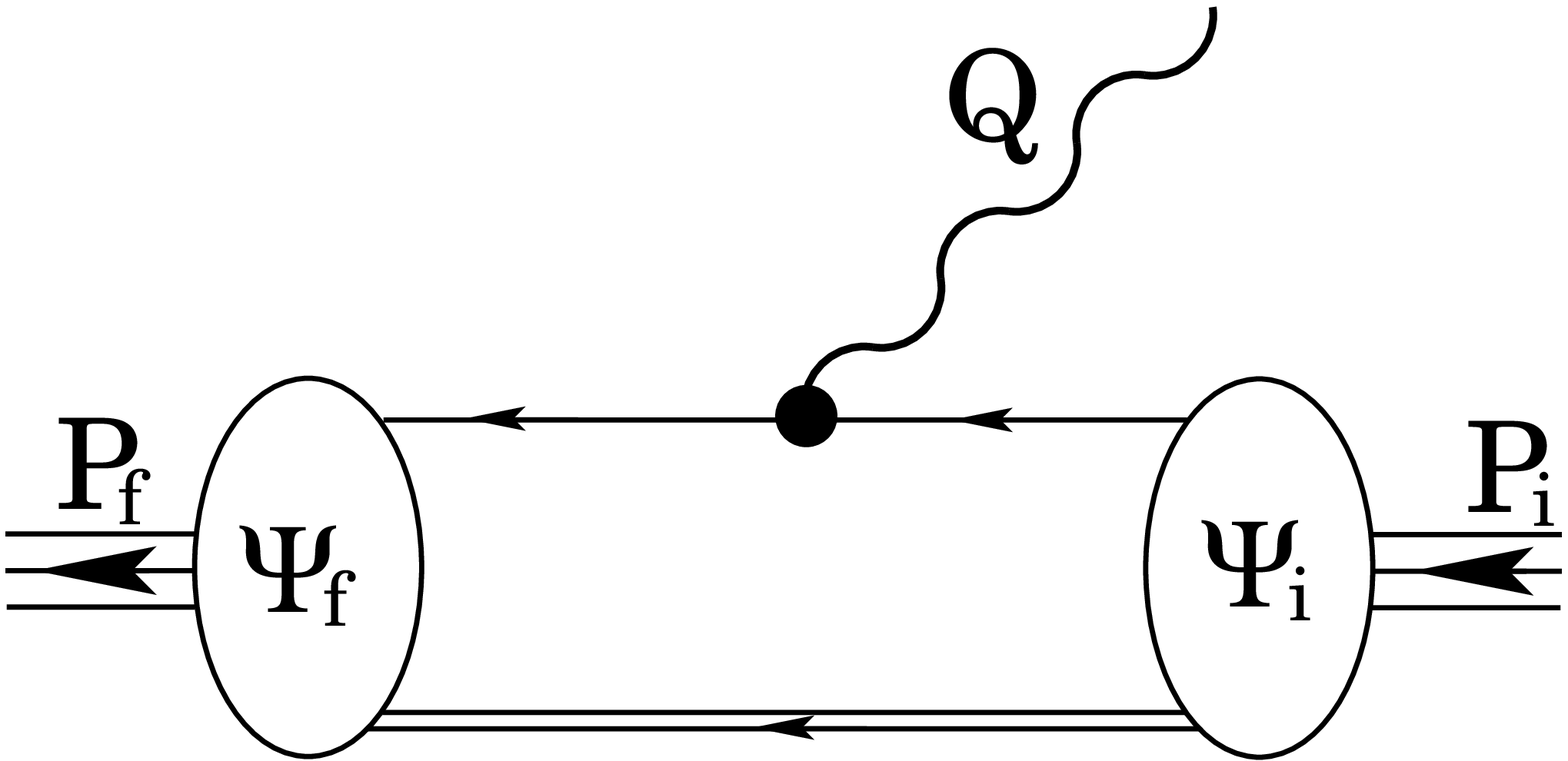}}
\end{minipage}
\begin{minipage}[t]{0.45\textwidth}
\rightline{\includegraphics[width=0.90\textwidth]{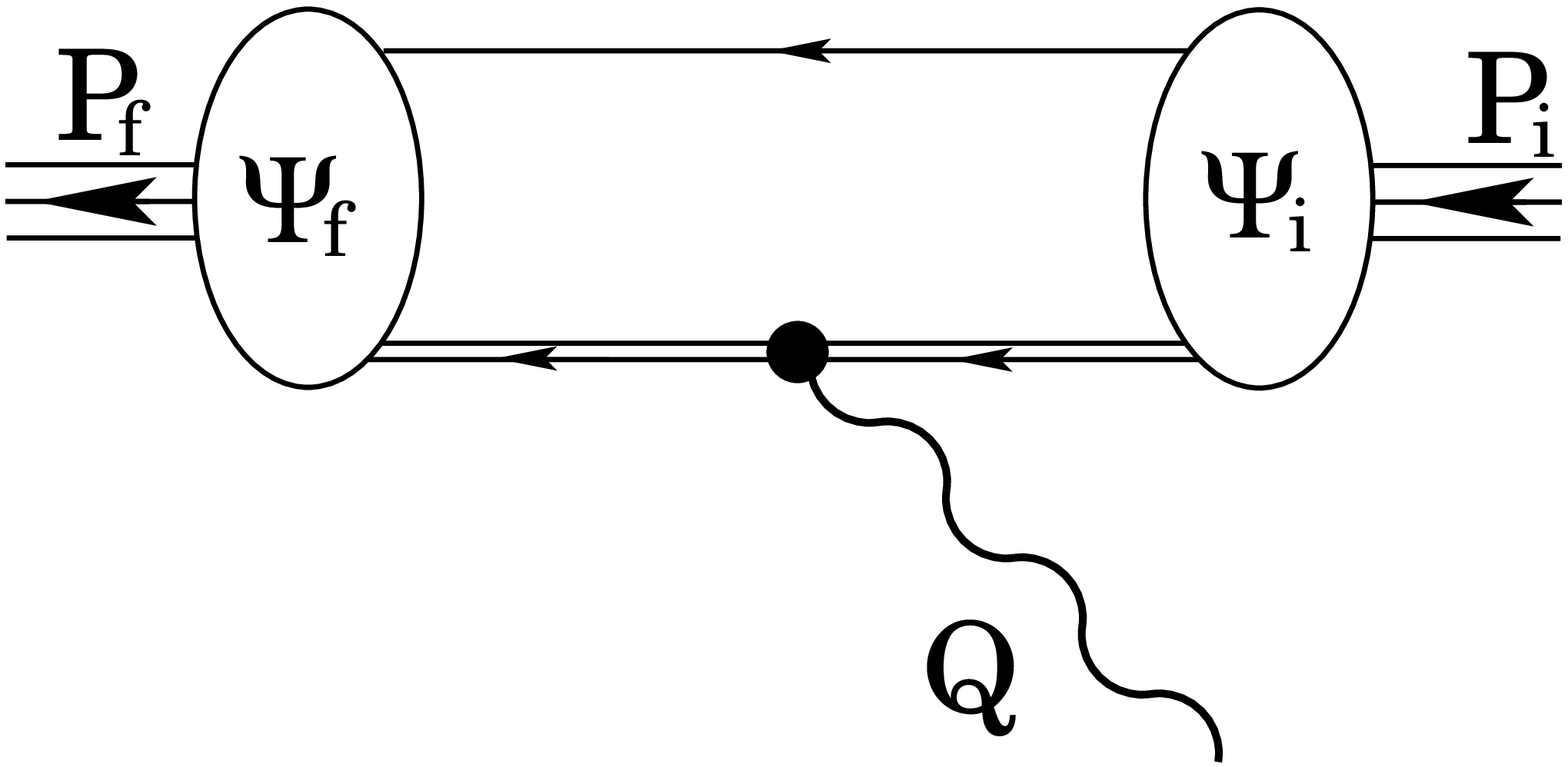}}
\end{minipage}\vspace*{3ex}

\begin{minipage}[t]{0.45\textwidth}
\leftline{\includegraphics[width=0.90\textwidth]{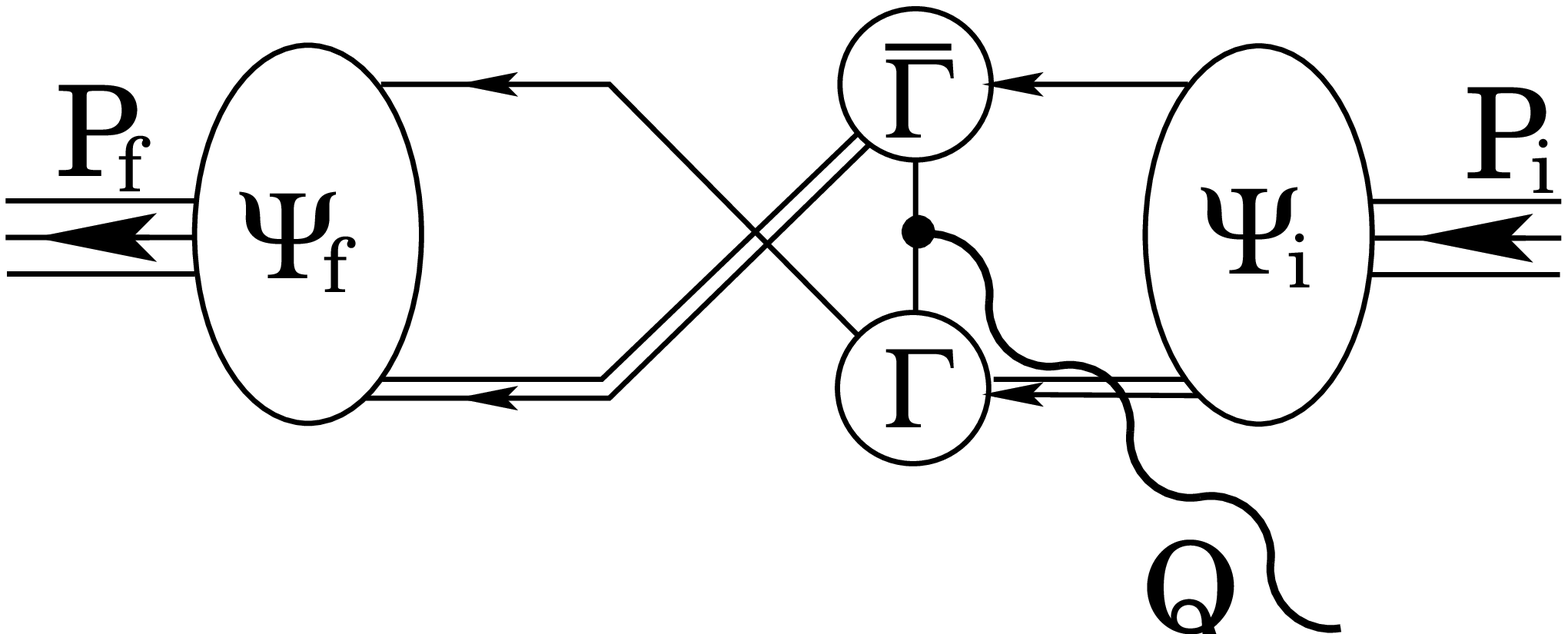}}
\end{minipage}
\begin{minipage}[t]{0.45\textwidth}
\rightline{\includegraphics[width=0.90\textwidth]{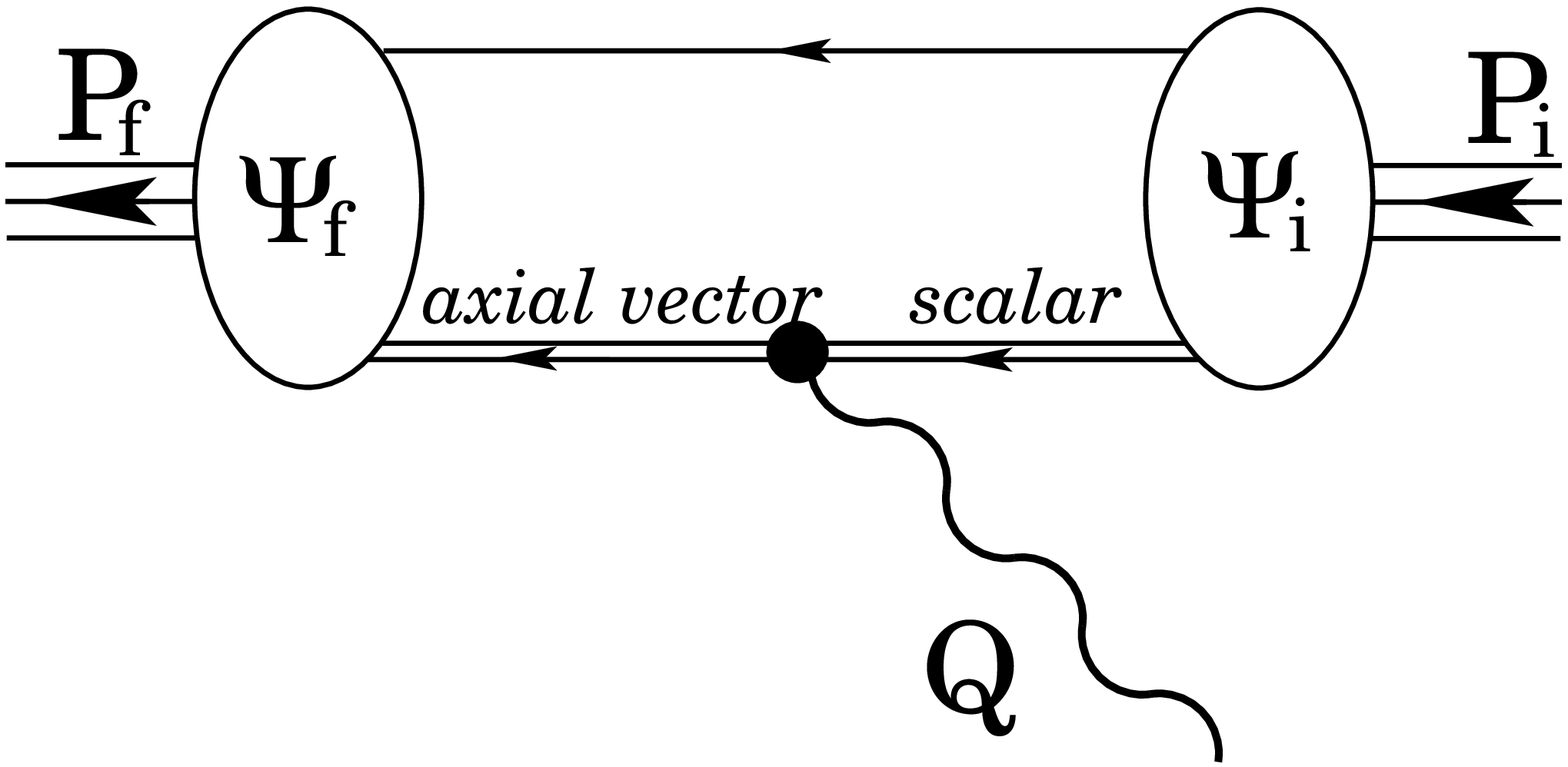}}
\end{minipage}\vspace*{3ex}

\begin{minipage}[t]{0.45\textwidth}
\leftline{\includegraphics[width=0.90\textwidth]{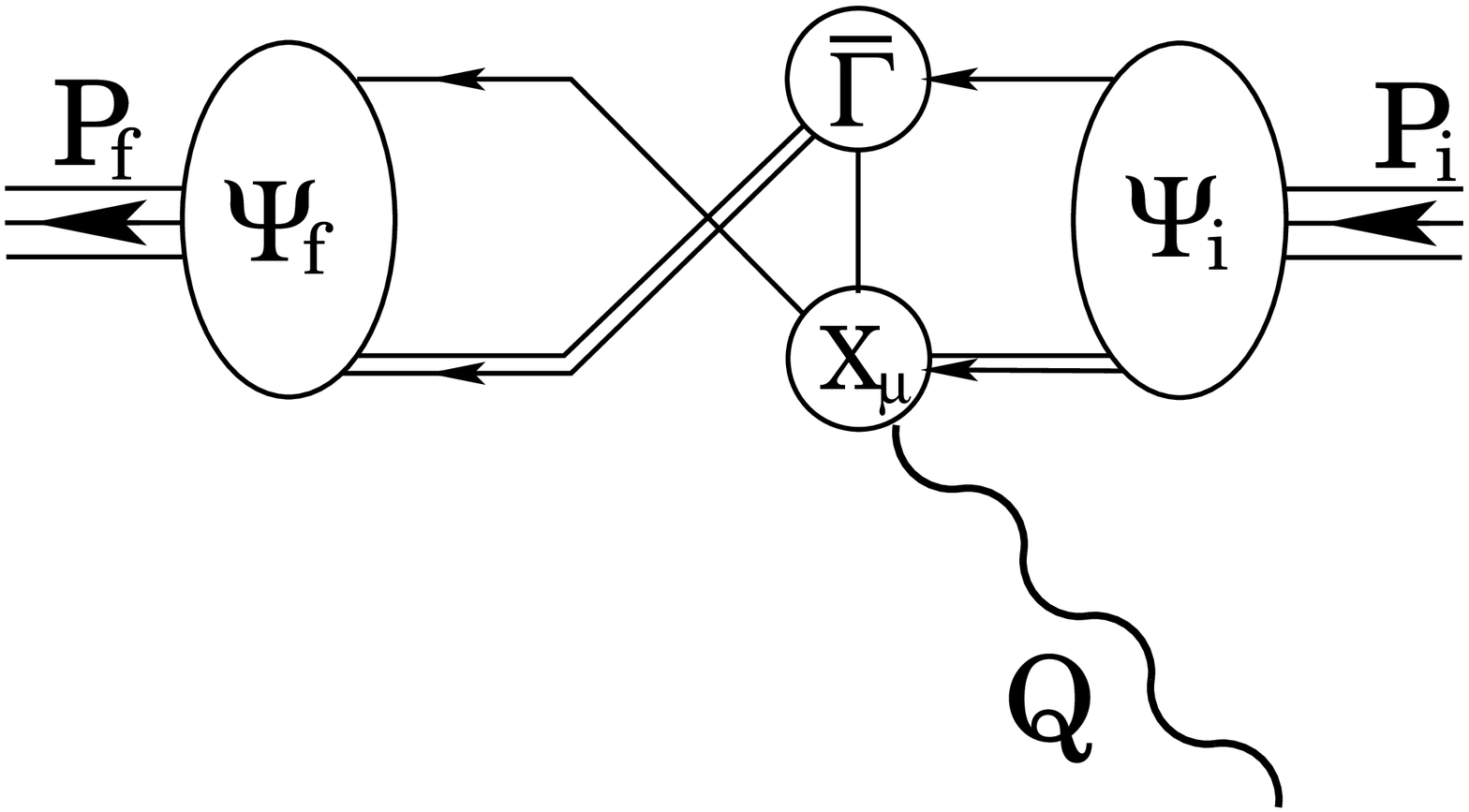}}
\end{minipage}
\begin{minipage}[t]{0.45\textwidth}
\rightline{\includegraphics[width=0.90\textwidth]{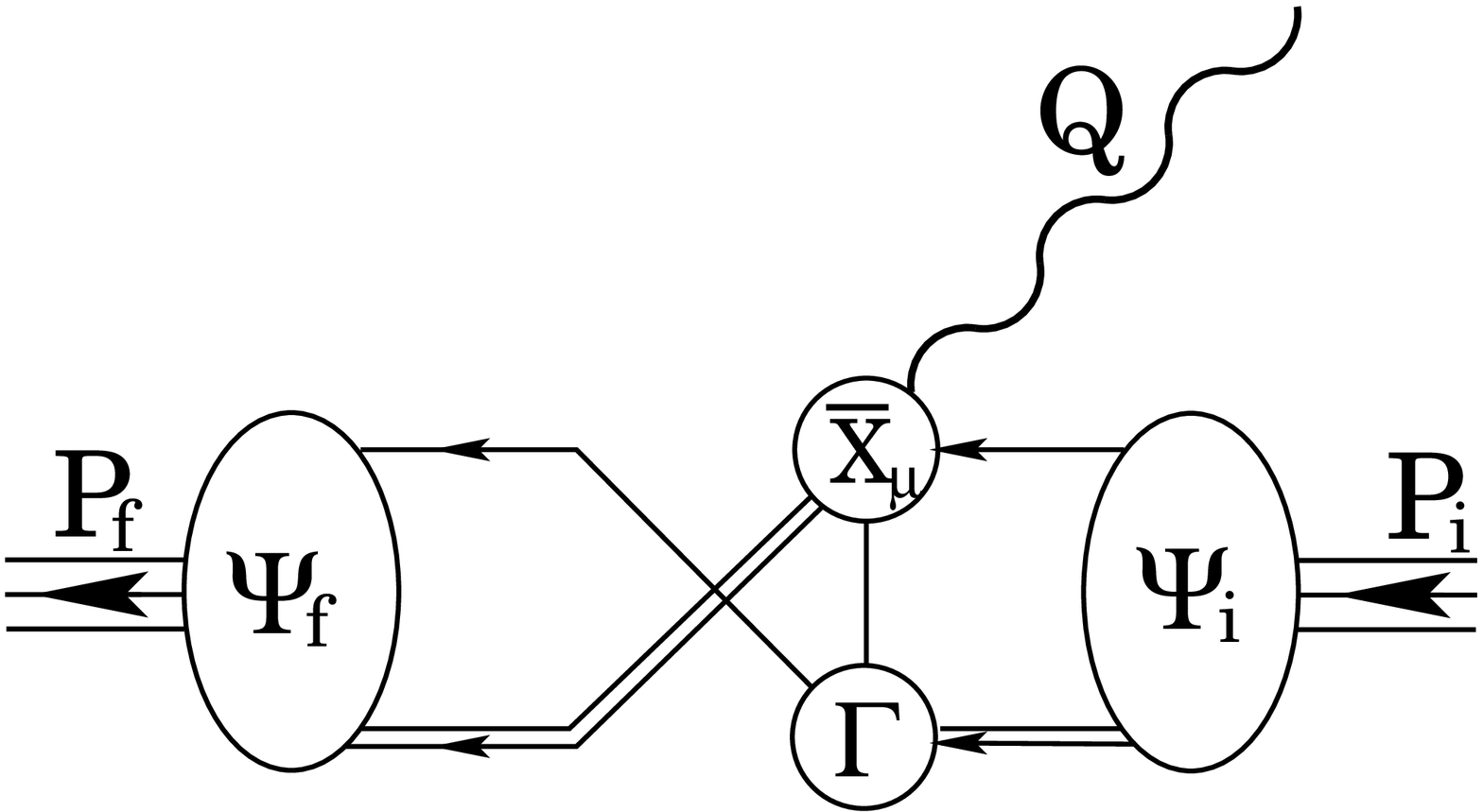}}
\end{minipage}
\end{minipage}
\caption{\label{vertex} Nucleon-photon vertex which ensures a conserved current for on-shell nucleons described by the Faddeev amplitudes, $\Psi_{i,f}$, calculated in Sect.\,\protect\ref{faddeevsolution}.  The single line represents $S(p)$, the dressed-quark propagator, the double line, the diquark propagator, and $\Gamma$ is the diquark Bethe-Salpeter amplitude, all of which are described in Sect.\,\protect\ref{completing}.  Aspects of the remaining vertices are described in Sect.\,\protect\ref{nucleonform}: the top-left image is diagram~1; the top-right, diagram~2; and so on,  with the bottom-right image, diagram~6.  (Adapted from Ref.\,\protect\cite{arneJ}.)}
\end{figure}

\paragraph{Diagram~1}
This represents the photon coupling directly to the bystander quark.  It is a necessary condition for current conservation that the dressed-quark-photon vertex satisfy the \index{Ward-Takahashi identity: vector} Ward-Takahashi identity:
\begin{equation}
\label{vwti}
Q_\mu \, i\Gamma_\mu(\ell_1,\ell_2) = S^{-1}(\ell_1) - S^{-1}(\ell_2)\,,
\end{equation}
where $Q=\ell_1-\ell_2$ is the photon momentum flowing into the vertex.  Since the quark is dressed, Sect.\,\ref{completing}, the vertex cannot be bare; i.e., $\Gamma_\mu(\ell_1,\ell_2) \neq \gamma_\mu$.  It can be obtained by solving Eq.\,(\ref{photonvertex}), which was the procedure employed in Sec.\,\ref{modelindependent}.  However, since $S(p)$ is parametrised, Ref.\,\cite{blochff} can be followed and the vertex written\,\cite{bc80}
\begin{equation}
\label{bcvtx}
i\Gamma_\mu(\ell_1,\ell_2)  =  
i\Sigma_A(\ell_1^2,\ell_2^2)\,\gamma_\mu +
2 k_\mu \left[i\gamma\cdot k \,
\Delta_A(\ell_1^2,\ell_2^2) + \Delta_B(\ell_1^2,\ell_2^2)\right] \!;
\end{equation}
with $k= (\ell_1+\ell_2)/2$ and
\begin{equation}
\Sigma_F(\ell_1^2,\ell_2^2) = \sfrac{1}{2}\,[F(\ell_1^2)+F(\ell_2^2)]\,,\;
\Delta_F(\ell_1^2,\ell_2^2) =
\frac{F(\ell_1^2)-F(\ell_2^2)}{\ell_1^2-\ell_2^2}\,,
\label{DeltaF}
\end{equation}
where $F= A, B$; viz., the scalar functions in Eq.\,(\ref{Sgeneral}).  It is
critical that $\Gamma_\mu$ in Eq.\ (\ref{bcvtx}) satisfies Eq.\ (\ref{vwti})
and very useful that it is completely determined by the dressed-quark
propagator.  Moreover, the \textit{Ansatz} possesses the essential feature required by \index{Asymptotic freedom} asymptotic freedom that at deep spacelike momenta the dressed-quark-photon-vertex evolves to the bare vertex.

\paragraph{Diagram~2}
This represents the photon coupling directly to a diquark correlation.  In the case of a scalar correlation, the general form of the diquark-photon vertex is
\begin{equation}
\Gamma_\mu^{0^+}(\ell_1,\ell_2) = 2\, k_\mu\, f_+(k^2,k\cdot Q,Q^2) + Q_\mu  \, f_-(k^2,k\cdot Q,Q^2)\,,
\end{equation}
and it must satisfy the Ward-Takahashi identity: 
\begin{equation}
\label{VWTI0}
Q_\mu \,\Gamma_\mu^{0^+}(\ell_1,\ell_2) = \Pi^{0^+}(\ell_1^2)  - \Pi^{0^+}(\ell_2^2)\,,\; \Pi^{J^P}(\ell^2) = \{\Delta^{J^P}(\ell^2)\}^{-1}.
\end{equation} 
The adaption of Eq.\,(\ref{bcvtx}) to this case is
\begin{equation}
\label{Gamma0plus}
\Gamma_\mu^{0^+}(\ell_1,\ell_2) =  k_\mu\,
\Delta_{\Pi^{0^+}}(\ell_1^2,\ell_2^2)\,,
%
\end{equation}  
with the definition of $\Delta_{\Pi^{0^+}}(\ell_1^2,\ell_2^2)$ apparent from Eq.\,(\ref{DeltaF}).  Equation~(\ref{Gamma0plus}) is the minimal \textit{Ansatz} that: satisfies Eq.\,(\ref{VWTI0}); is completely determined by quantities introduced already; and is free of kinematic singularities.  It also guarantees a valid normalisation of electric charge; viz., 
\begin{equation}
\lim_{\ell^\prime\to \ell} \Gamma_\mu^{0^+}(\ell^\prime,\ell) = 2 \, \ell_{\mu} \, \frac{d}{d\ell^2}\, \Pi^{0^+}(\ell^2) \stackrel{\ell^2\sim 0}{=} 2 \, \ell_{\mu}\,,
\end{equation}
owing to Eq.\,(\ref{DQPropConstr}).  NB.\ The fractional diquark charge has been factored.  It therefore appears subsequently as a simple multiplicative factor. 

For the case in which the struck diquark correlation is axial-vector, the vertex structure is more involved.  Nonetheless, there are many constraints that may be employed to build a realistic \textit{Ansatz}.  That composed in Ref.\,\cite{arneJ} has two parameters: the magnetic dipole and electric quadrupole moments of the axial-vector diquark, $\mu_{1^+}$ and $\chi_{1^+}$, respectively. 

\paragraph{Diagram~3}
This image depicts a photon coupling to the quark that is exchanged as one diquark breaks up and another is formed.  While this is the first two-loop diagram in the current, no new elements appear in its specification: the quark-photon vertex was described above.  It is noteworthy that the process of quark exchange provides the attraction necessary in the Faddeev equation to bind the baryon.  It also guarantees that the Faddeev amplitude has the correct antisymmetry under the exchange of any two dressed-quarks.  This key feature is absent in models with elementary (noncomposite) diquarks.

\paragraph{Diagram~4}This differs from Diagram~2 in expressing the contribution to the nucleons' form factors owing to an electromagnetically induced transition between scalar and axial-vector diquarks.  The transition vertex is a rank-2 pseudotensor, kindred to the matrix element describing the $\rho\, \gamma^\ast \pi^0$  transition \cite{maristandy4}, and can therefore be expressed 
\begin{equation}
\label{SAPhotVertex}
\Gamma_{SA}^{\gamma\alpha}(\ell_1,\ell_2) = -\Gamma_{AS}^{\gamma\alpha}(\ell_1,\ell_2) 
= \frac{i}{M_N} \, {\check T}(\ell_1,\ell_2) \, \varepsilon_{\gamma\alpha\rho\lambda}\ell_{1\rho} \ell_{2 \lambda}\,,
\end{equation}
where $\gamma$, $\alpha$ are, respectively, the vector indices of the photon and axial-vector diquark.  For simplicity, Ref.\,\cite{arneJ} proceeded under the assumption
\begin{equation}
{\check T}(\ell_1,\ell_2) = \kappa_{\check T}\,;
\end{equation}
viz., a constant, for which a typical value is\,\cite{oettel2}: 
\begin{equation}
\label{kTbest}
\kappa_{\check T} \sim 2\,.
\end{equation}

In the nucleons' rest frame, a conspicuous piece of the Faddeev wave function that describes an axial-vector diquark inside the bound state can be characterised as containing a bystander quark whose spin is antiparallel to that of the nucleon, with the axial-vector diquark's parallel.  The interaction pictured in this diagram does not affect the bystander quark but the transformation of an axial-vector diquark into a scalar effects a flip of the quark spin within the correlation.  After this transformation, the spin of the nucleon must be formed by summing the spin of the bystander quark, which is still aligned antiparallel to that of the nucleon, and the orbital angular momentum between that quark and the scalar diquark.\footnote{A less prominent component of the amplitude has the bystander quark's spin parallel to that of the nucleon while the axial-vector diquark's is antiparallel: this $q^\uparrow \oplus (qq)_{1^+}^{\downarrow} $ system has one unit of angular momentum.  That momentum is absent in the $q^\uparrow \oplus (qq)_{0^+}$ system.  Other combinations also contribute via Diagram~3 but all mediated processes inevitably require a modification of spin and/or angular momentum.}   Diagram~4 can therefore be expected to impact strongly on the nucleons' magnetic form factors.

\paragraph{Diagram~5 and 6}
These two-loop diagrams are the so-called ``seagull'' terms, which appear as partners to Diagram~3 and arise because binding in the nucleons' Faddeev equations is effected by the exchange of \textit{nonpointlike} diquark correlations\,\cite{oettelpichowsky}.  The new elements in these diagrams are the couplings of a photon to two dressed-quarks as they either separate from (Diagram~5) or combine to form (Diagram~6) a diquark correlation.  As such they are components of the five point Schwinger function which describes the coupling of a photon to the quark-quark scattering kernel.  This Schwinger function could be calculated, as is evident from the recent computation of analogous Schwinger functions relevant to meson observables \cite{mariscotanch}.  However, such a calculation provides valid input only when a uniform truncation of the DSEs has been employed to calculate each of the elements described hitherto.  In the present context, it is appropriate instead to employ the algebraic parametrisation of Ref.\,\cite{oettelpichowsky}, which is simple, completely determined by the elements introduced already, and guarantees current conservation for on-shell nucleons.

\subsubsection{Calculated results}
\label{numerical}
In order to place the calculation of baryon observables on the same footing as the study of mesons, the proficiency evident in Ref.\,\cite{pmcdr} will need to be applied to every line and vertex that appears in Fig.\,\ref{vertex}.  While that is feasible, it remains to be done.  In the meantime, one can relate a study whose merits include a capacity to: explore the potential of the Faddeev equation truncation of the baryon three-body problem; and elucidate the role of additional correlations, such as those associated with pseudoscalar mesons.

It is worthwhile to epitomise the input before presenting results.  One element is the dressed-quark propagator, Sect.\,\ref{completing}.  The form used\,\cite{mark} both anticipated and expresses the features that are now known to be true \cite{alkoferdetmold,bhagwat}.  It carries no free parameters, because its behaviour was fixed in analyses of meson observables, and is basic to a description of light- and heavy-quark mesons that is accurate to better than 10\% \cite{mishaSVY}.

The nucleon is supposed at heart to be composed of a dressed-quark and nonpointlike diquark with binding effected by an iterated exchange of roles between the bystander and diquark-participant quarks.  The picture is realised via a Poincar\'e covariant Faddeev equation, Sect.\,\ref{faddeev}, which incorporates scalar and axial-vector diquark correlations.  There are two parameters, Sect.\,\ref{faddeevsolution}: the mass-scales associated with these correlations.  They were fixed by fitting to specified nucleon and $\Delta$ masses.  There are no free parameters at this point.

With the constituents and the bound states' structure defined, only a specification of the nucleons' electromagnetic interaction remained.  Its formulation, sketched in Sect.\,\ref{Ncurrent}, is guided almost exclusively by a requirement that the nucleon-photon vertex satisfy a Ward-Takahashi identity.  Since the scalar diquark's electromagnetic properties are readily resolved, the result, Fig.\,\ref{vertex}, depends on three parameters that are all tied to properties of the axial-vector diquark correlation: $\mu_{1^+}$ \& $\chi_{1^+}$, respectively, the axial-vector diquarks' magnetic dipole and electric quadrupole moments; and $\kappa_{\check T}$, the strength of electromagnetic axial-vector $\leftrightarrow$ scalar diquark transitions.  Hence, the study of Ref.\,\cite{arneJ} exhibits and interprets the dependence of the nucleons' form factors on these three parameters, and also on the nucleons' intrinsic quark structure as expressed in the Poincar\'e covariant Faddeev amplitudes. 

\begin{table}[t]
\caption{%
Magnetic moments \index{Magnetic moment: nucleon} calculated with the diquark mass-scales in Table~\protect\ref{ParaFix} for a range of axial-vector-diquark--photon vertex parameters, centred on the point-particle values of $\mu_{1^+}=2$ \& $\chi_{1^+}=1$, and $\kappa_{\check T} = 2$, Eq.\,(\protect\ref{kTbest}).  Columns labelled $\sigma$ give the percentage-difference from results obtained with the reference parameters.
Experimental values are: $\kappa_p:=\mu_p-1 = 1.79$ \& $\mu_n = -1.91$.  The magnetic moments are expressed in units of nuclear magnetons defined with the calculated nucleon mass. (Adapted from Ref.\,\protect\cite{arneJ}.)
}
{\normalsize
\begin{tabular*}{1.0\textwidth}{c@{\extracolsep{0ptplus1fil}}c@{\extracolsep{0ptplus1fil}}
c@{\extracolsep{0ptplus1fil}}|c@{\extracolsep{0ptplus1fil}} 
c@{\extracolsep{0ptplus1fil}}c@{\extracolsep{0ptplus1fil}}
c@{\extracolsep{0ptplus1fil}}|c@{\extracolsep{0ptplus1fil}}
c@{\extracolsep{0ptplus1fil}}c@{\extracolsep{0ptplus1fil}}
c@{\extracolsep{0ptplus1fil}}c@{\extracolsep{0ptplus1fil}}}
%
\multicolumn{3}{c}{} & \multicolumn{4}{c}{Set~A} & \multicolumn{4}{c}{Set~B} \\
\multicolumn{3}{c}{} & \multicolumn{4}{c}{\rule{10em}{0.1ex}} & \multicolumn{4}{c}{\rule{10em}{0.1ex}} \\
$\mu_{1^+}$ & $\chi_{1^+}$ & $\kappa_{\check T}$ & $\kappa_p$ & \rule{0em}{2.5ex}$\sigma_{\kappa_p}^A$ & $|\mu_n|$ & $\sigma_{|\mu_n|}^A$ & $\kappa_p$ & $\sigma_{\kappa_p}^B$ & $|\mu_n|$ & $\sigma_{|\mu_n|}^B$ \\\hline
1 & 1 & 2 & 1.79 & -15.3  & 1.70 & -5.1 & 2.24 & -21.9 & 2.00 & -6.2 \\
2 & 1 & 2 & 2.06 & ~~~~~  & 1.79 & ~~~~ & 2.63 & ~~~~~ & 2.13 & ~~~~ \\
3 & 1 & 2 & 2.33 & ~15.4  & 1.88 & ~5.1 & 3.02 & ~21.9 & 2.26 & ~6.1 \\\hline
2 & 0 & 2 & 2.06 & ~~0.0  & 1.79 & ~0.0 & 2.63 & ~~0.0 & 2.13 & ~0.0 \\
2 & 2 & 2 & 2.06 & ~~0.0  & 1.79 & ~0.0 & 2.63 & ~~0.0 & 2.13 & ~0.0 \\\hline
2 & 1 & 1 & 1.91 & ~-8.4  & 1.64 & -8.4 & 2.45 & -10.1 & 1.95 & -8.5 \\
2 & 1 & 3 & 2.21 & ~~8.4  & 1.85 & ~8.3 & 2.82 & ~10.1 & 2.31 & ~8.5 \\\hline
\end{tabular*}\label{moments}}
\end{table}

Table~\ref{moments} lists results for the nucleons' magnetic moments, $G_M^N(0)$, where $N=n,p$ \cite{arneJ}.  It indicates that the moments are insensitive to the axial-vector diquarks' quadrupole moment but react to the diquarks' magnetic moment, increasing quickly in magnitude as $\mu_{1^+}$ increases.  As anticipated in connection with Eq.\,(\ref{kTbest}), the nucleons' moments respond strongly to alterations in the strength of the scalar $\leftrightarrow$ axial-vector transition, increasing rapidly as $\kappa_{\check T}$ is increased.   Set~A, which is fitted to the experimental values of $M_N$ \& $M_\Delta$, describes the nucleons' moments quite well: $\kappa_p$ is 15\% too large; and $|\mu_n|$, 16\% too small.  On the other hand, Set~B, which is fitted to baryon masses that are inflated so as to make room for pion cloud effects, overestimates $\kappa_p$ by 47\% and $|\mu_n|$ by 18\%.  

The nucleons' charge and magnetic radii \index{Charge radii: nucleon}
\begin{equation}
r_{N}^2:= \left.- \, 6\,\frac{d}{ds} \ln G_E^{N}(s) \right|_{s=0},\;
(r_{N}^\mu)^2:= \left.- \, 6\,\frac{d}{ds} \ln G_M^{N}(s) \right|_{s=0},
\end{equation}
were also reported in Ref.\,\cite{arneJ}.  The charge radii, particularly that of the neutron, were most sensitive to changes in the axial-vector diquarks' electric quadrupole moment, $\chi_{1^+}$.  That is not surprising given that $\chi_{1^+}$ is the only model parameter which speaks directly of the axial-vector diquarks' electric charge distribution.  With the point-particle reference values of $\mu_{1^+}=2$ \& $\chi_{1^+}=1$, and $\kappa_{\check T} = 2$, Set~A underestimates the proton radius by 30\% and the magnitude of the neutron radius by 43\%, while for Set~B these differences are 32\% and 50\%, respectively.  The magnetic radii are insensitive to the axial-vector diquarks' quadrupole moment but react to the diquarks' magnetic moment as one would anticipate: increasing in magnitude as $\mu_{1^+}$ increases.  Moreover, again consistent with expectation, these radii respond to changes in $\kappa_{\check T}$, decreasing as this parameter is increased.  With the reference parameter values both Sets~A \& B underestimate $r_N^\mu$ by approximately 40\%.

\begin{figure}[t]
\begin{minipage}{0.45\textwidth}
\centerline{\hspace*{2.5em}%
\includegraphics[width=0.95\textwidth,angle=270]{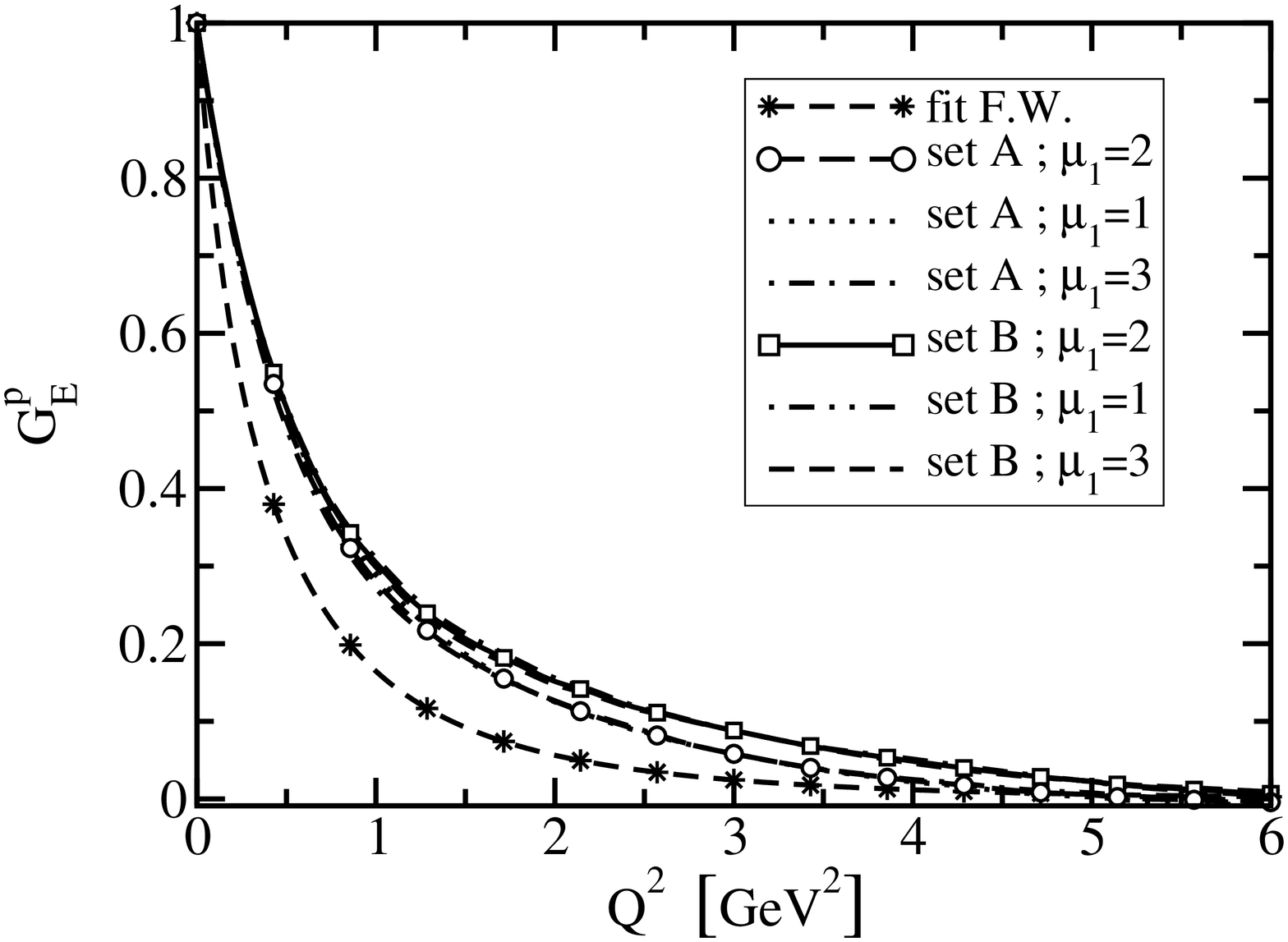}}
\end{minipage}
\hfill
\begin{minipage}{0.45\textwidth}
\centerline{\includegraphics[width=0.95\textwidth,angle=270]{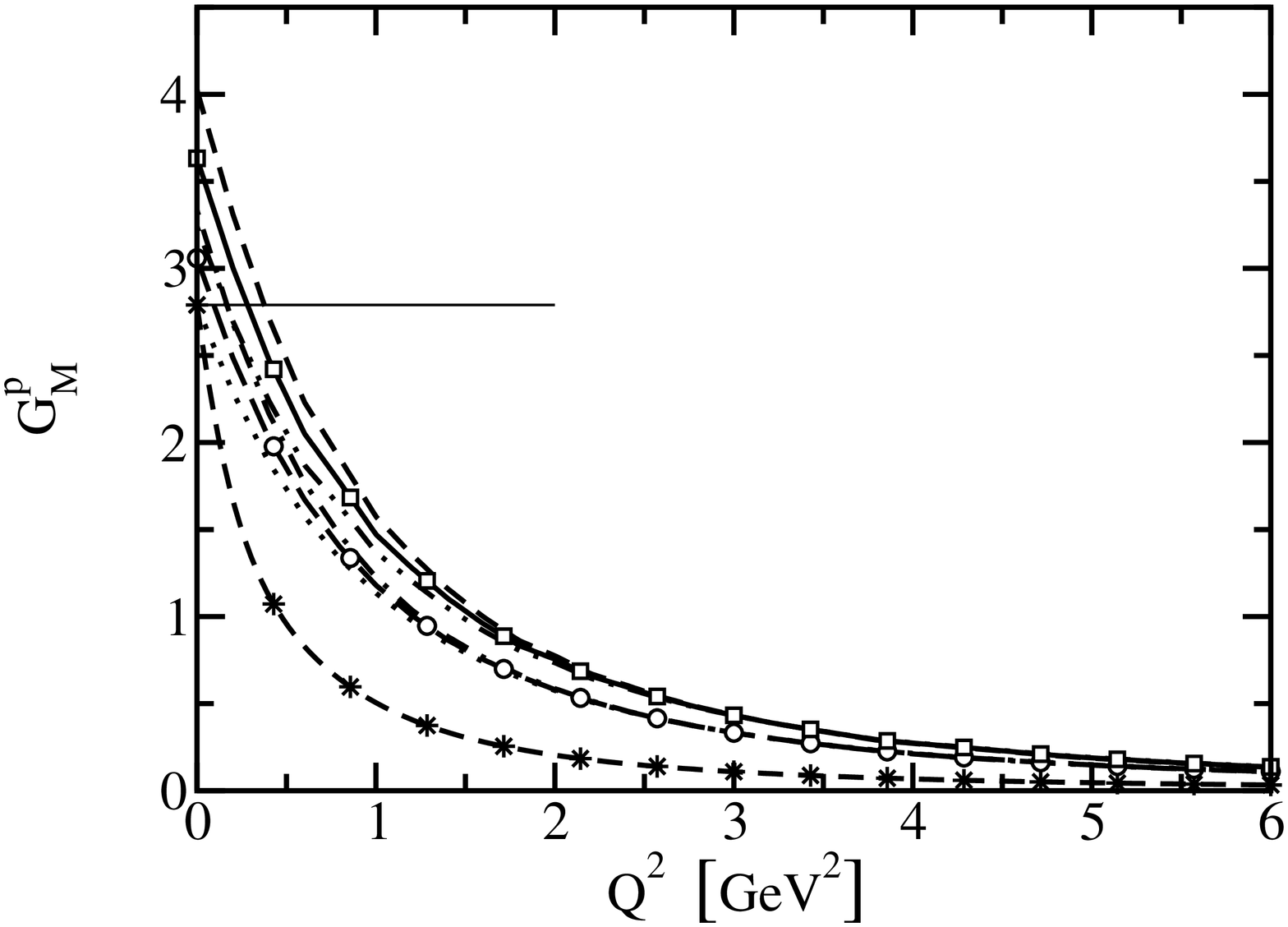}%
\hspace{1em}}
\end{minipage}
\\
\begin{minipage}{0.45\textwidth}
\centerline{\hspace*{2.5em}%
\includegraphics[width=0.95\textwidth,angle=270]{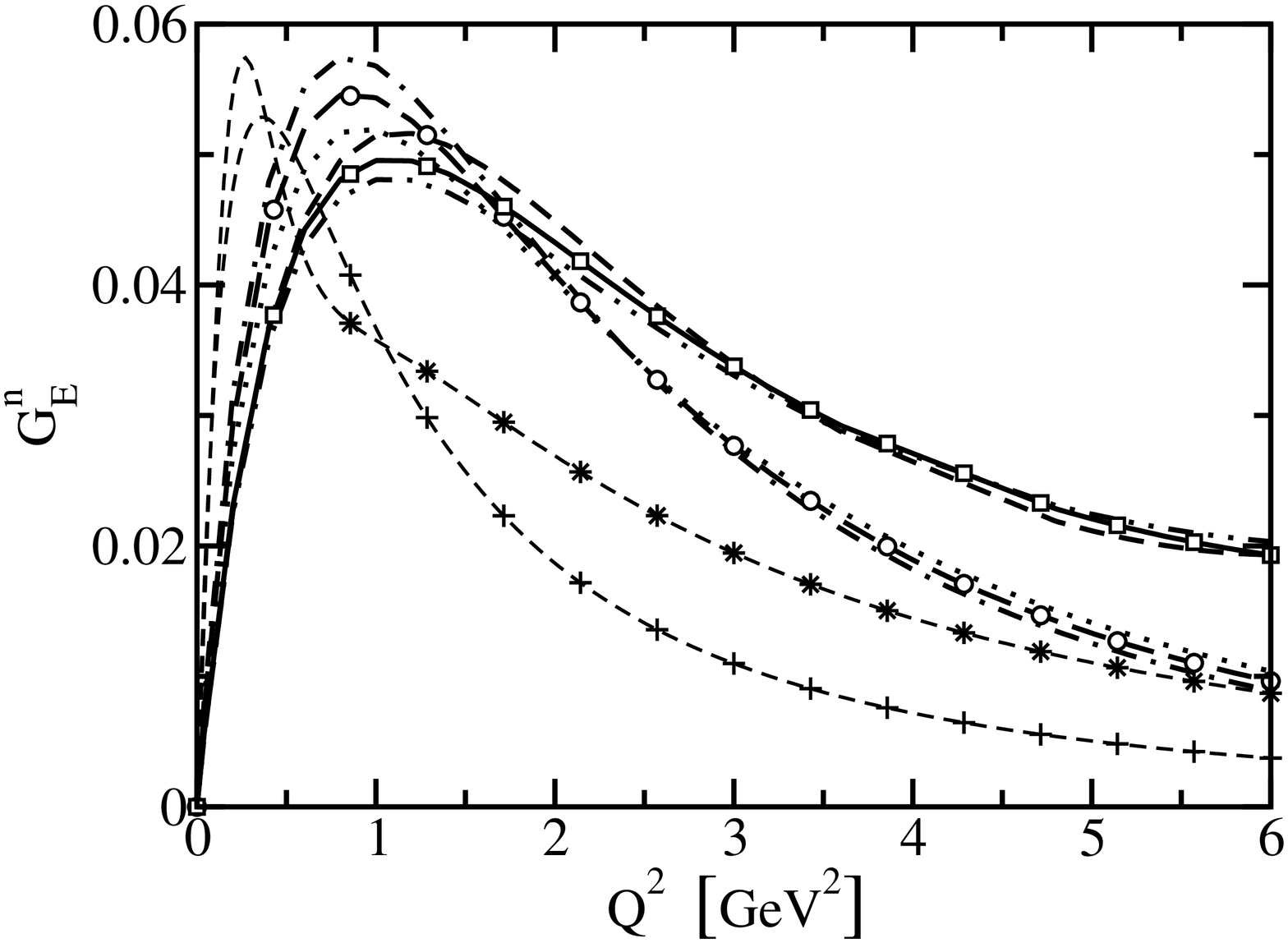}}
\end{minipage}
\hfill
\begin{minipage}{0.45\textwidth}
\centerline{\includegraphics[width=0.95\textwidth,angle=270]{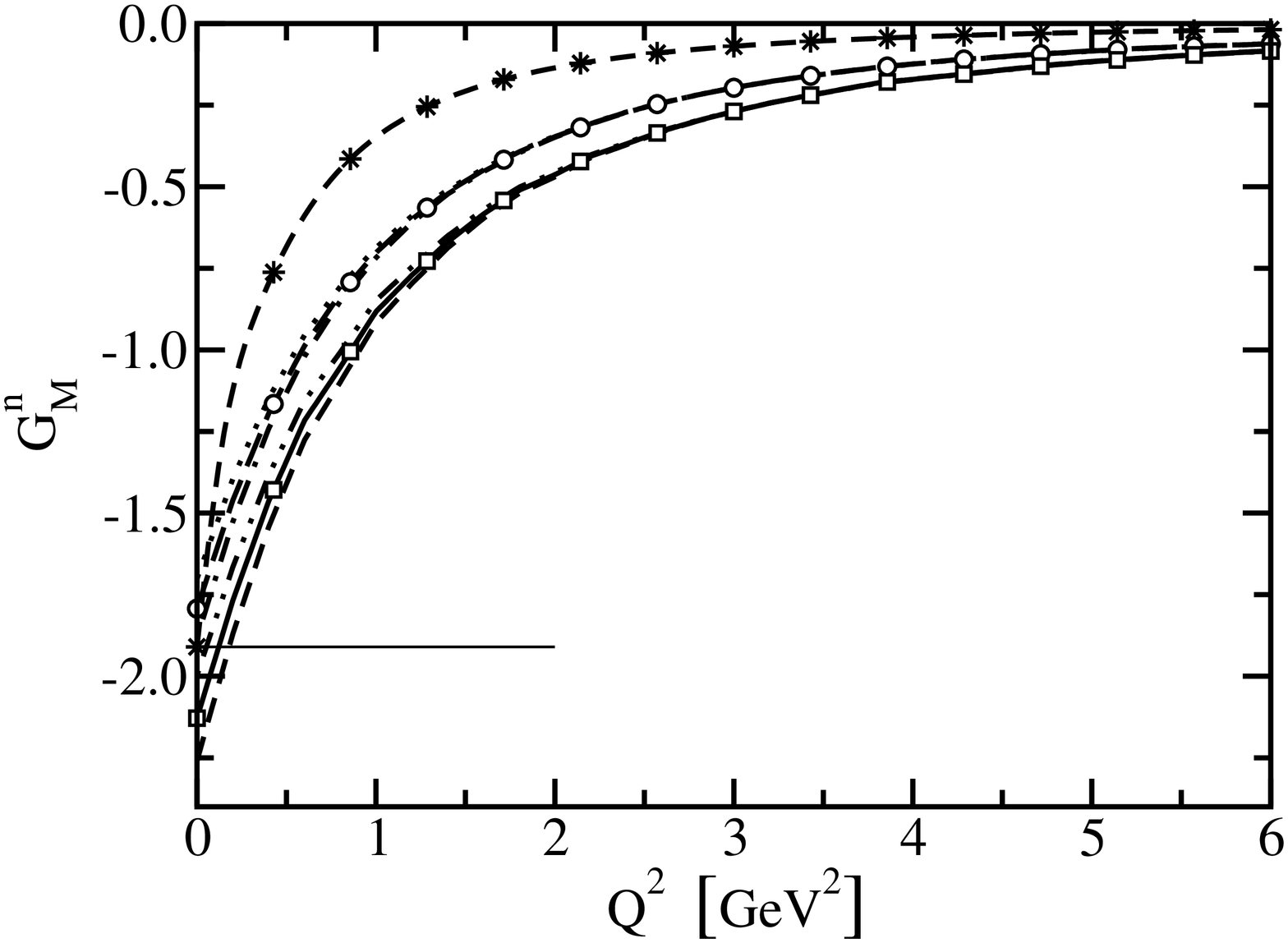}%
\hspace*{1em}}
\end{minipage}
\caption{\label{plot1}
Response of nucleon form factors \index{Form factors: nucleon} to variations in the magnetic moment of the axial-vector diquark: $\mu_{1^+}=1,2,3$; with $\chi_{1^+}=1$, $\kappa_{\check T}=2$.  
The legend in the top-left panel applies to all; the dashed-line marked by ``$\ast$'' is a fit to experimental data\,\protect\cite{Walcher03} and the dashed-line marked by ``$+$'' in the lower-left panel is the fit to $G_E^n(Q^2)$ of Ref.\,\protect\cite{galster}; and the horizontal lines in the right panels mark the experimental value of the nucleon's magnetic moment.  (Adapted from Ref.\,\protect\cite{arneJ}.)}
\end{figure}

Nucleon electromagnetic form factors associated with the tabulated values of static properties are presented in Fig.\,\ref{plot1}.  The figure confirms and augments the information in Table~\ref{moments}.  Consider, e.g., the electric form factors.  One observes that the differences between results obtained with Set~A and Set~B generally outweigh those delivered by variations in the parameters characterising the axial-vector diquark's electromagnetic properties.  The proton's electric form factor, in particular, is largely insensitive to these parameters.  The nucleons' magnetic form factors exhibit the greatest sensitivity to the axial-vector diquarks's electromagnetic properties but in this case, too, the differences between Set~A and Set~B are more significant.  For $Q^2 \gsim 4\,$GeV$^2$ there is little sensitivity to the diquarks' electromagnetic parameters in any curve because the model expresses the diquark current's perturbative limit \cite{brodskyhiller92}.  It is thus apparent from these plots that the behaviour of the nucleons' form factors is primarily determined by the information encoded in the Faddeev amplitudes.

The results show that the nucleons' electromagnetic properties are sensitive to the strength of axial-vector diquark correlations in the bound state and react to the electromagnetic properties of these correlations.  In all cases the dependence is readily understood intuitively.  However, taken together the results indicate that one cannot readily tune the model's parameters to provide a uniformly good account of nucleon properties: something more than dressed-quark and -diquark degrees of freedom is required.

\subsubsection{Chiral corrections}
\index{Chiral loops and nucleon electromagnetic properties}
\label{chiralem}
It is appropriate now to examine effects that arise through coupling to pseudoscalar mesons.  As with baryon masses, Sect.\,\ref{loopmass}, there are two types of contributions to electromagnetic form factors from meson loops: regularisation-scheme-dependent terms, which are analytic in the neighbourhood of $\hat m = 0$; and nonanalytic scheme-independent terms.  The leading-order scheme-independent contributions for the nucleon static properties presented are\,\cite{kubis}
\begin{eqnarray}
\label{rpnpion}
\langle r_{p\atop n}^2\rangle^{1-loop}_{NA} &=& \mp\,\frac{1+5 g_A^2}{32 \pi^2 f_\pi^2} \,\ln (\frac{m_\pi^2}{M_N^2}) \,, \\
\label{rpnmpion}
\langle (r_{N}^\mu)^2\rangle^{1-loop}_{NA} &=& -\,\frac{1+5 g_A^2}{32 \pi^2 f_\pi^2} \,\ln (\frac{m_\pi^2}{M_N^2})+ \frac{g_A^2\, M_N}{16 \pi f_\pi^2 \mu_v} \frac{1}{m_\pi} \,,\\
\label{mpnpion}
(\mu_{p\atop n})_{NA}^{1-loop} & = & \mp \, \frac{g_A^2\, M_N}{4\pi^2 f_\pi^2}\, m_\pi\,,
\end{eqnarray}
where $g_A=1.26$, $f_\pi=0.0924$\,GeV\,$=1/(2.13 \,{\rm fm})$, $\mu_v=\mu_p-\mu_n$.  Clearly, the radii diverge in the chiral limit, a much touted aspect of chiral corrections.  While these scheme-independent terms are immutable, at physical values of the pseudoscalar meson masses they do not usually provide the dominant contribution to observables: that is provided by the regularisation-parameter-dependent terms.  (We saw this with the baryon masses in Sect.\,\ref{loopmass}.  See also footnote~\ref{fnchiral}.)

Since regularisation-parameter-dependent parts of the chiral loops are important it is sensible to follow Ref.\,\cite{ashley} and estimate the corrections using modified formulae that incorporate a single parameter which mimics the effect of regularisation-dependent contributions from the integrals.  Thus Eqs.\,(\ref{rpnpion}) -- (\ref{mpnpion}) are rewritten
\begin{eqnarray}
\label{rpnpionR}
\langle r_{p\atop n}^2\rangle^{1-loop^R}_{NA} &=& \mp\,\frac{1+5 g_A^2}{32 \pi^2 f_\pi^2} \,\ln (\frac{m_\pi^2}{m_\pi^2+\lambda^2}) \,, \\
\nonumber \langle (r_{N}^\mu)^2\rangle^{1-loop^R}_{NA} &=& -\,\frac{1+5 g_A^2}{32 \pi^2 f_\pi^2} \,\ln (\frac{m_\pi^2}{m_\pi^2+\lambda^2}) + \frac{g_A^2\, M_N}{16 \pi f_\pi^2 \mu_v} \frac{1}{m_\pi} \,
\frac{2}{\pi}\arctan(\frac{\lambda}{m_\pi})\,,\\
\label{rpnmpionR}\\
\label{mpnpionR}
(\mu_{p\atop n})_{NA}^{1-loop^R}& = & \mp \, \frac{g_A^2\, M_N}{4\pi^2 f_\pi^2}\, m_\pi\, \frac{2}{\pi}\arctan(\frac{\lambda^3}{m_\pi^3})\,,
\end{eqnarray}
wherein $\lambda$ is a regularisation mass-scale.  NB.\ The loop contributions vanish when the pion mass is much larger than the regularisation scale, as required: very massive states must decouple from low-energy phenomena.  

\begin{table}[t]
\caption{Row~1 -- static properties calculated with Set~B diquark masses, Table~\protect\ref{ParaFix}, and $\mu_{1^+}=2$, $\chi_{1^+}=1$, $\kappa_{\check T}=2$: charge radii in fm, with $r_n:= -\sqrt{-\langle r_n^2\rangle}$; and magnetic moments in nuclear magnetons.  Row~2 adds the corrections of Eqs.\,(\protect\ref{rpnpionR})--(\protect\ref{mpnpionR}) with $\lambda=0.3\,$GeV.  $\varsigma$ in row~$n$, is the rms relative-difference between the entries in rows $n$ and 3.  (Adapted from Ref.\,\protect\cite{arneJ}.)}
{\normalsize
\begin{tabular*}{1.0\textwidth}{l@{\extracolsep{0ptplus1fil}}
c@{\extracolsep{0ptplus1fil}}
c@{\extracolsep{0ptplus1fil}}c@{\extracolsep{0ptplus1fil}} 
c@{\extracolsep{0ptplus1fil}}c@{\extracolsep{0ptplus1fil}}
c@{\extracolsep{0ptplus1fil}}c@{\extracolsep{0ptplus1fil}}
c@{\extracolsep{0ptplus1fil}}}\\\hline
  & $r_p$ & $r_n$ & $r_p^\mu$ & $r_n^\mu$ & $\mu_p$ & $-\mu_n$ && $\varsigma$\\\hline
$q$-$(qq)$ core & 0.595 & 0.169 & 0.449 & 0.449 & 3.63 & 2.13 && 0.39\\
$+\pi$-loop correction & 0.762 & 0.506 & 0.761 & 0.761 & 3.05 & 1.55 && 0.23 \\\hline
experiment & 0.847 & 0.336 & 0.836 & 0.889 & 2.79& 1.91 && \\\hline
\end{tabular*}\label{picorrected}}
\end{table}

One may now return to the calculated values of the nucleons' static properties.  Consider the Set~B results obtained with $\mu_{1^+}=2$, $\chi_{1^+}=1$, $\kappa_{\check T}=2$.  Set~B was chosen to give inflated values of the nucleon and $\Delta$ masses in order to make room for chiral corrections, and therefore one may consistently apply the corrections in Eqs.\,(\ref{rpnpionR}) -- (\ref{mpnpionR}) to the static properties.  With $\lambda=0.3\,$GeV this yields the second row in Table~\ref{picorrected}: the regularised chiral corrections reduce the rms relative-difference significantly.  

This crude analysis, complementing Sect.\,\ref{loopmass}, suggests strongly that a veracious description of baryons can be obtained using dressed-quark and -diquark degrees of freedom augmented by a sensibly regulated pseudoscalar meson cloud.  The inverse of the regularisation parameter is a length-scale that may be viewed as a gauge of the distance from a nucleon's centre-of-mass to which the pseudoscalar meson cloud penetrates: $1/\lambda \approx \mbox{\small $\frac{2}{3}$}\,{\rm fm}$ is an intuitively reasonable value that indicates the cloud is expelled from the nucleon's core but materially affects its properties at distances $\gsim r_p^{\rm core}$; viz., in the vicinity of the nucleon's quark-core surface and farther out.

\begin{figure}[t]
\centerline{\includegraphics[width=0.7\textwidth,angle=270]{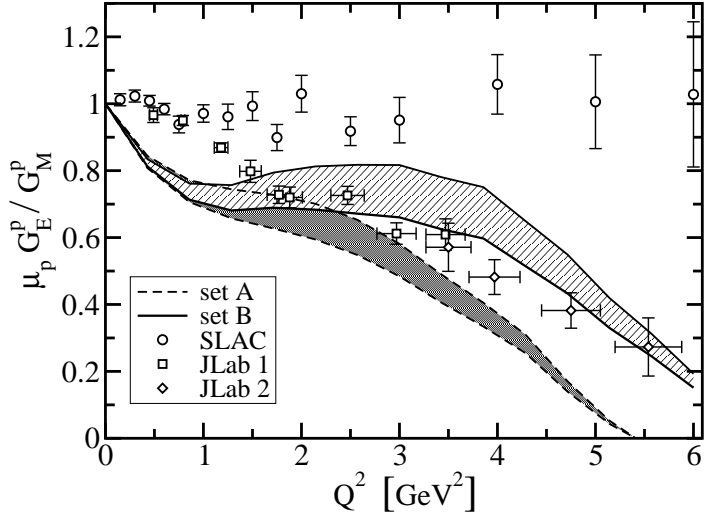}}
\caption{\label{plotGEpGMp} Form factor ratio: $\mu_p\, G_E^p(Q^2)/G_M^p(Q^2)$.  Calculated results: \textit{lower band} - Set~A in Table~\protect\ref{ParaFix}; and \textit{upper band} - Set~B.  For both, $G_E^p(Q^2)$ was calculated using the point-particle values: $\mu_{1^+}=2$ \& $\chi_{1^+}=1$, and $\kappa_{\check T} = 2$; i.e., the reference values in Table~\ref{moments}.  Changes in the axial-vector diquark parameters used to evaluate $G_E^p$ have little effect on the plotted results.  The width of the bands reflects the variation in $G_M^p$ with axial-vector diquark parameters.  In both cases, the upper border is obtained with $\mu_{1^+}=3$, $\chi_{1^+}=1$ and $\kappa_{\check T}= 2$, while the lower has $\mu_{1^+}= 1$.  The data are: \textit{circles} - Ref.\,\protect\cite{walker}; \textit{squares} - Ref.\,\protect\cite{jones}; and \textit{diamonds} - Ref.\,\protect\cite{gayou}.  (Adapted from Ref.\,\protect\cite{arneJ}.)}
\end{figure}

\subsubsection{Form factor ratios}
\index{Form factor: proton ratios}
\label{FFratios}
One is now in a position to return to our discussion of Fig.\,\ref{gepgmpdata}, and in Fig.\,\ref{plotGEpGMp} plot the calculated ratio $\mu_p\, G_E^p(Q^2)/G_M^p(Q^2)$.  The behaviour of the experimental data at small $Q^2$ is now readily understood.  In the neighbourhood of $Q^2=0$, 
\begin{equation}
\mu_p\,\frac{ G_E^p(Q^2)}{G_M^p(Q^2)} \stackrel{Q^2\sim 0}{=} 1 - \frac{Q^2}{6} \,\left[ (r_p)^2 - (r_p^\mu)^2 \right]\,,
\end{equation}
and because experimentally $r_p\approx r_p^\mu$ the ratio varies by less than 10\% on $0<Q^2< 0.6\,$GeV$^2$, if the form factors are approximately dipole.  In the calculation described herein, $r_p> r_p^\mu$ without chiral corrections.  Hence the ratio must fall immediately with increasing $Q^2$.  Incorporating pion loops, one obtains the results in Row~2 of Table\,\ref{picorrected}, which have $r_p\approx r_p^\mu$.  The small $Q^2$ (long-range) behaviour of this ratio is thus materially affected by the proton's pion cloud.

It has been emphasised that true pseudoscalar mesons are not pointlike and therefore pion cloud contributions to form factors diminish in magnitude with increasing $Q^2$.  To further exemplify this, it is notable that in a study of the $\gamma N \to \Delta$ transition \cite{sato}, pion cloud contributions to the $M1$ form factor fall from 50\% of the total at $Q^2=0$ to $\lsim 10$\% for $Q^2\gsim 2\,$GeV$^2$.  Hence, the evolution of $\mu_p\, G_E^p(Q^2)/G_M^p(Q^2)$ on $Q^2\gsim 2\,$GeV$^2$ is primarily determined by the quark core of the proton.  This is evident in Fig.\,\ref{plotGEpGMp}, which illustrates that, on $Q^2\in (1,5)\,$GeV$^2$, $\mu_p\, G_E^p(Q^2)/G_M^p(Q^2)$ is sensitive to the parameters defining the axial-vector-diquark--photon vertex.  The response diminishes with increasing $Q^2$ because the representation of the diquark current expresses the perturbative limit.  

The behaviour of $\mu_p\, G_E^p(Q^2)/G_M^p(Q^2)$ on $Q^2\gsim 2\,$GeV$^2$ is determined either by correlations expressed in the Faddeev amplitude, the electromagnetic properties of the constituent degrees of freedom, or both.  The issue is decided by the fact that the magnitude and trend of the results are not materially affected by the axial-vector-diquarks' electromagnetic parameters.  This observation suggests strongly that the ratio's evolution is due primarily to spin-isospin correlations in the nucleon's Faddeev amplitude.  It is notable that while Set~A is ruled out by the data, Set~B, which anticipates pion cloud effects, is in reasonable agreement with both the trend and magnitude of the polarisation transfer data \cite{jones,roygayou,gayou}.  NB.\ Neither this nor the Rosenbluth\,\cite{walker} data played any role in developing the Faddeev equation and nucleon current.  The agreement between calculation and experiment therefore yields clear understanding.  (See footnote \ref{fnrosenbluth}.)

\begin{figure}[t]
\centerline{%
\includegraphics[clip,height=0.5\textwidth]{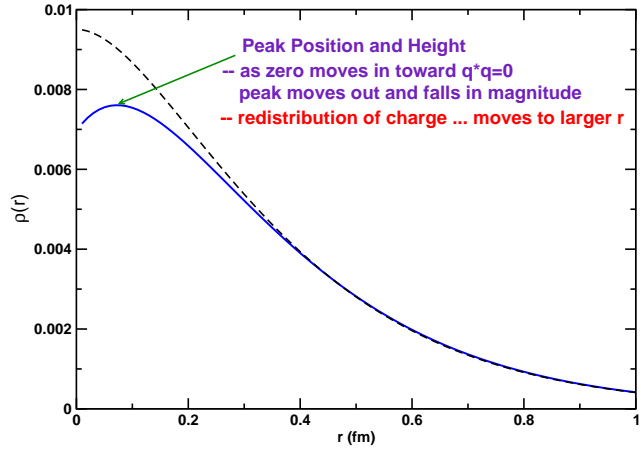}}
\caption{\label{GEzero} Three dimensional Fourier transform of a dipole charge distribution compared with that of a distribution which exhibits a zero; viz., $(1- a Q^2/m_D^2)/(1+Q^2/m_D^2)^3$, where $m_D$ is the dipole mass and $a=0.1$.  The zero indicates a depletion of charge at $r=0$, and its relocation to larger $r$.  This profile is familiar from nuclear physics: correlations in the proton wave functions of nuclei effect a similar redistribution of charge \protect\cite{wiringa}.}
\end{figure}

The calculation is extended to larger $Q^2$ in Ref.\,\cite{arneJ2}, with a prediction that the ratio will pass through zero at $Q^2\approx 6.5\,$GeV$^2$.  An experiment is planned at JLab that will acquire data on this ratio to $Q^2=9.0\,$GeV$^2$ \cite{E04108}.  It is expected to begin running around the beginning of 2008.  If one adheres to a simple interpretation of $G_E^p(Q^2)$ as a Fourier-transform of the electric charge distribution within a proton, then the meaning of a zero in this form factor is depicted in Fig.\,\ref{GEzero}: it corresponds to a depletion of electric charge at the heart of the nucleon.  The distribution of magnetic current exhibits no such effect.  While this picture is not truly valid because the zero appears far into the relativistic domain, the parallel it draws between the effects of correlations in the wave functions of nuclei and those within the nucleon's Faddeev amplitude are useful.

\begin{figure}[t]
\centerline{
\includegraphics[width=0.63\textwidth,angle=270]{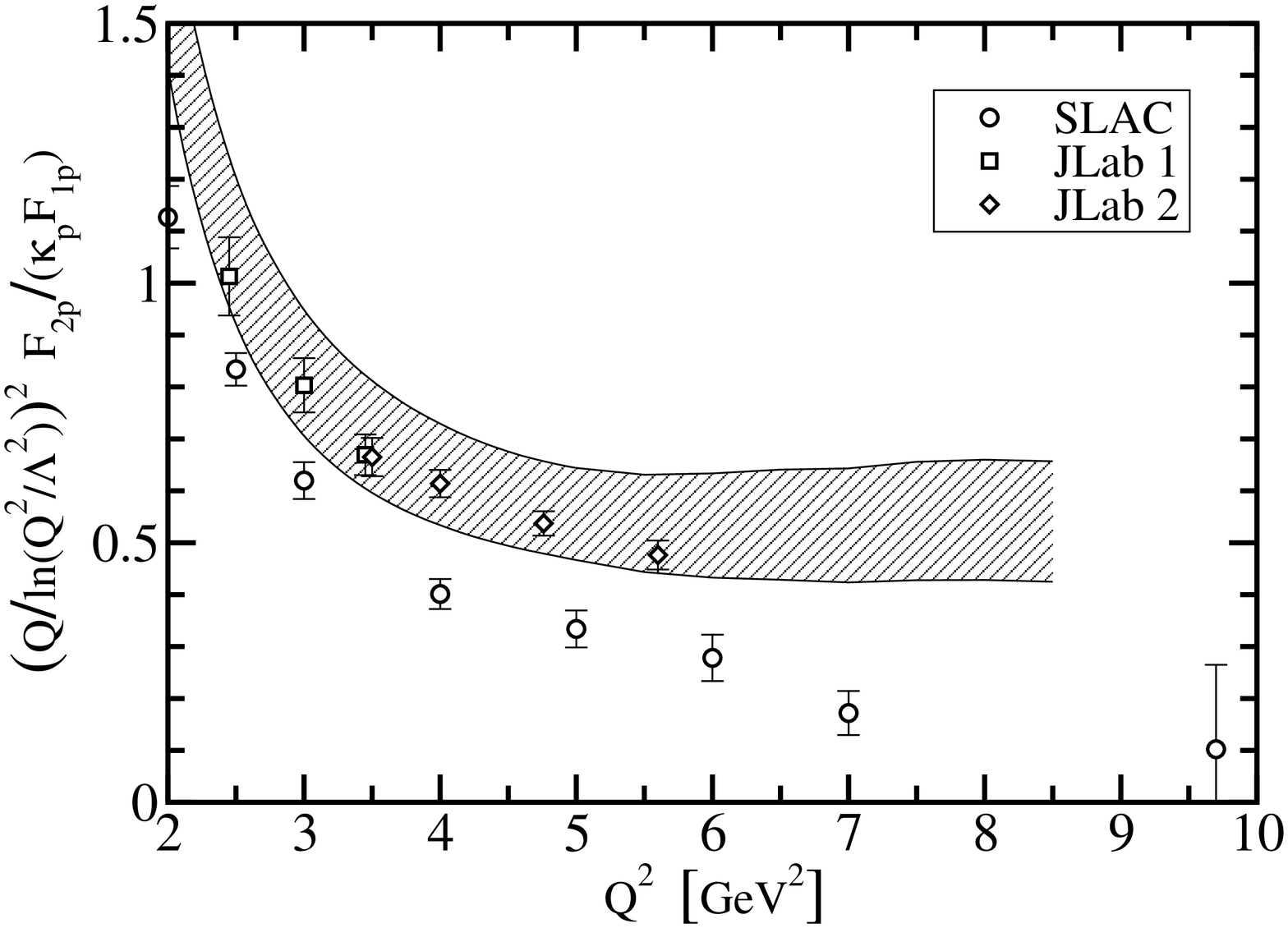}}
\caption{\label{plotF2F1log}  Weighted proton Pauli$/$Dirac form factor ratio, calculated with $\Lambda= 0.94\,$GeV.  The data are described in Fig.\,\protect\ref{plotGEpGMp}.  The band was calculated using the point-particle values: $\mu_{1^+}=2$ and $\chi_{1^+}=1$, and $\kappa_{\check T} = 2$.  Here the upper border is obtained with $\mu_{1^+}=1$, $\chi_{1^+}=1$ and $\kappa_{\check T}= 2$, and the lower with $\mu_{1^+}=3$.  (Adapted from Ref.\,\protect\cite{arneJ2}.)}
\end{figure}

Figure~\ref{plotF2F1log} depicts a weighted ratio of Pauli and Dirac form factors.  A perturbative QCD analysis\,\cite{belitsky} that considers effects arising from both the proton's leading- and subleading-twist light-cone wave functions, the latter of which represents quarks with one unit of orbital angular momentum, suggests
\begin{equation}
\label{scaling}
\frac{Q^2}{[\ln Q^2/\Lambda^2]^2} \, \frac{F_2(Q^2)}{F_1(Q^2)} =\,{\rm constant,}\;\; Q^2\gg \Lambda^2\,,
\end{equation}  
where $\Lambda$ is a mass-scale that corresponds to an upper-bound on the domain of soft momenta.  An argument may be made that a judicious estimate of the least-upper-bound on this domain is\,\cite{arneJ} $\Lambda = M$.  The figure hints that Eq.\,(\ref{scaling}) may be valid for $Q^2 \gsim 6\,$GeV$^2$.  NB.\ The model for the nucleon and its current reviewed herein is consistent with quark counting rules, albeit neglecting the anomalous dimensions that arise via renormalisation.  However, they were also omitted in deriving Eq.\,(\ref{scaling}).

It is worth reiterating that orbital angular momentum is not a Poincar\'e invariant.  However, if absent in a particular frame, it will in general appear in another frame related via a Poincar\'e transformation.  Nonzero quark orbital angular momentum is a necessary outcome of a Poincar\'e covariant description, and a nucleon's covariant Faddeev amplitude possesses structures that correspond in the rest frame to $s$-wave, $p$-wave and even $d$-wave components.  The result in Fig.\,\ref{plotF2F1log} is not significantly influenced by details of the diquarks' electromagnetic properties.  Instead, the behaviour is primarily governed by correlations expressed in the proton's Faddeev amplitude and, in particular, by the amount of intrinsic quark orbital angular momentum \cite{blochff}.  NB.\ This phenomenon is analogous to that observed in connection with the pion's electromagnetic form factor.  In that instance the so-called axial-vector components of the pion's Bethe-Salpeter amplitude, Eqs.\,(\ref{fwti}) \& (\ref{gwti}), are responsible for the large $Q^2$ behaviour of the form factor: they alone ensure $Q^2 F_\pi(Q^2) \approx \,$constant for truly ultraviolet momenta \cite{farrar,mrpion}, which is the result anticipated from perturbative QCD \cite{pQCDpionFF,pQCDpionFF1}.  These components are required by covariance\,\cite{mrt98} and signal the presence of quark orbital angular momentum in the pseudoscalar pion.

\subsection{Nucleon weak and strong form factors}
\index{Weak and strong form factors of the nucleon}
\label{FFweakstrong}
This framework can naturally be applied to calculate weak and strong form factors of the nucleon.  Preliminary studies of this type are reported in Refs.\,\cite{oettel2,jacquesmyriad}.  Such form factors are sensitive to different aspects of quark-nuclear physics and should prove useful, e.g., in constraining coupled-channel models for medium-energy production reactions on the nucleon.  This is important to the search for the so-called \textit{missing nucleon resonances}\footnote{This refers to the fact that numerous excited states of the nucleon ($N^*$), which are predicted by the $SU(6)\otimes O(3)$ constituent quark model ($CQM_{6\times 3}$), are not seen in the baryon spectrum that is obtained from amplitude analyses of $\pi N$ elastic scattering.  At this time more than half of the low-lying states predicted by $CQM_{6\times 3}$ are missing \cite{leeburkert}.}  and the related problem of identifying exotic baryons.\footnote{The ``pentaquark'', with a putative valence-quark structure $uudd\bar s$, would be an exotic baryon.  Confirmation of its existence would likely be a great fillip for hadron physics.  A perspective on the current status of pentaquark searches is provided in Ref.\,\protect\cite{schumacher}.}

For the purpose of illustration of the methods and questions it is worth describing briefly the results currently available for three such form factors: the axial-vector and pseudoscalar nucleon form factors, which appear in the axial-vector--nucleon current
\begin{eqnarray}
\nonumber
\lefteqn{J_{5\mu}^j(P^\prime,P) = i \bar u(P^\prime) \frac{\tau^j}{2} \Lambda_{5\mu}(q;P) u(P)}\\
& =&  \bar u(P^\prime) \gamma_5 \frac{\tau^j}{2} \left[ \gamma_\mu \, g_A(q^2) + q_\mu\, g_P(q^2) \right] u(P)\,,
\label{5muNN}
\end{eqnarray}
where $q=P^\prime - P$, $j=1,2,3$ is the isospin index, and the nucleon spinor, $u(P)$, is defined in Eq.\,(\ref{DiracN}); and the pion-nucleon coupling
\begin{equation}
\label{Jpi}
J^j_\pi(P^\prime,P) = \bar u(P^\prime) \Lambda_{\pi}^j(q;P) u(P)
= g_{\pi NN}(q^2) \bar u(P^\prime) i \gamma_5 \tau^j u(P)\,.
\end{equation} 

\begin{figure}[t]
\centerline{\includegraphics*[width=0.55\textwidth,angle=-90]{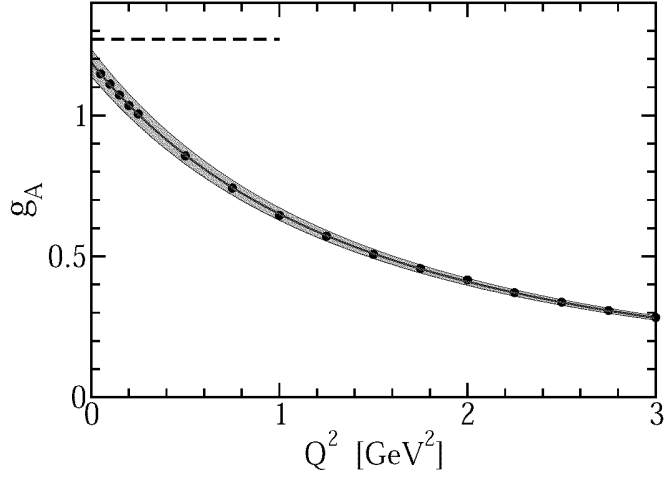}}
\caption{\label{F1} \textit{Filled circles}: $g_A(Q^2)$ in Eq.\,(\protect\ref{5muNN}) calculated in the chiral limit using the nucleon Faddeev amplitudes and the axial-vector-nucleon vertex obtained from Eqs.\,(\protect\ref{avansatz}), (\protect\ref{vA1}) \& (\protect\ref{vA01}).  \textit{Solid line}: dipole fit to the calculation, with mass-scale $m_D^A=1.69\,$GeV.  The shaded band delimits the result's variation subject to $10\,$\% changes in the parameter values in Eq.\,(\protect\ref{paramsK}).  The experimental value of the nucleon's axial coupling ($g_A \approx 1.27$) is marked by a dashed line.  (Adapted from Ref.\,\protect\cite{morelia}.)}
\end{figure}

In the chiral limit the pseudovector vertex of Eq.\,(\ref{5muNN}) takes the following form in the neighbourhood of $q^2=0$ \cite{mrt98}
\begin{equation}
\Lambda^j_{5\mu}(q;P) \stackrel{q^2\sim 0}{=} \mbox{regular}\; + \frac{q_\mu}{q^2} f_\pi \Lambda_\pi^j(q;P)\,,
\end{equation}
where $\Lambda_\pi^j(q;P)$ is the pion-nucleon vertex and ``regular'' denotes non-pole terms.  In addition, $q_\mu J_{5\mu}^j(P^\prime,P) =0$.  From these observations ensues the Goldberger-Treiman relation: \index{Goldberger-Treiman relation: nucleon}
\begin{equation}
M \, g_A(q^2=0) = f_\pi \,g_{\pi NN}(q^2=0)\,,
\end{equation}
where $M$ is the calculated nucleon mass and $g_A(q^2)$ is solely associated with the regular part of the axial-vector vertex.

\begin{figure}[t]
\begin{center}
\begin{minipage}[c]{0.7\textwidth}
\hspace*{-5ex}
\includegraphics*[clip,width=0.95\textwidth,angle=-90]{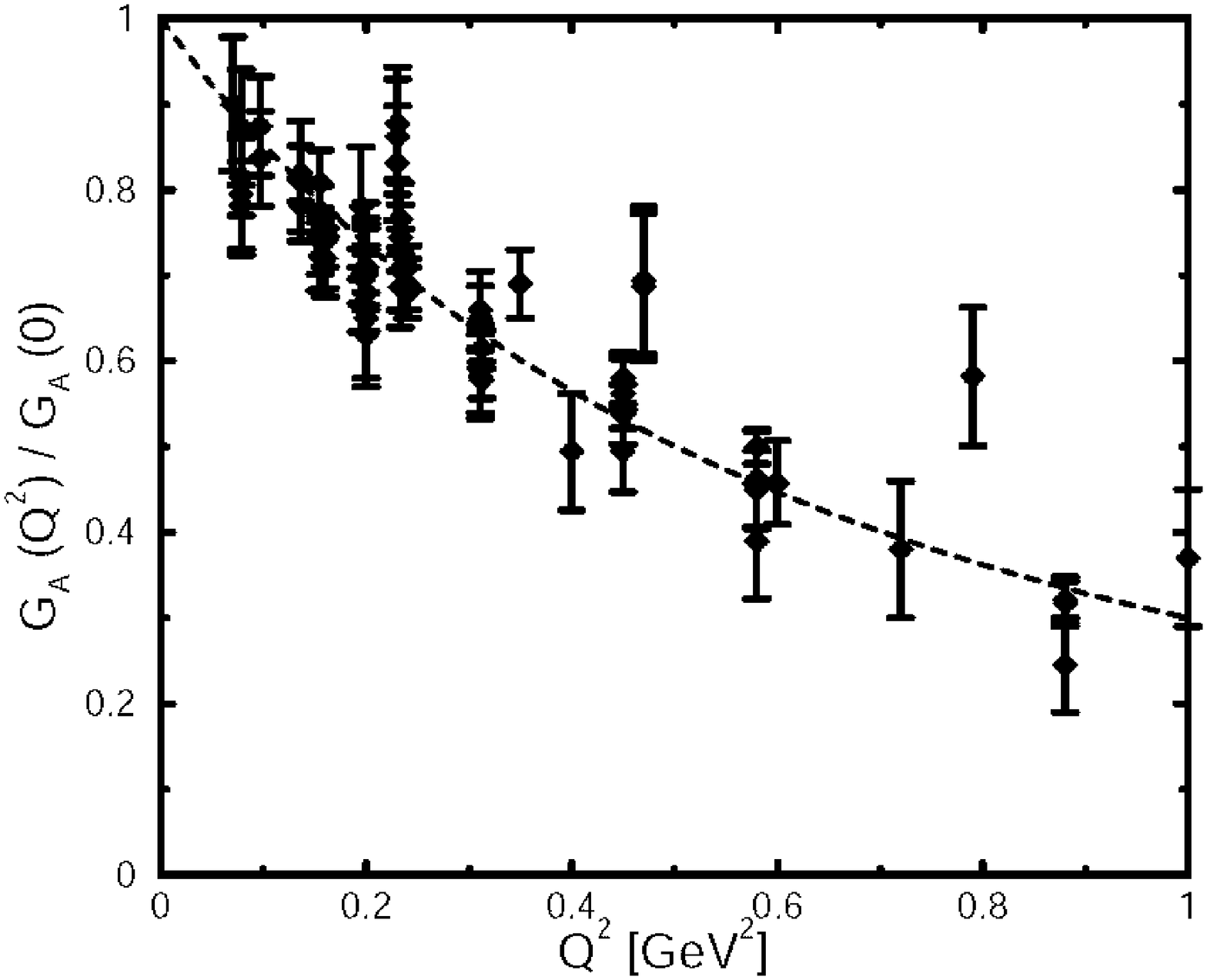}\\[-58ex]
\hspace*{-1.5ex}
\includegraphics*[width=1.01\textwidth,angle=-90]{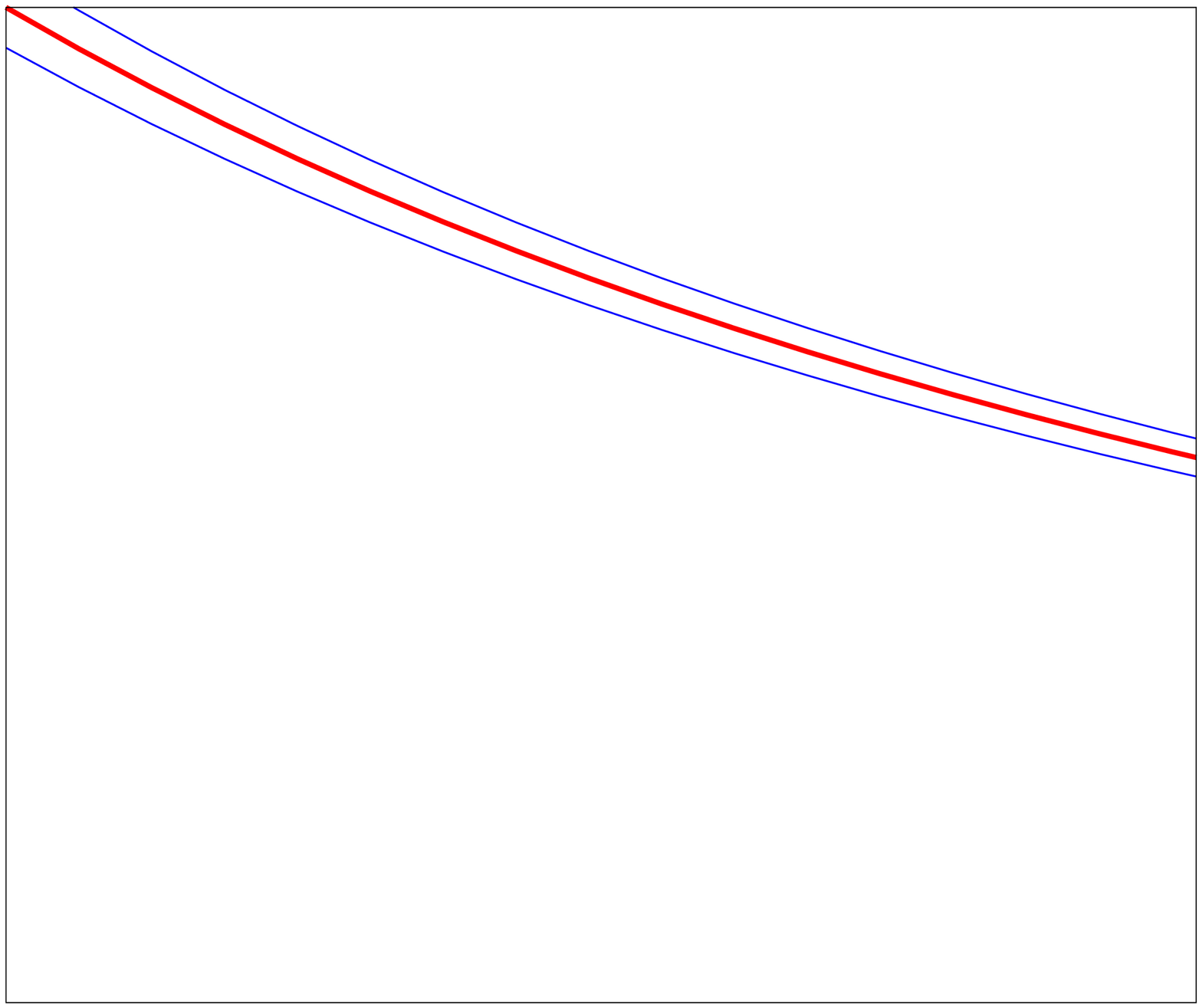}
\end{minipage}
\end{center}
\caption{\label{F2} Calculated chiral-limit result for $g_A(Q^2)/g_A(0)$, \textit{solid line}, compared with data obtained via pion electroproduction in the threshold region, as described in Ref.\,\protect\cite{bernard}.  The band delimits the result's variation subject to $10\,$\% changes in the parameter values in Eq.\,(\protect\ref{paramsK}).  \textit{Dashed line}: dipole fit to data with mass-scale $m_D^{AE}=1.1\,$GeV.  (Adapted from Ref.\,\protect\cite{morelia}.)}
\end{figure}

The calculation of electromagnetic form factors outlined in Sect.\,\ref{nucleonform} sets a pattern for determining $g_A(q^2)$, $g_P(q^2)$ and $g_{\pi NN}(q^2)$.  To follow this one needs to know how a dressed-quark couples to an axial-vector probe.  In the chiral limit the dressed-quark--axial-vector vertex satisfies Eq.\,(\ref{avwtim}) with the mass-dependent term omitted.  Subject to the assumption of isospin symmetry, $S_u=S_d$, in which case the chiral-limit \index{Ward-Takahashi identity: axial-vector} axial-vector Ward-Takahashi identity is solved by, recall Eq.\,(\ref{DeltaF}), 
\begin{eqnarray}
\nonumber \lefteqn{\Gamma_{5\mu}^j(k;Q) = \gamma_5 \frac{\tau^j}{2} \bigg[ \gamma_\mu \Sigma_A(k^2_+,k_-^2)  }  \\
&+ & 2 k_\mu \gamma\cdot k \Delta_A(k^2_+,k_-^2) + 2\,i\, \frac{Q_\mu}{Q^2}\Sigma_B(k^2_+,k_-^2)   \bigg].
\label{avansatz}
\end{eqnarray}
Naturally, Eq.\,(\ref{avansatz}) is not a unique \textit{Ansatz} for the dressed-quark--axial-vector vertex but it is an adequate starting point.  

\begin{figure}[t]
\centerline{\includegraphics*[width=0.60\textwidth,angle=-90]{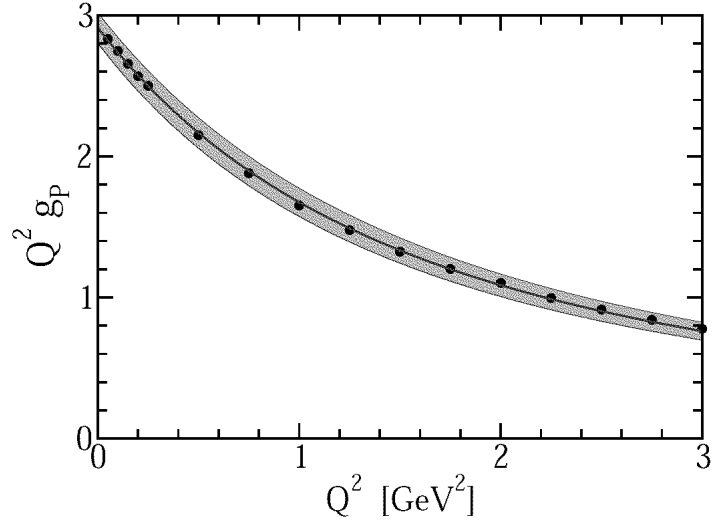}}
\caption{\label{F3} \textit{Filled circles}: Chiral limit result for $Q^2 g_P(Q^2)$ in Eq.\,(\protect\ref{5muNN}) calculated as described in the caption of Fig.\,\protect\ref{F1}.  \textit{Solid line}: dipole fit to the calculation, with mass-scale $m_D^P=1.77\,$GeV.  The shaded band delimits the result's variation subject to $10\,$\% changes in the parameter values in Eq.\,(\protect\ref{paramsK}).  (Adapted from Ref.\,\protect\cite{morelia}.)}
\end{figure}

For the pion-nucleon coupling, one needs the pion's Bethe-Salpeter amplitude and a definition for its extension off pion mass-shell.  In chiral QCD one has Eqs.\,(\ref{bwti}), upon which one can base the \textit{Ansatz}
\begin{equation}
\label{piBSA}
\Gamma_\pi^j(k;Q) =  i \gamma_5 \tau^j \, \sfrac{1}{{\cal N}_\pi}\, \Sigma_B(k_+^2,k_-^2)\,,
\end{equation}
where ${\cal N}_\pi$ is the canonical normalisation constant calculated with this amplitude via the analogue of Eq.\,(\ref{BSEnorm}) (see footnotes~\ref{fnnorm}, \ref{fnnorm2}).

\begin{figure}[t]
\centerline{\includegraphics*[width=0.6\textwidth,angle=-90]{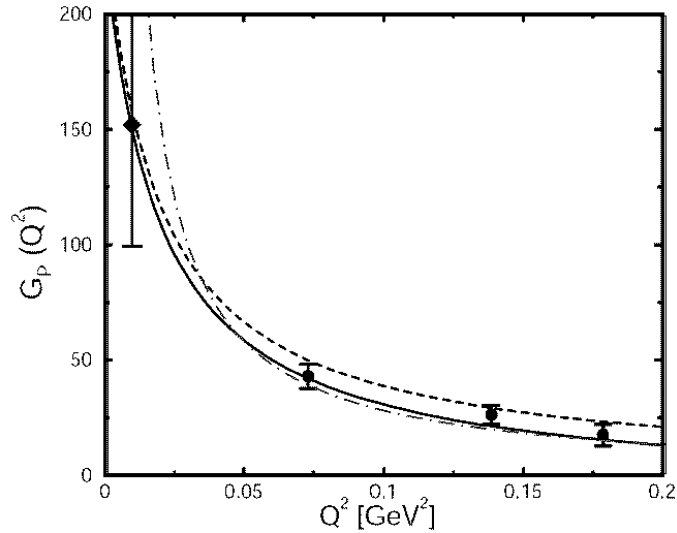}}
%
\caption{\label{F4} Chiral-limit result for $g_P(Q^2)$, \textit{dash-dot curve}.  Data obtained via pion electroproduction (\textit{filled circles}) \protect\cite{choi} and world average for muon capture at $Q^2=0.88 m_\mu^2$ (\textit{filled diamond}).  \textit{Dashed curve} -- current-algebra result; and  \textit{solid curve} -- next-to-leading order chiral
perturbation theory result \protect\cite{bernard}.  (Adapted from Ref.\,\protect\cite{morelia}.)}
\end{figure}

One also needs to know the following vertices: pion--axial-vector-diquark;  axial-vector-probe--axial-vector-diquark; and the pion- and axial-vector-probe-induced scalar-diquark $\leftrightarrow$ axial-vector-diquark transitions.  For these one may follow Ref.\,\cite{oettel2}:
\begin{eqnarray}
\Gamma^{\pi 1}_{\alpha\beta}(p^\prime,p) & = & \frac{\kappa^{\pi 1}}{2 M_N} \, \frac{M_Q^E}{f_\pi}\, \epsilon_{\alpha\beta\mu\nu} (p^\prime+p)_\mu Q_\nu\,, 
\label{pi1}\\
\Gamma^{A 1}_{\mu\alpha\beta}(p^\prime,p) & = & \sfrac{1}{2} \kappa^{A 1} \, \epsilon_{\mu\alpha\beta\nu} (p^\prime+p)_\nu + 2 f_\pi \frac{Q_\mu}{Q^2} \, \Gamma^{\pi 1}_{\alpha\beta}(p^\prime,p)\,,\label{vA1}\\
\Gamma^{\pi 01}_\beta(p^\prime,p) & = & -i \kappa^{\pi 01} \, \frac{M_Q^E}{f_\pi}\, Q_\beta\,,\label{pi01}\\
\Gamma_{\mu\beta}^{A 01}(p^\prime,p) & = & i M_N \kappa^{A 01} \delta_{\mu\beta} + 2 f_\pi \frac{Q_\mu}{Q^2} \, \Gamma^{\pi 01}_{\beta}(p^\prime,p)\,,
\label{vA01}
\end{eqnarray}
where $M_Q^E$ is the Euclidean light-quark constituent-mass, Eq.\,(\ref{MEq}), $p$ \& $p^\prime$ are the incoming and outgoing diquark momenta and $Q=(p^\prime-p)$.  Each  \textit{Ansatz} introduces one parameter, for which typical values are \cite{oettel2}:
\begin{equation}
\label{paramsK}
\kappa^{\pi 1} \simeq  \kappa^{A 1} \simeq 4.5\,,\; \kappa^{\pi 01} \simeq 3.9\,, \; \kappa^{A 01} \simeq 2.1\,.
\end{equation}
These parameters were used to obtain the results recapitulated herein, with the bands representing a variation of $\pm 10\,$\%.  NB.\ A scalar diquark does not couple to a single pseudoscalar or axial-vector probe.

With the elements heretofore described one has an analogue for the nucleon's chiral current which corresponds to the top four diagrams in Fig.\,\ref{vertex}.  This is necessary but not sufficient to guarantee that the axial-vector--nucleon vertex automatically fulfills the chiral Ward-Takahashi identity for on-shell nucleons.  This situation must be improved and work on that is underway.

Figure~\ref{F1} displays the result for the nucleon's axial-vector form factor.  A comparison with extant data is provided in Fig.\,\ref{F2}.  The mismatch can be attributed to a failure of the axial-vector-nucleon vertex obtained from Eqs.\,(\protect\ref{avansatz}), (\protect\ref{vA1}) \& (\protect\ref{vA01}) to properly express the diquarks' nonpointlike nature: the momentum evolution thus describes a distribution of axial-charge that is too hard.

In Fig.\,\ref{F3} the result for the nucleon's induced-pseudoscalar form factor is depicted.  A comparison with data is provided in Fig.\,\ref{F4}.  The form factor is dominated by the pion pole in the neighbourhood of $q^2= -m_\pi^2$, which for this chiral-limit calculation is $q^2\sim 0$.  The comparison with data is more favourable in this instance, particularly once one allows for a shift of the pion pole to $q^2=0$ in the chiral-limit calculation.  This can be attributed to Eqs.\,(\ref{piBSA}), (\ref{pi1}) \& (\ref{pi01}); viz., as it is based on Eqs.\,(\ref{bwti}), the calculation incorporates a fairly accurate representation of pion structure and the pion nucleon coupling.

\begin{figure}[t]
\centerline{\includegraphics*[width=0.6\textwidth,angle=-90]{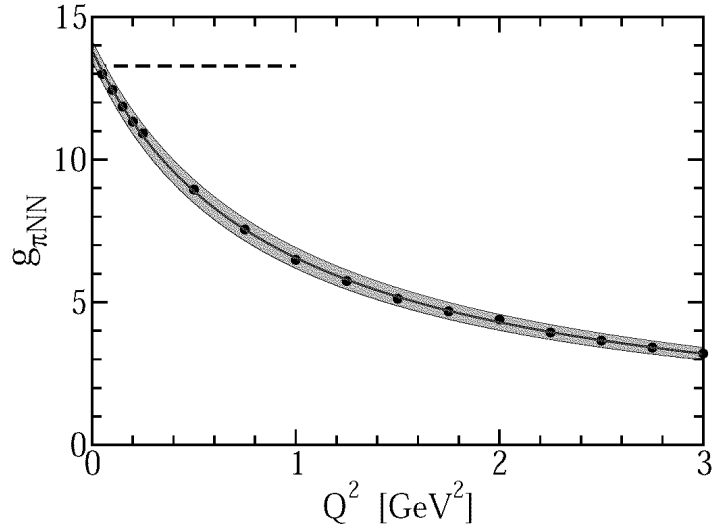}}
\caption{\label{F5} \textit{Filled circles}: Chiral limit result for $g_{\pi NN}(Q^2)$ in Eq.\,(\protect\ref{Jpi}) calculated using the nucleon's Faddeev amplitude and the $\pi NN$ vertex constructed from Eqs.\,(\ref{piBSA}), (\ref{pi1}) \& (\ref{pi01}).  \textit{Solid line}: monopole fit to the calculation, with mass-scale $m_M^\pi=0.95\,$GeV.  The shaded band delimits the result's variation subject to $10\,$\% changes in the parameter values in Eq.\,(\protect\ref{paramsK}).  The experimental value of the $\pi NN$ coupling ($g_{\pi NN} \approx 13.4$) is marked by a dashed line.}
\end{figure}

This view is supported by the result for $g_{\pi NN}(q^2)$, which is depicted in  Fig.\,\ref{F5}.  Within reasonable variation of the parameters that characterise the pion-nucleon vertex, the calculated value of $g_{\pi NN}^0(0)$ is consistent with standard phenomenology.  The result yields a chiral-limit value $r_{\pi NN}^0 \simeq 0.51\pm 0.02 \,$fm.  For comparison, a massive-quark value of $r_{\pi NN}\sim 0.3\,$fm appears in Ref.\,\protect\cite{machleidt}, while $r_{\pi NN}\sim 0.93\,$--$\,1.06\,$fm is employed in Ref.\,\cite{lee1}. 

In order to improve upon these preliminary results, construction must be completed of an axial-vector--nucleon vertex that automatically fulfills the chiral Ward-Takahashi identity for on-shell nucleons described by the solution of the Faddeev equation based on confined-quark and -nonpointlike-diquark degrees of freedom.  This will subsequently lead to an improved pion-nucleon vertex.  In addition, as is known to be necessary for an accurate description of nucleon electromagnetic properties, Sect.\,\ref{chiralem}, the effect of pseudoscalar meson loops on the axial-vector and pseudoscalar couplings must be incorporated.  These steps are prerequisites for a veracious description of weak and pionic processes.

\section{Epilogue}
\label{epilogue}
\setcounter{equation}{0}
Protons and neutrons are the seeds of all the universe's observable matter.  The standard model of particle physics is supposed to explain their properties.  However, this theory's perturbative formulation fails spectacularly to account for even the simplest bulk properties.  Two fundamental, emergent phenomena are responsible: confinement and dynamical chiral symmetry breaking.  Their importance is difficult to overestimate.  For example, they determine which chemical elements are stable and hence influence even the existence of life.  

Dynamical chiral symmetry breaking (DCSB) is a singularly effective mass generating mechanism.  It can take the almost massless light-quarks of perturbative QCD and turn them into the massive constituent-quarks whose mass sets the scale which characterises the spectrum of the strong interaction accessible at modern experimental facilities.  The phenomenon is understood via QCD's gap equation, the solution of which delivers a quark mass function with a momentum-dependence that connects the perturbative and nonperturbative, constituent-quark domains.  

Despite the fact that light-quarks are made heavy, the mass of the pseudoscalar mesons remains unnaturally small.  That, too, owes to DCSB, expressed this time in a remarkable relationship between QCD's gap equation and those colour singlet Bethe-Salpeter equations which have a pseudoscalar projection.  Goldstone's theorem is a natural consequence of this connection.

These features may only be veraciously understood in relativistic quantum field theory.  They can be viewed as an essential consequence of the presence and role of particle-antiparticle pairs in an asymptotically free theory.  This is apparent via a self-consistent solution of the appropriate Dyson-Schwinger equations (DSEs), which wrap each of QCD's elementary excitations in a cloud of virtual particles that is exceedingly dense at low momentum.

Indeed, the DSEs provide a natural framework for the exploration of QCD's emergent phenomena.  They are a generating tool for perturbation theory and thus give a clean connection with processes that are well understood.  Moreover, they admit a systematic, symmetry preserving and nonperturbative truncation scheme, and thereby give access to 
strong QCD in the continuum.  On top of this, a quantitative comparison and feedback between DSE and lattice-QCD studies is today proving fruitful.

The existence of a sensible truncation scheme enables the proof of exact results using the DSEs.  That the truncation scheme is also tractable provides a means by which the results may be illustrated, and furthermore a practical tool for the prediction of observables that are accessible at contemporary experimental facilities.  The consequent opportunities for rapid feedback between experiment and theory brings within reach an intuitive understanding of nonperturbative strong interaction phenomena.

Modern, high-luminosity experimental facilities employ large momentum transfer reactions to probe the structure of hadrons.  They are providing remarkable and intriguing new information.  For an example one need only look so far as the discrepancy between the ratio of electromagnetic proton form factors extracted via Rosenbluth separation and that inferred from polarisation transfer.  This discrepancy is marked for $Q^2\gsim 2\,$GeV$^2$ and grows with increasing $Q^2$.  At such values of momentum transfer, $Q^2 > M^2$, where $M$ is the nucleon's mass, a true understanding of these and other contemporary data require a Poincar\'e covariant description of the nucleon.  This can be obtained with a Faddeev equation that describes a baryon as composed primarily of a quark core, constituted of confined quark and confined diquark correlations, but augmented by pseudoscalar meson cloud contributions that are sensed by long wavelength probes.  Short wavelength probes pierce the cloud, and expose spin-isospin correlations and quark orbital angular momentum within the baryon.  The veracity of the elements in this description makes plain that a picture of baryons as a bag of three constituent-quarks in relative $s$-waves is profoundly misleading.

While there are indications that confinement may be expressed in the analyticity properties of the dressed propagators, one cannot say that it is understood.  Consequently, one pressing task, to which the methods described herein can be applied, is the drawing of an accurate map of the confinement force between light-quarks within mesons.  This will enable a clear connection to be established between this force and the realisation of dynamical chiral symmetry breaking, and an accounting of the distribution of mass within mesons.  In addition one may then begin to determine whether the confinement force-field can be excited to produce exotic systems of light-quarks and glue.  The same must be done for baryons, but here the difficulties are greater because one is confronted with a Poincar\'e covariant three-body problem.  New and improved tools must be developed, which may extend beyond a rigorous grounding of the Faddeev equation.  

It is also crucial to develop the tools necessary for charting the pointwise distribution of quarks and gluons within hadrons.  With such in hand, one might lay out and apportion the pointwise distribution of mass within the hadron, and its evolution with the resolving scale of the probe.  That knowledge could be used to elucidate the impact of confinement and DCSB on these distributions; e.g., by exhibiting the effects of quenching these emergent phenomena.  

It should now be plain that modern nuclear physics poses numerous challenges and may reasonably be expected to provide many new surprises.

\section*{Acknowledgments}
Conversations with the following people were of benefit in preparing this article: L.~Chang, P.~Jaikumar, A.~Krassnigg, Y.-X.~Liu, P.~Maris, G.\,A.~Miller, D.~Nicmorus, S.\,M.~Schmidt and P.\,C.~Tandy.
This work is supported by: 
Department of Energy, Office of Nuclear Physics, contract no.\ W-31-109-ENG-38; 
%
%
%
and benefited from the facilities of ANL's Computing Resource Center.

\appendix
\renewcommand{\theequation}{\Alph{section}.\arabic{equation}}
\setcounter{equation}{0}
\section{Euclidean Space}
\index{Euclidean metric}
\label{Appendix1}
It is possible to view the Euclidean formulation of a quantum field theory as  definitive \cite{glimm,symanzik,streater,seiler}.  That decision is crucial when a consideration of nonperturbative effects becomes important.  In addition, the discrete lattice formulation in Euclidean space has allowed some progress to be made in attempting to answer existence questions for interacting gauge field theories.  NB.\ A lattice formulation is impossible in Minkowski space because the integrand is not non-negative and hence does not provide a probability measure.  An heuristic exposition of probability measures in quantum field theory can be found in Ref.\,\cite{rivers}, Chap.\,6, while Ref.\,\cite{glimm}, Chaps.\,3 and 6, provides a more rigorous discussion in the context of quantum mechanics and quantum field theory.

Our Euclidean conventions are easily made plain.  For $4$-vectors $a$, $b$:
\begin{equation}
a\cdot b := a_\mu\,b_\nu\,\delta_{\mu\nu} := \sum_{i=1}^4\,a_i\,b_i\,,
\end{equation}
where $\delta_{\mu\nu}$ is the Kronecker delta and the metric tensor.  Hence, a spacelike vector, $Q_\mu$, has $Q^2>0$.  The Dirac matrices are Hermitian and defined by the algebra 
\begin{equation}
\{\gamma_\mu,\gamma_\nu\} = 2\,\delta_{\mu\nu}\,.
\end{equation}
We use 
\begin{equation}
\gamma_5 := -\,\gamma_1\gamma_2\gamma_3\gamma_4\,,
\end{equation}
so that 
\begin{equation}
{\rm tr}\left[ \gamma_5 \gamma_\mu\gamma_\nu\gamma_\rho\gamma_\sigma \right] = 
- 4 \,\varepsilon_{\mu\nu\rho\sigma}\,,\; \varepsilon_{1234}= 1\,.
\end{equation}

A Dirac-like representation of these matrices is: 
\begin{equation} 
\vec{\gamma}=\left( 
\begin{array}{cc} 
0 & -i\vec{\tau}  \\ 
i\vec{\tau} & 0 
\end{array} 
\right),\; 
\gamma_4=\left( 
\begin{array}{cc} 
\tau^0 & 0 \\ 
0 & -\tau^0 
\end{array} 
\right), 
\end{equation} 
where the $2\times 2$ Pauli matrices are: 
\begin{equation} 
\label{PauliMs} 
\rule{-4ex}{0ex}
\tau^0 = \left( 
\begin{array}{cc} 
1 & 0 \\ 
0 & 1 
\end{array}\right),\; 
\tau^1 = \left( 
\begin{array}{cc} 
0 & 1 \\ 
1 & 0 
\end{array}\right),\; 
\tau^2 = \left( 
\begin{array}{cc} 
0 & -i \\ 
i & 0 
\end{array}\right),\; 
\tau^3 = \left( 
\begin{array}{cc} 
1 & 0 \\ 
0 & -1 
\end{array}\right). 
\end{equation} 

It is possible to derive the Euclidean version of every equation introduced above assuming certain analytic properties of the integrands.  However, the derivations can be sidestepped using the following \textit{transcription rules}:
\begin{center} 
\parbox{30em}{ 
\parbox{16em}{Configuration Space 
\begin{enumerate} 
\item $\displaystyle \int^M \!d^4x^M \, \rightarrow \,-i \int^E \!d^4x^E$ 
\item $\slash\!\!\! \partial \,\rightarrow \, i\gamma^E\cdot \partial^E $ 
\item $\slash \!\!\!\! A \, \rightarrow\, -i\gamma^E\cdot A^E$ 
\item $A_\mu B^\mu\,\rightarrow\,-A^E\cdot B^E$ 
\item $x^\mu\partial_\mu \to x^E\cdot \partial^E$ 
\end{enumerate}}\hspace*{0.5em} 
\parbox{16em}{Momentum Space 
\begin{enumerate} 
\item $\displaystyle \int^M\! d^4k^M \, \rightarrow \,i \int^E\! d^4k^E$ 
\item $\slash\!\!\! k \,\rightarrow \, -i\gamma^E\cdot k^E $ 
\item $\slash \!\!\!\! A \, \rightarrow\, -i\gamma^E\cdot A^E$ 
\item $k_\mu q^\mu \, \rightarrow\, - k^E\cdot q^E$ 
\item $k_\mu x^\mu\,\rightarrow\,-k^E\cdot x^E$ 
\end{enumerate}}} 
\end{center} 
These rules are valid in perturbation theory; i.e., the correct Euclidean space integral for a given diagram will be obtained by applying these rules to the Minkowski integral.  The rules take account of the change of variables and Wick rotation of the contour.  When one begins with Euclidean space, as we do, the reverse is also true.  However, for diagrams that represent DSEs which involve dressed $n$-point functions, whose analytic structure is not known \textit{a priori}, the Minkowski space equation obtained using this prescription will have the right appearance but it's solutions may bear no relation to the analytic continuation of the solution of the Euclidean equation.  Any such differences will be nonperturbative in origin.  It is this fact that makes a choice of metric crucial at the outset.

To return to practical matters, a positive energy spinor satisfies 
\begin{equation} 
\label{DiracN}
\bar u(P,s)\, (i \gamma\cdot P + M) = 0 = (i\gamma\cdot P + M)\, u(P,s)\,, 
\end{equation} 
where $M$ is the mass obtained by solving the Faddeev equation and $s=\pm$ is the spin label.  The spinor is normalised: 
\begin{equation} 
\bar u(P,s) \, u(P,s) = 2 M \,,
\end{equation} 
and may be expressed explicitly: 
\begin{equation} 
u(P,s) = \sqrt{M- i {\check{E}}}\left( 
\begin{array}{l} 
\chi_s\\ 
\displaystyle \frac{\vec{\sigma}\cdot \vec{P}}{M - i \check{E}} \chi_s 
\end{array} 
\right)\,, 
\end{equation} 
with $\check{E} = i \sqrt{\vec{P}^2 + M^2}$, 
\begin{equation} 
\chi_+ = \left( \begin{array}{c} 1 \\ 0  \end{array}\right)\,,\; 
\chi_- = \left( \begin{array}{c} 0\\ 1  \end{array}\right)\,. 
\end{equation} 
For the free-particle spinor, $\bar u(P,s)= u(P,s)^\dagger \gamma_4$. 
 
The spinor can be used to construct a positive energy projection operator: 
\begin{equation} 
\label{Lplus} \Lambda_+(P):= \frac{1}{2 M}\,\sum_{s=\pm} \, u(P,s) \, \bar 
u(P,s) = \frac{1}{2M} \left( -i \gamma\cdot P + M\right). 
\end{equation} 
 
A negative energy spinor satisfies 
\begin{equation} 
\bar v(P,s)\,(i\gamma\cdot P - M) = 0 = (i\gamma\cdot P - M) \, v(P,s)\,, 
\end{equation} 
and possesses properties and satisfies constraints obtained via obvious analogy 
with $u(P,s)$. 
 
A charge-conjugated Bethe-Salpeter amplitude is obtained via 
\begin{equation} 
\label{chargec}
\bar\Gamma(k;P) = C^\dagger \, \Gamma(-k;P)^{\rm T}\,C\,, 
\end{equation} 
where ``T'' denotes a transposing of all matrix indices and 
$C=\gamma_2\gamma_4$ is the charge conjugation matrix, $C^\dagger=-C$. 
 
In describing the $\Delta$ resonance we employ a Rarita-Schwinger spinor to 
unambiguously represent a covariant spin-$3/2$ field.  The positive energy 
spinor is defined by the following equations: 
\begin{equation} 
\label{rarita}
(i \gamma\cdot P + M)\, u_\mu(P;r) = 0\,,\;
\gamma_\mu u_\mu(P;r) = 0\,,\;
P_\mu u_\mu(P;r) = 0\,, 
\end{equation} 
where $r=-3/2,-1/2,1/2,3/2$.  It is normalised: 
\begin{equation} 
\bar u_{\mu}(P;r^\prime) \, u_\mu(P;r) = 2 M\,\delta^{r^\prime r}, 
\end{equation} 
and satisfies a completeness relation 
\begin{equation} 
\frac{1}{2 M}\sum_{r=-3/2}^{3/2} u_\mu(P;r)\,\bar u_\nu(P;r) = 
\Lambda_+(P)\,R_{\mu\nu}\,, 
\end{equation} 
where 
\begin{equation} 
R_{\mu\nu} = \delta_{\mu\nu} I_{\rm D} -\frac{1}{3} \gamma_\mu \gamma_\nu + 
\frac{2}{3} \hat P_\mu \hat P_\nu I_{\rm D} - i\frac{1}{3} [ \hat P_\mu 
\gamma_\nu - \hat P_\nu \gamma_\mu]\,, 
\end{equation} 
with $\hat P^2 = -1$, which is very useful in simplifying the positive energy 
$\Delta$'s Faddeev equation. 

%
%
%

%
%


\printindex

\end{document}